\documentclass[12pt,aps,prc,tightenlines,showpacs,
showkeys,superscriptaddress]{revtex4}
\usepackage{graphicx,color}

\def\simge{%
    \mathrel{\rlap{\raise 0.511ex
        \hbox{$>$}}{\lower 0.511ex \hbox{$\sim$}}}}
\def\simle{%
    \mathrel{\rlap{\raise 0.511ex
        \hbox{$<$}}{\lower 0.511ex \hbox{$\sim$}}}}
\def\lbar{\mathrel{\rlap{
\raise3pt\hbox{\hskip-1.75pt-}}
\hbox{$\lambda_e$}}}

\def\be{\begin{eqnarray}}
\def\ee{\end{eqnarray}}
\def\pmb#1{\setbox0=\hbox{$#1$}
\kern-.025em\copy0\kern-\wd0
\kern.05em\copy0\kern-\wd0
\kern-.025em\raise.0433em\box0}

\begin{document}
\title{Neutron Star Observations: \\ Prognosis for Equation of State
Constraints}

\author{James M. Lattimer}
\affiliation{Department of Physics and Astronomy, SUNY at Stony Brook,
Stony Brook, New York 11794-3800}

\author{Madappa Prakash}
\affiliation{Department of Physics and Astronomy, Ohio University,
Athens, OG 45701-2979}


\begin{abstract}
We investigate how current and proposed observations of neutron stars
can lead to an understanding of the state of their interiors and the
key unknowns: the typical neutron star radius and the neutron star
maximum mass.  We consider observations made not only with photons,
ranging from radio waves to X-rays, but also those involving neutrinos
and gravity waves.  We detail how precision determinations of
structural properties would lead to significant restrictions on the
poorly understood equation of state near and beyond the equilibrium
density of nuclear matter.  

To begin, a theoretical analysis of neutron star structure, including
general relativistic limits to mass, compactness, and spin rates is
made.  A review is the made of recent observations such as pulsar
timing (which leads to mass, spin period, glitch and moment of inertia
estimates), optical and X-ray observations of cooling neutron stars
(which lead to estimates of core temperatures and ages and inferences
about the internal composition), and X-ray observations of accreting
and bursting sources (which shed light on both the crustal properties
and internal composition).  Next, we discuss neutrino emission from
proto-neutron stars and how neutrino observations of a supernova, from
both current and planned detectors, might impact our knowledge of the
interiors, mass and radii of neutron stars.  We also explore the
question of how superstrong magnetic fields could affect the equation
of state and neutron star structure.  This is followed by a look at
binary mergers involving neutron stars and how the detection of
gravity waves could unambiguously distinguish normal neutron stars
from self-bound strange quark matter stars.  


\end{abstract}
\maketitle
\newpage
\setcounter{tocdepth}{3}
\tableofcontents
\newpage
\section{Introduction}

\setcounter{section}{1}
\setcounter{subsection}{0}

Neutron stars are the most compact known objects without event
horizons and therefore serve as extraordinary laboratories for dense
matter physics.  The internal composition of the cores of neutron stars is
currently poorly understood.  Most models of dense matter predict that
above twice the equilibrium density of nuclear matter,
$\rho_s\simeq2.7\times10^{14}$ g cm$^{-3}$ or $n_s\simeq0.16$ baryons
fm$^{-3}$, exotica in the form of hyperons, a Bose condensate of pions
or kaons, or deconfined quark matter, will eventually appear.
However, whether the threshold density for such exotica is around
twice $\rho_s$, or much larger, is unclear.

It has been suggested that if strange quark matter is the ultimate
ground state of matter (i.e., has a lower energy at zero pressure than
iron) \cite{Witten84}, compression of matter to sufficiently high density
would trigger a phase transition converting virtually the entire
star into strange quark matter.  Such a star is
self-bound as opposed to being gravitationaly bound as is the case of
a normal neutron star.

It has so far proved very difficult to find venues from astrophysical
observations that could unambiguously distinguish strange quark stars
from normal neutron stars.  This is because self-bound stars have
similar radii, moments of inertia, and neutrino emissivities and
opacities to that of moderate mass normal neutron stars.  Therefore,
it may be unlikely that photon or neutrino observations, or radio
binary pulsar timing measurements, will be able to differentiate these
cases, especially if strange quark stars have a small hadronic crust,
supported perhaps by electrostatic forces or a mixed phase
\cite{Alford06}.  In that case, the effective temperatures and radii
of solar-mass-sized strange quark stars and normal neutron stars would
tend to be similar.  Even during the proto-neutron star stage, which
is observable through neutrino emissions \cite{Burrows86}, these two
types of stellar configurations likely yield similar neutrino signals
until such late times that the low luminosities prevent an unambiguous
discrimination \cite{Prakash01}.  However, major differences in the
evolutions of normal neutron stars and strange quark matter stars
emerge during the final stages of binary mergers if stable mass
transfer occurs. These differences would be prominent in both the
amplitudes and frequencies of gravitational wave emisions.
Except for discussing this latter topic, we will not discuss
self-bound strange quark matter stars in detail further in this paper.

Possibly the two most important properties of neutron stars - their
maximum masses and typical radii -- are not yet well known. These
properties reflect rather different aspects of the dense matter
equation of state (EOS).  The neutron star maximum mass, which is a
consequence of general relativity and does not exist in Newtonian
gravity, has a limit of, at most, 3 M$_\odot$, assuming causality
\cite{RR74}.  The maximum mass is controlled by the stiffness of the
dense matter EOS at densities in excess of a few fimes times
$n_s$. The introduction of non-nucleonic degrees of freedom at
supra-nuclear densities generally implies a softening of the EOS.  The
largest precisely known neutron star mass is only 1.44 M$_\odot$,
which is too small to provide an effective constraint on the
composition of dense matter and the structural properties neutron
stars.  However, some recent mass measurements, detailed below, from
timing of pulsars in binaries with white dwarf companions suggest that
the maximum mass might be about 2 M$_\odot$ or even larger.  In
addition, large masses are also suggested from observations of QPO's,
or quasi-periodic oscillations, observed in X-ray emission from
neutron stars accreting from binary companions.  If verified, a
maximum mass in the vicinity of 2 M$_\odot$ would fundamentally alter
our perception of the properties of matter in neutron star cores.
As we demonstrate below, sufficiently large observed neutron star
masses set interesting upper limits to the densities possible in
neutron stars.  It could be the case that the maximum density allowed
is not large enough for exotic forms of matter to play significant
roles.

On the other hand, the neutron star radius is controlled by properties
of the nuclear force in the immediate vicinity of $n_s$, in particular
by the density dependence of the nuclear symmetry energy \cite{LP01}
(the symmetry energy is the difference, at a given density, between
the baryon energies of pure neutron matter and symmetric nucleonic
matter).  Measurements of the neutron star radius are far less precise
than mass measurements.  Among the observations recently utilized
include radius upper limits from rapidly rotating neutron stars,
estimates from the thermal emission of cooling neutron stars,
including redshifts, estimates obtained from the properties of sources
with bursts or thermonuclear explosions on their surfaces, and
estimates of crustal properties from (a) glitches of pulsars, (b)
``star-quakes'' occurring in the aftermath of giant flares from
soft-gamma ray repeaters (SGR's), and (c) cooling timescales during
periods of quiescence in-between X-ray bursts from accreting neutron
stars in low-mass X-ray binaries (LMXRB's).

Aside from direct mass and radius determinations, one of our best
windows into the interior is through observations of their thermal
properties.  Neutron star cooling curves ({\it i.e.}, their
luminosities or temperatures as a function of age) are sensitive to
the internal composition and the superfluid characteristics of its
components. The interior of the star cools through the emission of
neutrinos, which is very sensitive to composition and possible
superfluidity.  The crucial question is whether or not the neutrino
emission processes are {\it rapid} or {\it slow}, terms having
historical significance.  The surface temperature of the neutron star,
apart from an initial period of several to one hundred years, the
thermal timescale of the crust, is intimately coupled to that of the
core.  After the crust reaches thermal equilibrium with the
core, a rapidly cooling core would be revealed through relatively low
surface temperatures for a given age.  It is now realized
that the behavior of the symmetry energy of dense matter is crucial in
determining the relative rate of neutrino cooling.  This is because
the symmetry energy not only controls the rate of cooling for a
mixture of nucleons, but also determines at what densities exotic
material, such as hyperons, Bose condensates, or deconfined quark
matter appear, all of which could allow relatively rapid cooling even
if neutrino processes involving nucleons were slow.

There are other, less traditional, ways in which information about
neutron stars can be gleaned.  These involve neutrinos and gravity
waves.  Indeed, the fewer than two dozen neutrinos observed from SN
1987A are only the harbinger of the thousands expected in current
detectors, or millions from planned detectors, from a galactic
supernova.  These neutrinos are emitted in two phases.  The first is a
tremendous burst accompanying core bounce, with peak luminosities of
nearly 1000 bethes s$^{-1}$ (1 bethe $\equiv 10^{51}$ erg), exceeding
the photon luminosity of all the stars in the visible universe during
the same instant.  The second is a continuous flux, quickly ramping
down from the peak of the initial burst, from the deleptonizing
proto-neutron star that will remain visible for perhaps fifty
seconds.  Nearly all ($\sim$ 99\%) of the neutron star's binding
energy of about $300$ bethes is eventually radiated away in neutrinos
of all flavors .  The neutrino luminosities and the emission timescale
are controlled by several factors including the total mass of the
proto-neutron star and the opacity of neutrinos at supranuclear
density, which depend on the star's composition and the EOS of
strongly interacting dense matter.  The details of the neutrino light
curve and neutrino energy distributions could reveal details of their
internal compositions.

Last, but perhaps not least, prodiguous amounts of gravity waves are
expected to be emitted in the death throes of binary mergers
involving neutron stars.  Compact binaries, containing neutron stars
and/or black holes, continuously radiate gravitational waves which
cause their orbits to decay.  Observed neutron star binaries have
times-to-merger ranging from about 85 million years to tens of
billions of years.  Current estimates imply that there are about 1
merger per 10,000 years in our galaxy, and detectors such as the
advanced LIGO might observe 40--400 events per year
\cite{Kalogera04}.  The observed patterns of gravitational waves could
yield information about the masses of the components as well as
estimates of the radii of inspiralling neutron stars.  Furthermore, as
we discuss below, there should be a pronounced difference in the
observed wave pattern between mergers involving normal and self-bound
strange stars.  In contrast to other observational differences between
normal and strange stars, which could have multiple sources, the
differences observed in gravitational radiation would be relatively
unambiguous and therefore, possibly, unique.

Complementary to observational constraints on neutron star structure
are laboratory measurements which can restrict the range of nuclear
parameters such as the incompressibility, the symmetry energy and its
density dependence, nucleonic effective masses and nuclear specific
heats, the high-density pressure-density relation, and hyperon-nucleon
couplings.  We briefly review laboratory work involving dense matter
and its relation to neutron star structure and compositions (we
provide references to more detailed summaries).  This experimental
information ranges from measurements of nuclear masses (including
high-N and low-Z nuclei in Radioactive Isotope Accelerators), nuclear
charge radii, neutron skin thicknesses, frequencies of giant monopole
and dipole resonances, and collective flow, multifragmentation and
isospin diffusion information from heavy ion collisions.

\subsection{Personal note (James Lattimer)}

I would like to take this opportunity to add some
personal recollections of interactions with Hans Bethe.  I was
introduced to Hans in the summer of 1978 when I was an Urbana postdoc
visiting Gerry Brown in Copenhagen.  Gerry and Hans were interested in
my work with Geoff Ravenhall on finite temperature nuclear phase
coexistence \cite{LR78} and Hans had developed his ``low entropy''
thesis that was so valuable to understanding gravitational collapse
supernovae.  During this time, I was invited to become a co-author of
the paper that we, at least, have since referred to as BBAL
\cite{BBAL}.  I was greatly honored, of course.  But when I was given
a draft of the paper and saw it began with the phrase ``Massive stars
live for eons and eons ...'', I had second thoughts.  I conveyed my
misgivings about the astronomy to Gerry, but he assured me that Hans
was very happy to accept criticism.  I agreed to collaborate and began
rewriting, much to Gerry's chagrin.  After my return to Urbana, and
Hans to Ithaca, Gerry would distribute my rewrites to Hans and Hans'
to me.  This iteration did not seem to rapidly converge, the paper
developed into a 46-page journal article with seven appendices, and
did not get submitted until the following Spring.  Gerry still blames
me for the delay, probably because a competing letter (with not
dissimilar conclusions) I wrote at Urbana with Don Lamb, Chris Pethick
and Geoff Ravenhall \cite{LLPR78} had appeared shortly beforehand.

Another memorable occasion occurred in 1981, a year or two after I
came to Stony Brook (partly, at least, on the back of BBAL).  I had
frequently cooked breakfast and dinner for Hans during summer trips to
Copenhagen over the previous few years; he especially loved meat and
always liked the rare steaks I cooked for him in Denmark (at least,
what passed for steaks in Denmark).  Gerry decided I should host a
dinner for Hans at which all the graduate students in our group would
attend.  I decided to cook a leg of lamb on the barbecue.  Around this
time, Hans had become interested in neutrino diffusion in supernovae,
and was intrigued by comments from Don Lamb that the boundary in
collapsing material between trapped and freely-streaming neutrinos was
particularly important in supernovae.  Hans referred to this theory as
the joint of Lamb.  So, of course, when I took my leg of lamb off the
grill and put it on the table before Hans, I pronounced it ``the joint
of lamb''.  Hans, who loved puns, had a great laugh over this.

\subsection{Personal note (Madappa Prakash)}
I distinctly recall my first meeting with Hans Bethe. 
One morning during my early years as a postdoc at Stony Brook, 
Gerry Brown came into my office and said ``Come and tell Hans
what you're doing now; we're in the common room'', and left. 
Expecting to see Hans  Hansson, a friendly particle physicist, 
I grabbed my coffee cup and went into the  common room. 
Sitting there was Hans Bethe, eating raisins one by one. 
Gerry said, ``Go  on, write something on the board and entertain us.'' 
I recall feeling very ill. 
I said, ``Could I bring my notes?'' 
Gerry said, ``Ok, if you really need them.'' 
I ran  into my room, grabbed my notes and returned to the 
common room where 
the two of them were waiting. 
I started writing on the board feeling more and more  ill, 
but fortunately I was interrupted. 
Hans Bethe said, ``Why are you keeping so many  
terms in the expansion?'' 
(The physics issue was the influence of  
effective masses on the pressure of supernova matter.) 
I somehow found the courage to  say, ``I also have an 
exact and short expression.'' Bethe seemed to like that 
and said  ``If you have to keep more than one term in 
the expansion, you're expanding about the  wrong point.''  
Later, Gerry told me that Hans liked what I was doing, 
which was very  gratifying.  The work I was reporting to Bethe was 
later written up (with Tom  Ainsworth, Jean-Paul Blaizot 
and Herman Wolter as co-authors) for Gerry Brown's 60th  Birthday 
Conference Proceedings ``Windsurfing the Fermi Sea'' \cite{pabw}.

My most memorable meeting with Bethe took place in Seattle during the
``Supernova Physics'' program, organized by Gerry Brown during the
early years of the U.S.  Institute for Nuclear Theory and held in a
building some distance away from the old Physics Department.  I was in
awe that Bethe had solved two coupled differential equations to three
decimal point accuracy using his slide rule, while others were still
cajoling their computers.  One afternoon, Gerry asked me to bring
Bethe to the brown-bag lunch in the Physics Department, where Stan
Woosley was to speak.  We started walking toward the Physics building.
Bethe walked slowly, so we were getting late. To make matters worse,
he stopped now and then to pop raisins into his mouth. We arrived at a
street crossing light, which turned green as we approached it.  I
hurried Bethe as best as I could, but when we were half-way across the
lights turned red again. To my utter horror, the perpendicular traffic
was accelerating toward us! I stood in the middle of the road with my
arms widespread to stop the racing cars.  Fortunately, Bethe ambled
across to safety, while I was busy stopping impatient drivers in the
middle of the road! Imagine the thoughts in my head! (I would be
forever known as the man who got Bethe killed!)  When we arrived at
the lunch meeting, I was really angry at Gerry, who just smiled and
said ``Why did you think I asked you to bring him here?''

I wish there were more people like Hans Bethe. He was very kind to me, 
always made me  feel good, and I learned a whole lot from him. 
I am very fortunate to have known him.  
    
\section{Some maximal attributes of neutron stars}
\setcounter{section}{2}
\setcounter{subsection}{0}
\subsection{The maximum mass}

General relativity introduces a profound change to Newtonian
hydrostatic equilibrium: the existence of a maximum mass.  In general
relativity, hydrostatic equilibrium is expressed by
\begin{equation}
{dp(r)\over dr}=-{G\over c^2}~{[p(r)+\epsilon(r)][M(r)+4\pi r^3p(r)/c^2]\over
r(r-2GM(r)/c^2)},\qquad {dM(r)\over dr}=4\pi ^2\rho(r)\,,
\label{tov}
\end{equation}
where $\epsilon=\rho c^2$ is the total mass-energy density.  Rhoades
\& Ruffini \cite{RR74} derived an upper limit of 3.2 M$_\odot$ to
neutron star masses under the assumptions that the EOS
(1) is less stiff than that of non-interacting degenerate neutron
matter up to a fiducial mass (energy) density $(\rho_f) \epsilon_f$,
which they chose to be $\rho_f=\epsilon_f/c^2=4.6\times10^{14}$ g
cm$^{-3}$, and (2) is limited by causality at higher densities.  In
other words, they assumed the EOS
\begin{equation}
p(\epsilon)=p_{neutron}(\epsilon), \qquad\epsilon\le\epsilon_f; 
\qquad\qquad p(\epsilon)=\epsilon-\epsilon_f+p_f, 
\qquad \epsilon\ge\epsilon_f\,,
\label{rr}
\end{equation}
where $p_f$ is the density at $\epsilon_f$.
Hartle \& Sabbadini \cite{Hartle77} later showed this limit almost perfectly
scales with $\epsilon_f$ such that
\begin{equation}
M_{max}=4.2\sqrt{\epsilon_f/\epsilon_s}{\rm~M}_\odot\,.  
\end{equation}
\subsection{Maximum compactness}
Concerning the maximum compactness, Lindblom \cite{Lindblom84}
determined that the maximum redshift, for $\rho_f\ge3\times10^{14}$ g
cm$^{-3}$ and assuming causality, is
\begin{equation}
z={1\over\sqrt{1-2GM/Rc^2}}-1\le0.863 \,, 
\label{z}
\end{equation}
which is equivalent to
\begin{equation}
R\ge2.83GM/c^2 \,. 
\label{rm}
\end{equation}
This result is relatively insensitive to the value of $\rho_f$ and is
consistent with the empirical limit established by
Glendenning \cite{Glendenning92}.

Koranda et al. \cite{Koranda97} have given a very elegant and rigorous
causal limit that is independent of any other assumptions about the
EOS.  They emphasized the special nature of the ``minimum period''
EOS (so named for reasons made evident below)
\begin{equation}
p(\epsilon)=0,\qquad\epsilon\le\epsilon_c;\qquad
p(\epsilon)=\epsilon-\epsilon_c,\qquad\epsilon\ge\epsilon_c\,.
\label{causal1}
\end{equation}
This EOS contains a single parameter, $\epsilon_c$ or $\rho_c$, and as
a result the general relativistic structure equations,
Eq. (\ref{tov}), then contain only one free parameter and one
dimensional parameter, $\epsilon_c$, which corresponds to the surface
energy density.  All properties of spherical ({\it i.e.}
non-rotating) stars with this EOS must therefore scale as $M\propto R$
and $\epsilon\propto M^{-2}$.  These considerations must apply, in
particular, to the maximum mass configuration, whose radius, mass and
central density are $R_{sph}^{max}, M_{sph}^{max}$ and
$\epsilon_{cent,sph}^{max}$ (or $\rho_{cent,sph}^{max}$),
respectively.  Furthermore, this being the maximally stiff EOS at high
density, and the minimally stiff EOS at low density, it will support
the largest mass with the smallest radius.  For a given value of
$\epsilon_c$, the maximum mass configuration obviously is the most
compact configuration.  Integration of the dimensionless structure
equations then yields the scaling relations shown in the first line of
Table \ref{tableone} for the maximum mass configurations.

\begin{table}
\centerline{Scaling results for ``minimum period'' EOSs}
\begin{tabular}{|c|c|c|c|c|c|c|c|}
\hline
Eq. & $R_{sph}^{max}c^2/GM_{sph}^{max}$ & 
\multicolumn{2}{c|}{$(M_{sph}^{max})^2\rho_c$} & 
\multicolumn{2}{c|}{$(M_{sph}^{max})^2\rho_{cent,sph}^{max}$} & 
$\rho_{cent,sph}^{max}/\rho_c$ & $P_{min}/M_{sph}^{max}$\\ \hline 
{} & {}    & $c=G=1$ & $\quad$ M$_\odot^2$ g cm$^{-3}\quad$ &
$c=G=1$ & $\quad$ M$_\odot^2$ g cm$^{-3}\quad$ & --- &
$\quad$ M$_\odot^{-1}$ ms$\quad$\\ \hline
(\ref{causal1}) & 2.825 & 0.007247 & $4.476\times10^{15}$ & 0.02193 & 
$13.55\times10^{15}$ & 3.026 & 0.200 \\ \hline 
(\ref{causal2}) & 2.87 & 0.00725 & $4.478\times10^{15}$ & 0.0219 & 
$13.54\times10^{15}$ & 3.026 & 0.203 \\ \hline 
\end{tabular}
\caption{The quantities $R_{sph}^{max}, M_{sph}^{max}$ and
$\rho_{cent,sph}^{max}$ refer to the radius, mass and central
mass density of the maximum mass configuration.  $P_{min}$ is the
Keplerian rotational limit of the maximum mass configuration. The first
line of results corresponds to Eq. (\ref{causal1}); the second line is
for Eq. (\ref{causal2}).}
\label{tableone}
\end{table}

Of course, the EOS in Eq. (\ref{causal1}) is unrealistic in that the
density at the surface of the star is finite, with a value generally
well above the nuclear equilibrium density.  A less rigorous, but more
realistic, causally limited EOS consists of using a reasonably
understood model (here chosen to be the FPS EOS \cite{PR89}) for
$p(\epsilon)$ up to a matching energy density $\epsilon_m$ where the
pressure is $p_m$:
\begin{equation}
p(\epsilon)=p_{FPS}(\epsilon),\quad\epsilon\le\epsilon_m;\qquad
p(\epsilon)=p_{FPS}(\epsilon_m),\quad\epsilon_m\le\epsilon\le\epsilon_c;
\qquad p(\epsilon)=\epsilon-\epsilon_c+p_m,\quad\epsilon\ge\epsilon_c\,.
\label{causal2}
\end{equation}
Between $\epsilon_m$ and $\epsilon_c$, the pressure is fixed, and
above $\epsilon_c$, the EOS is causal.  Choosing a matching baryon
number density $n_m=0.1$ fm$^{-3}$, one has for the FPS EOS
$\epsilon_m=1.262\times10^{-4}$ km$^{-2}$ and $p_m=5.69\times10^{-7}$
km$^{-2}$ (units where $c=G=1$).  The maximum mass configurations
retain, to a high accuracy, the scalings of the EOS
Eq. (\ref{causal1}), but the integration constants are slightly
altered, as shown in the second line of Table \ref{tableone}.  These
stars have slightly larger maximum masses and radii for a given value
of $\epsilon_c$.  Also note that the revised compactness limit is practically
the same as established by Lindblom \cite{Lindblom84}.

\subsection{Maximum central density}
\begin{figure}
\hspace*{-1.5cm}
  \includegraphics[width=.6\textwidth,angle=90]{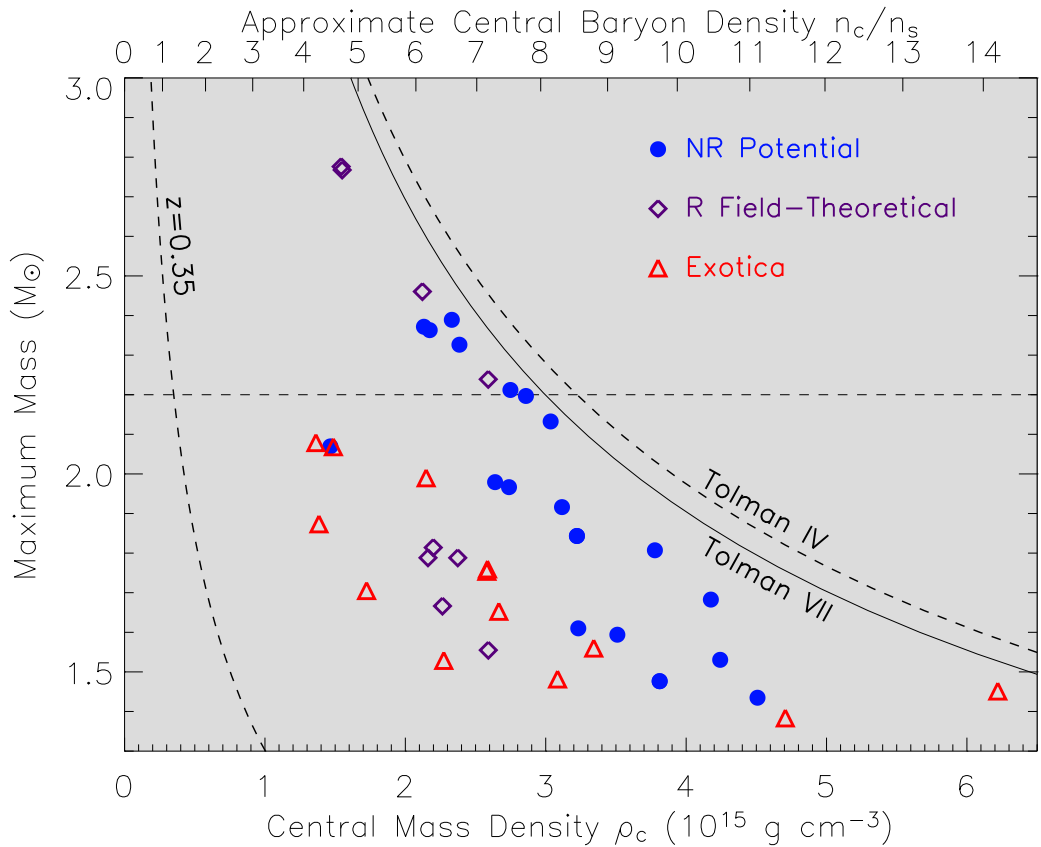}
\caption{The maximum mass - central density relation predicted by
  causality coupled with the Tolman VII and Tolman IV analytic GR
  solutions are compared with structure integration results for a
  variety of EOSs.  NR refers to non-relativistic
  potential EOSs, R refers to relativistic
  field-theoretical EOSs, and Exotica refers to
  EOSs with considerable softening at high density due
  to kaon condensation or strange quark matter deconfinement.  A
  possible redshift measurement of $z=0.35$ is also shown.
Figure taken from. Ref. \cite{LP05}.}
\label{maxdens}
\end{figure}
The ``minimum period'' EOS predicts the
smallest radius possible for a given mass, with $R\ge2.87GM/c^2$.
Dimensionally, $\rho\propto MR^{-3}$, so a plausible conjecture
would be that this EOS also predicts the largest
possible value of the central density for a given mass. However, this
is not the case.   Lattimer \&
Prakash \cite{LP05} investigated the maximum density
question, with an eye toward using measured masses to set upper limits to
the density of static matter in the universe.  Their empirical results
demonstrate that the predicted value of $M^2\rho_{cent}$ for the
``minimum period'' EOS is close to, but less than, the
actual limiting value.  They found
that an empirical limit can be found by combining the causality limit
$R\ge2.87GM/c^2$ with the central density-mass relation implied by the
Tolman VII analytic solution \cite{Tolman39} of the general
relativistic structure equations.  The Tolman VII solution has the
explicit energy density-radius relation
\begin{equation}
\rho=\rho_{cent}\left[1-\left({r\over R}\right)^2\right]\,,
\label{tolmanvii}
\end{equation}
which, combined with the causality limit, results in
\begin{equation}
\rho_{cent}M^2={15\over8\pi}\left({M\over
  R}\right)^3\le{15\over8\pi}\left({c^2\over2.87G}\right)^3=
15.3\times10^{15}{\rm~M}_\odot^2 {\rm~g~cm}^{-3}\,.
\label{rholim}
\end{equation}
This relation, and results for various EOSs, are
displayed in Fig. \ref{maxdens}.  From Table \ref{tableone}
note that the ``minimal period'' EOS predicts a maximum central
density with a coefficient of $13.5\times10^{15}$ in
Eq. (\ref{rholim}), about 10\% smaller than the empirical limit.

It should be emphasized that Eq. ({\ref{rholim}) represents the upper
limit to the central density of the measured star.  Since the maximum
neutron star mass must be larger than that of any measured star, and
the central density increases with mass for a given EOS, the
central density of the actual maximum mass star must be smaller than
the value given by Eq. (\ref{rholim}) using the largest measured mass.
With a mass measurement of 2.1 M$_\odot$, for example, the limiting
mass density is $3.4\times10^{15}$ g cm$^{-3}$.  If the maximum mass
was in fact about 10\% larger, the limiting mass density becomes
$2.8\times10^{15}$ g cm$^{-3}$.  Note that this mass density
corresponds to a baryon density of only about $7n_s$.  
This could be small enough to call into question the applicability of
perturbative QCD in invoking the presence of a quark phase in neutron
star cores. It must be stressed, however, that precocious
perturbativeness may indeed occur in the quark matter sector.

\subsection{Maximum spin rate}
The compactness limits above are also intimately connected to the
Keplerian, or mass-shedding, rotational limit obtained when the equatorial
surface velocity equals the orbital speed just above the surface.  The
most compact star, assuming uniform rotation, has the highest
rotational frequency.  For a uniform rigid sphere of mass $M$ and
radius $R$, the mass-shedding limit in Newtonian gravity is
\begin{equation}
P_{min}^N=2\pi\sqrt{R^3\over GM}=0.545\left({M_\odot\over M}\right)^{1/2}
\left({R\over10{\rm~km}}\right)^{3/2}{\rm~ms}\,.
\label{rotnewt}
\end{equation}
It is interesting and fortunate that the minimum spin period in fully
relativistic calculations \cite{Haensel89, FPI} employing
realistic hadronic EOSs is given to a good approximation by a
similar formula
\begin{equation}
P_{min}\simeq0.83\left({M_\odot\over M_{sph}^{max}}\right)^{1/2}
\left({R_{sph}^{max}\over10{\rm~km}}\right)^{3/2}{\rm~ms}\,.
\label{pmin}
\end{equation}
This spin period obtains for the maximum mass configuration, which has
the smallest radius for hadronic EOSs.  This formula takes into
account not only general relativity, but also the deformation of
rotating stars, yet is still expressed in terms of the non-rotating
values of the maximum mass and radius.  Therefore, the same scaling
$P_{min}\propto\sqrt{R^3/M}$ exists for both Newtonian and GR
gravitation.
\begin{figure}
\hspace*{-1.35cm}
\includegraphics[angle=90,width=.9\textwidth]{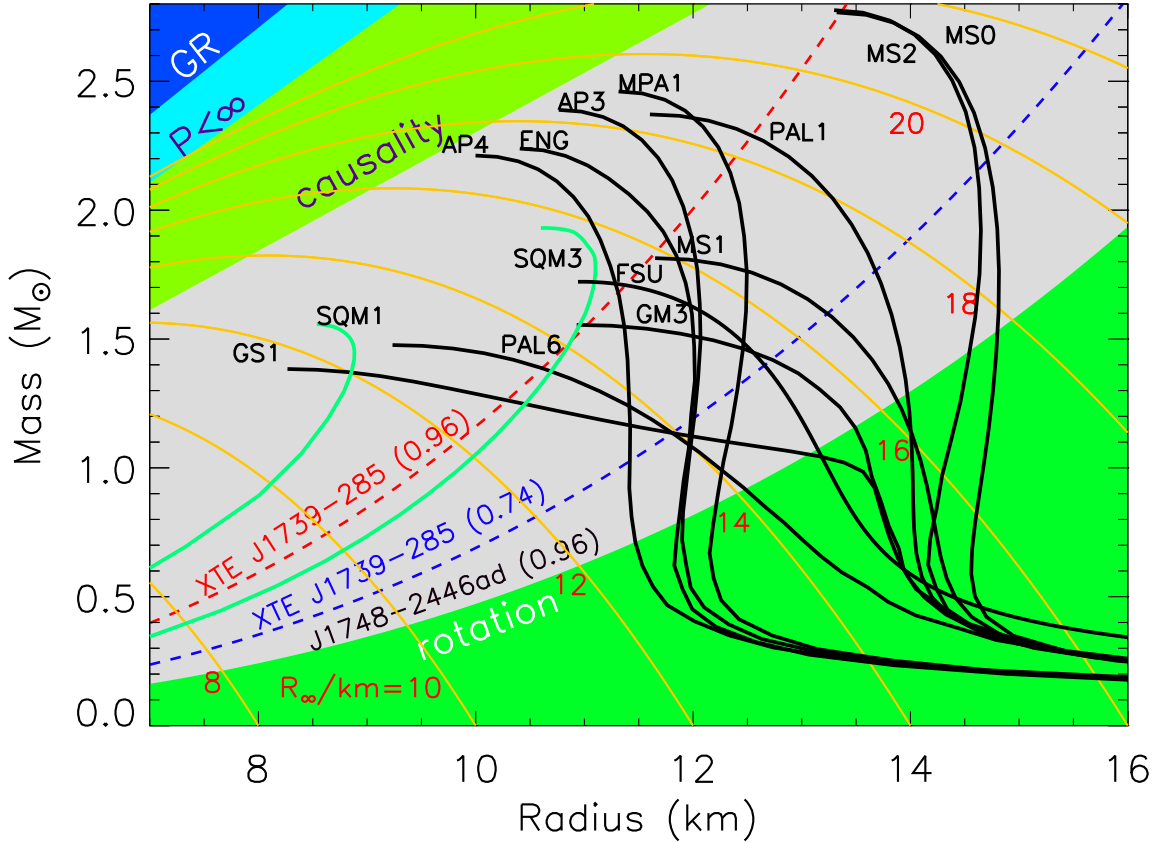}
\caption{Mass-radius trajectories for typical EOSs (see \cite{LP01}
for notation) are shown as black curves.  Green curves (SQM1, SQM3)
are self-bound quark stars.  Orange lines are contours of radiation
radius, $R_\infty=R/\sqrt{1-2GM/Rc^2}$.  The dark blue region is
excluded by the GR constraint $R>2GM/c^2$, the light blue region is
excluded by the finite pressure constraint $R>(9/4)GM/c^2$, and the
green region is excluded by causality, $R>2.9GM/c^2$.  The light green
region shows the region $R>R_{max}$ excluded by the 716 Hz pulsar
J1748-2446ad \cite{Hessels06} using Eq. (\ref{pmin1}).  The upper red
dashed curve is the corresponding rotational limit for the 1122 Hz
X-ray source XTE J1739-285 \cite{Kaaret06}; the lower blue dashed
curve is the rogorous causal limit using the coefficient 0.74 ms in
Eq. (\ref{pmin1}).}
\label{mr}
\end{figure}

The same scaling exists for the ``minimum period'' EOS
\cite{Koranda97}: the scaling $R_{sph}^{max}\propto M_{sph}^{max}$ for
maximum mass stars implies that $P_{min}\propto M_{sph}^{max}$.
Koranda et al. \cite{Koranda97} determined the proportionality
constant 0.200 ms M$_\odot^{-1}$ for this case (see Table
\ref{tableone}).  This translates into a coefficient of approximately
0.74 in Eq. (\ref{pmin}).  The ultra-compactness of stars constructed
with the ``minimum period'' EOS accounts for the considerably smaller
coefficient compared to that originally calculated for more realistic
EOSs.

The relation in Eq. (\ref{pmin}) applies only to the maximum mass
configuration.  Lattimer \& Prakash \cite{LP04} empirically found a
somewhat more useful result which applies to an {\it arbitrary}
neutron star mass, {\it so long as that mass is not close to the
maximum mass}:
\begin{equation}
P_{min}\simeq(0.96\pm0.03)\left({M_\odot\over M_{sph}}\right)^{1/2}
\left({R_{sph}\over10{\rm~km}}\right)^{3/2}{\rm~ms}\,.
\label{pmin1}
\end{equation}
In this equation, $M_{sph}$ and $R_{sph}$ refer to the non-rotating
mass and radius.
Inasmuch as an observed neutron star is likely to be at least slightly
smaller than the maximum mass, Eq. (\ref{pmin1}) can be used to limit 
masses and radii for observed stars:
\begin{equation}
R_{sph}<10.4 \left({1000{\rm~Hz}\over\nu}\right)^{2/3}
\left({M_{sph}\over{\rm M}_\odot}\right)^{1/3}{\rm~km}\,,
\label{pmin2}
\end{equation}
where the spin frequency $\nu=1/P$.

The most rapidly rotating pulsar is PSR J1748-2446ad with a spin rate
of 716 Hz \cite{Hessels06}.  With this value, Eq. (\ref{pmin2})
suggests that for a 1.4 M$_\odot$ star, the non-rotating radius would
be limited by $R<14.3$ km as shown in Fig. \ref{mr}.  However, there is
some liklihood that this star spun up through the accretion of a few
tenths of a solar mass.  If this star is, for example, 1.7 M$_\odot$,
the upper limit to the non-rotating radius would be 15.3 km.
Unfortunately, neither limit is very restrictive at present.

Recently, however, an 1122 Hz X-ray burst oscillation from the neutron
star X-ray transient XTE J1739-285 has been reported \cite{Kaaret06}.
The stability of this oscillation frequency strongly suggests that it
is the spin rate of the neutron star.  If true, this sets relatively
stringent limits to the radius that this star would have had if it was not
rotating.  According to the empirical relation Eq. (\ref{pmin2}), the
maximum radii are 10.8 and 12.2 km, assuming the mass of this star to be
1.4 and 2.0 M$_\odot$, respectively. Generally, these limits apply for
stars not in the immediate vicinity of their maximum mass or for quark
matter stars. As shown in Fig. \ref{mr} this constraint would rule
out a number of EOSs.  The rigorous causal limit, from Table \ref{tableone},
yields 12.6 and 14.2 km, respectively.  It will obviously be important to
confirm this observation.

\section{Recent mass measurements and their implications}
\label{sec:1}
Several recent observations of neutron stars have direct bearing on
the determination of the maximum mass.  The most accurately measured
masses are from timing observations of the radio binary pulsars. As
shown in Fig. \ref{masses}, which is compilation of the measured
neutron star masses as of November 2006, observations include pulsars
orbiting another neutron star, a white dwarf or a main-sequence star.
The compact nature of several binary pulsars permits detection of
relativistic effects, such as Shapiro delay or orbit shrinkage due to
gravitational radiation reaction, which constrains the inclination
angle and allows the measurement of each mass in the binary. A
sufficiently well-observed system can have masses determined to
impressive accuracy. The textbook case is the binary pulsar PSR
1913+16, in which the masses are $1.3867\pm0.0002$ and
$1.4414\pm0.0002$ M$_\odot$, respectively \cite{Weisberg05}.
\begin{figure}
\hspace*{-1.5cm}
  \includegraphics[width=.75\textwidth]{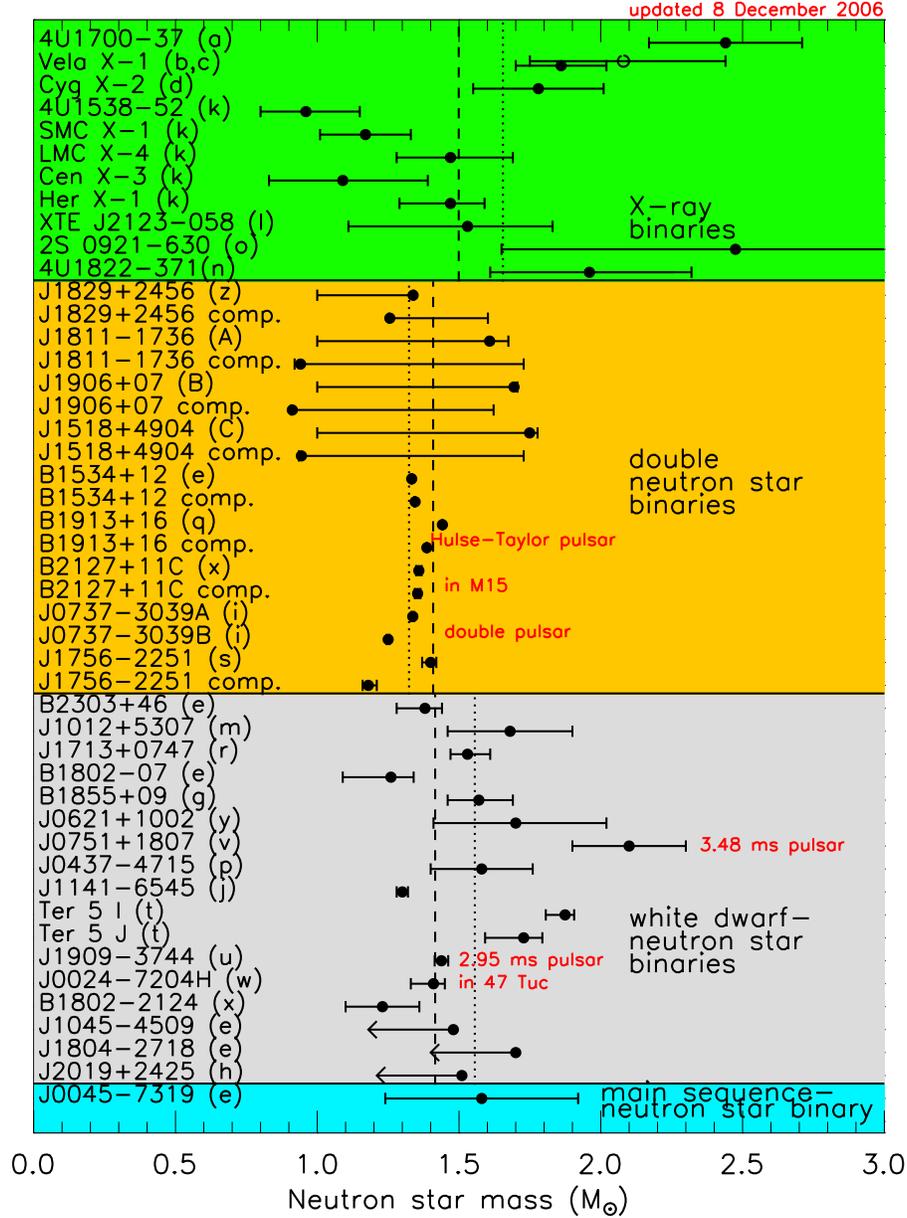}
\caption{Measured and estimated masses of neutron stars in radio
binary pulsars (gold, silver and blue regions) and in x-ray accreting
binaries (green).  For each region, simple averages are shown as
dotted lines; weighted averages are shown as dashed lines.  The
labels (a) = \cite{Clark02} through (C) = \cite{Nice99} are references cited in
the bibliography.  For the stars with references z-C, a lower limit to
the pulsar mass of 1 M$_\odot$ was assumed.}
\label{masses}
\end{figure}

One significant development concerns mass determinations in binaries
with white dwarf companions, which show a broader range of neutron
star masses than binary neutron star pulsars. Perhaps a rather narrow
set of evolutionary circumstances conspire to form double neutron star
binaries, leading to a restricted range of neutron star masses
\cite{Bethe98}.  This restriction is likely relaxed for other neutron
star binaries. Evidence is accumulating that a few of the white dwarf
binaries may contain neutron stars larger than the canonical 1.4
M$_\odot$ value, including the
intriguing case \cite{Nice05} of PSR
J0751+1807 in which the estimated mass with $1\sigma$ error bars is
$2.1\pm0.2$ M$_\odot$. In addition, to 95\%
confidence, one of the two pulsars Ter 5 I and J has a 
reported mass
larger than 1.68 M$_\odot$ \cite{Ransom05}.  

Whereas the observed simple mean mass of neutron stars with white
dwarf companions exceeds those with neutron star companions by 0.25
M$_\odot$, the weighted means of the two groups are virtually the
same.  The 2.1 M$_\odot$ neutron star, PSR J0751+1807, is about
$4\sigma$ from the canonical value of 1.4 M$_\odot$.
{\em It is furthermore the case that the $2\sigma$ errors of all but
two systems extend into the range below 1.45 M$_\odot$, so
caution should be exercised before concluding that firm evidence of
large neutron star masses exists.} Continued observations,
which will reduce the observational errors, are necessary to clarify
this situation.

Masses can also be estimated for another handful of binaries which
contain an accreting neutron star emitting x-rays, as shown in
Fig. \ref{masses}. Some of these systems are characterized by
relatively large masses, but the estimated errors are also large. The
system of Vela X-1 is noteworthy because its lower mass limit (1.6 to
1.7M$_\odot$) is at least mildly constrained by
geometry \cite{Quaintrell03}.

Raising the limit for the neutron star
maximum mass could eliminate entire families of EOSs, especially
those in which substantial softening begins around 2 to 3$n_s$. This
could be extremely significant, since exotica (hyperons, Bose
condensates, or quarks) generally reduce the maximum mass appreciably.

Assuming that the hyperon-nucleon couplings are comparable to the
nucleon-nucleon couplings typically results in the appearance of
$\Lambda$ and $\Sigma^-$ hyperons around 2 to 3 $n_s$ in neutron star
matter \cite{glenhyp,kaphyp,kpe,schaffner}.  In beta equilibrated
neutron star matter, the various chemical chemical potentals satisfy
the relations $\mu_n-\mu_p=\mu_e=\mu_{\Sigma^-}$ and
$\mu_n=\mu_\Lambda$.  As a consequence, the proton fraction in such
matter is quite small, of order 5-10\%.  Little is known about the
symmetry dependence of the hyperon-nucleon couplings as these
couplings are chiefly determined from hyperon binding energies in more
or less symmetric nuclei.  If hyperons indeed appear at as low a
density as 2-3 $n_s$, the maximum neutron star mass becomes relatively
small, typically less than 1.6 M$_\odot$ \cite{kpe}.

The suggestion of Kaplan and Nelson \cite{kapnel} that, above some
critical density, the preferred state of matter might contain a
Bose-Einstein condensate of negatively charged kaons has been examined
extensively \cite{bro2,pol,muto,maruy,bro3,tpl,bro4}. The astrophysical
consequences have been explored in some detail in Ref. \cite{tpl}.
The physics is that in medium, the strong attraction between $K^-$
mesons and baryons increases with density
and lowers the energy of the zero-momentum state. A condensate forms
when this this energy becomes equal to the kaon chemical potential,
$\mu$ which is related to the electron and nucleon chemical potentials
by $\mu = \mu_n-\mu_p=\mu_e=\mu_\mu$ due to chemical equilibrium in
the various reactions. Typically, the critical density for
condensation (which depends primarily on the symmetry energy of
nucleonic matter) is $\sim (3-4)n_s$, although it
is model and parameter dependent. Relative to matter without a
kaon-condensed state and depending upon the models employed, 
maximum masses only as high as 1.5 to 1.6 M$_\odot$ can be obtained with 
kaon condensation. 

If a different form of strangeness can appear prior to hyperons or
Bose condensates, for example, deconfined $u,~d,~{\rm and}~s$ quark
matter, maximum masses up to approximately 2.0 M$_\odot$ are possible
\cite{Alford05}, but only with special fine-tuning of the nucleonic
and quark matter parameters.  Therefore a confirmation of a neutron
star mass in excess of 2 M$_\odot$ would be especially interesting.

A parameter of increasing interest, as the database of neutron star
masses expands, concerns the {\it minimum} mass of neutron stars.  It
is believed that neutron stars are created in the aftermath of
gravitational collapse supernovae and must therefore pass through a
proto-neutron star state in which a large number of neutrinos are
trapped and a moderate entropy exists \cite{Burrows86}.   For such a
configuration, the range of stable masses is much narrower than for
cold, catalyzed systems.  The minimum mass of a cold star is about
0.09 M$_\odot$, only slightly sensitive to the EOS around $\rho_s$.  A
star with a trapped lepton fraction $Y_L\sim0.35-0.4$ and an entropy
per baryon $s\sim1-2$ has a minimum mass of about 0.85 M$_\odot$
\cite{Gondek98} for stability.  The smallest reliably estimated
neutron star mass is the companion of the binary pulsar J1756-2251
whose mass is $1.18\pm0.02$ M$_\odot$ \cite{Faulkner04}.  It will be
interesting to see if smaller neutron stars are found.

\section{The radius constraint}
\label{sec:2}
As previously mentioned, the radius of a neutron star is primarily
determined by the density dependence of the symmetry energy.  This
connection arises through the relation between the neutron star radius
and the intenral pressure of matter at intermediate densities ($1.5n_s
< n < 2-3~n_s$).  That such a relation exists can be seen most clearly
by considering Newtonian polytropes.  For the pressure-density
relation
\begin{equation}
p = K\rho^{1+1/n}\,,
\label{prho}
\end{equation}
where $K$ is a constant and $n$ is the polytropic index, hydrostatic
equilibrium implies that
\begin{equation}
R\propto K^{n/(3-n)}M^{(1-n)/(3-n)}\,.
\label{rmk}
\end{equation}
Realistic EOSs typically have $n\simeq1$, but $K$ is
uncertain by a factor of 5 or 6.  For some average density $\rho_*$,
in the vicinity of $(1-2) m_bn_s$, suppose the pressure is $p_*$.  For the
case $n=1$, one therefore has
\begin{equation}
R\propto p_*^{1/2}\rho_*^{-1}M^0\,.
\label{rp}
\end{equation}
\begin{figure}
\vspace*{-.5cm}
\hspace*{-1.5cm}
  \includegraphics[width=.7\textwidth, angle=90]{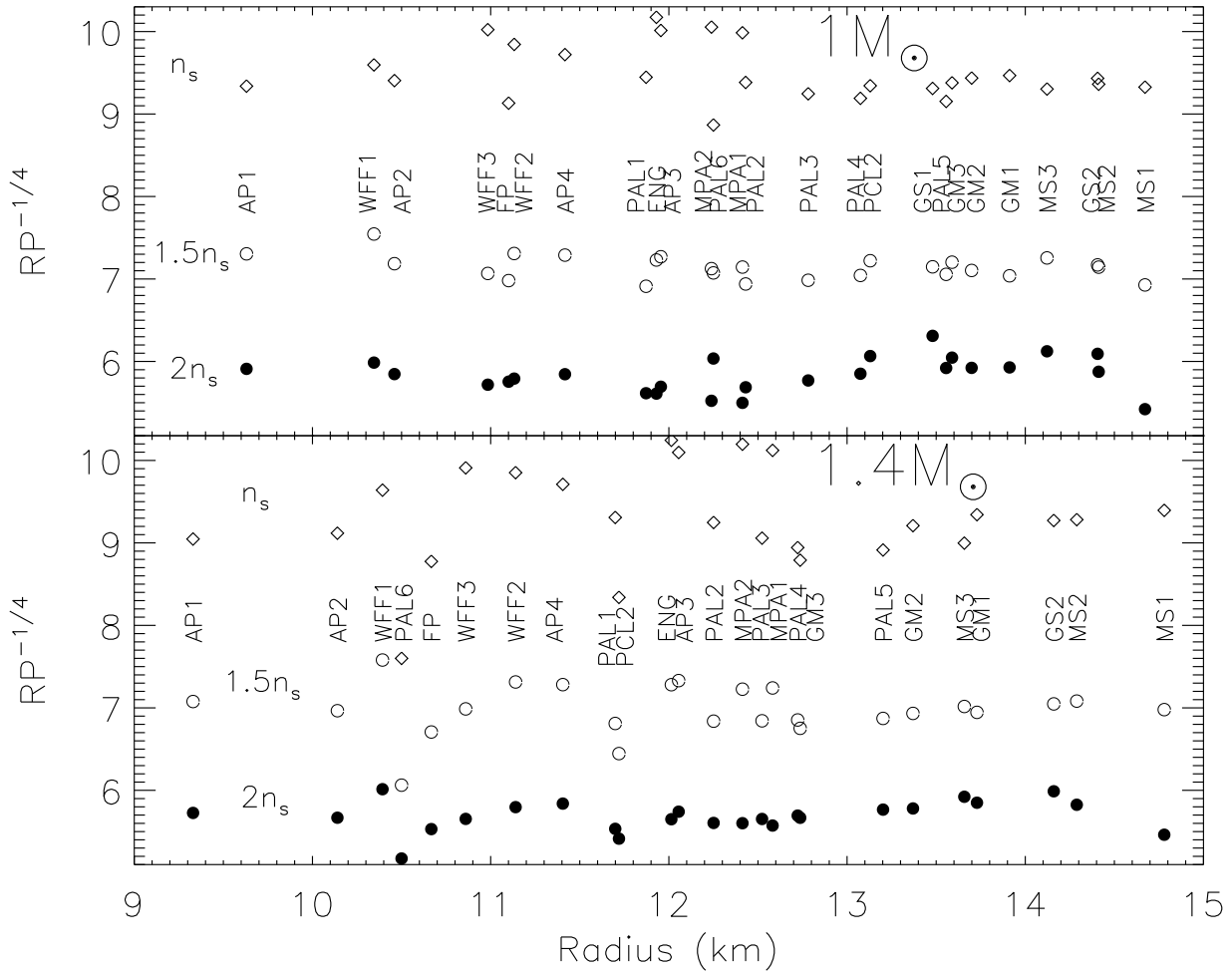}
\vspace*{.5cm}
\caption{Empirical demonstration of the constancy $Rp_*^{-1/4}$, for 1
  M$_\odot$ (upper panel) and 1.4 M$_\odot$ (lower panel) stars.  For
  each mass, 3 fiducial number densities are selected.  Figure and EOS
  labels are from reference \cite{LP01}.}
\label{p-r}
\end{figure}
The independence of $R$ from $M$ is a characteristic result of
the mass-radius trajectories for EOSs without extreme
low-density softening, as seen in Fig. \ref{mr}.  Equation (\ref{rp})
suggests that the radius scales with the square root of the fiducial
pressure $p_*$.  However, general relativity plays an important role,
and 
with its inclusion, the scaling is such that $R\propto
p_*^{1/4}$ \cite{LP01}.  This is an empirical result, and is
graphically illustrated in Fig. \ref{p-r}.  It is
important to note that it applies to nearly all neutron star EOSs, not
just those that have a portion of the $M-R$ curve with vertical
slope.  The exceptions to the $R\propto p_*^{1/4}$ rule are the EOSs
for strange quark stars or EOSs with extreme softening near $\rho_s$.
The correlation is more robust if the fiducial density is taken to be
$2n_s$ rather than $n_s$.

Using the only known analytic solution of Einstein's equations
with vanishing surface energy density and an explicit pressure -
density relation, this scaling can be analytically
demonstrated.  This solution, due to Buchdahl \cite{Buchdahl67},
assumes the EOS
\begin{equation}
\epsilon=12\sqrt{p_*p}-5p\,,
\label{buch}
\end{equation}
where $p_*$ is a parameter.  For low densities, when $p<<p_*$, one
sees that this EOS is that of an $n=1$ polytrope, so
this solution is a reasonable approximation for a neutron star.  For
this solution, the radius is an implicit function of $M$ and $p_*$:
\begin{equation}
R=(1-\beta)\sqrt{\pi\over288p_*(1-2\beta)}\,,
\label{buch1}
\end{equation}
where $\beta=GM/Rc^2$.  In turn, these relations lead to
\begin{equation}
{d\ln R\over d\ln p}\Biggr|_{n,M}={1\over2}{(1-\beta)(1-2\beta)\over
(1-3\beta+3\beta^2)}{6-5\sqrt{p/p_*}\over6+\sqrt{p/p_*}}.
\label{dlnrdlnp}
\end{equation}
Choosing for typical values $M=1.4$ M$_\odot, R=13$ km, and
$\epsilon=1.5m_bn_s=2.97\times10^{-4}$ km$^{-2}$, one finds
$\beta=0.318$ and, using Eq. (\ref{buch}), $p/p_*=0.141$.  Substitution
in Eq. (\ref{dlnrdlnp}) yields $d\ln R/d\ln p=0.230$, quite close to
the empirical result of about 1/4.

This correlation is significant because the pressure of degenerate
neutron-star matter near the nuclear saturation density $n_s$ is, in
large part, determined by the symmetry properties of the EOS. For the
present discussion, we introduce the incompressibility $K$ and the
skewness $K^\prime$, and expand the nucleonic energy per particle
about its values at $n_s$ and $x=1/2$, where $x$ is the proton
fraction:
\begin{eqnarray}
E(n,x) = -16 + \frac {K}{18} \left(1-\frac {n}{n_s}\right)^2
+\frac {K^\prime}{27}  \left(1-\frac {n}{n_s}\right)^3
+E_{sym}(n)(1-2x)^2\dots\,.\label{eexp}
\label{energy}
\end{eqnarray}
Here, $E_{sym}$ is the symmetry energy function, approximately the
energy difference at a given density between symmetric and pure
neutron matter.  The symmetry energy parameter is defined as
$S_v\equiv E_{sym}(n_s)$.  For the total energy, leptonic
contributions (mainly from electrons as that from
muons is small in the vicinity of the nuclear equilibrium density)
\begin{eqnarray}
E_e=(3/4)\hbar cx(3\pi^2nx^4)^{1/3} 
\label{eenergy}
\end{eqnarray}
must be added to $E(n,x)$.  Because catalyzed matter in neutron stars is in
beta equilibrium, i.e., $\mu_e = \mu_n - \mu_p = - \partial E/\partial
x$, the equilibrium proton fraction at $n_s$ is
\begin{eqnarray}
x_s\simeq(3\pi^2 n_s)^{-1}(4S_v/\hbar c)^3 \simeq 0.04\,.
\end{eqnarray}
This small value of $x_s$ enables the pressure at $n_s$ to be
expressed as 
\begin{eqnarray}
P(n_s,x_s)= \left(n^2{\partial [E(n,x)+E_e(n,x)]\over\partial
n}\right)_{n_s,x_s}=n_s(1-2x_s)[n_sS_v^\prime(1-2x_s)+S_v x_s] \simeq
n_s^2S_v^\prime\,,
\label{pexp}
\end{eqnarray}
where $S_v^\prime\equiv(\partial E_{sym}/\partial
n)_{n_s}$.  The pressure depends primarily upon $S_v^\prime$, because the
terms proportional to $x_s$ are relatively small.
The equilibrium pressure at moderately larger densities is similarly 
insensitive to $K$ and $K^\prime$.
Experimental constraints to the compression modulus $K$, most
importantly from analyses of giant monopole resonances
give $K\cong 220$ MeV \cite{Blaizot80}.  The
skewness parameter $K^\prime$ has been estimated to lie in the range
(1780--2380) MeV \cite{Pearson91}.  Evaluating the pressure for
$n=1.5n_s$,
\begin{eqnarray}
P(1.5n_s)= 2.25n_s [K/18-K^\prime/216 + n_s(1-2x)^2(\partial
E_{sym}/ \partial n)_{1.5n_s}]\,.
\label{keep}
\end{eqnarray} 
Note that the contributions from $K$ and $K^\prime$ largely cancel
leading to the result that the pressure is dominated by the term
involving the density
derivative of the symmetry energy.

The density dependence of the symmetry energy is constrained by
laboratory data such as nuclear mass fits, giant dipole resonances,
and neutron skin thicknesses of neutron-rich nuclei.  However,
significant difficulties are encountered in obtaining the quantity
$S_v^\prime$ in the necessary supra-nuclear density range.  First of
all, nuclei sample nucleonic matter in the density range $n<n_s$.
However, to determine neutron star radii, we need the symmetry energy
in the range $n_s$ to $2n_s$.  Secondly, nuclear mass fits and giant
dipole resonance data predict a strong correlation between the volume
symmetry energy $S_v$ and the surface symmetry energy $S_s$
parameters in a liquid-drop sense.  Neutron skin thickness is
predicted to be proportional to the ratio $S_s/S_v$.  If the
nuclear energy per baryon can be expressed as in Eq. (\ref{energy}), a
simplified Thomas-Fermi model for the nucleus shows that $S_s/S_v$
and the neutron skin thickness are proportional to the integral
quantity Refs.~\cite{Krivine84,Lattimer96}
\begin{equation}
\int_0^{n_s} {(S_v/E_{sym}(n)-1)\over\sqrt{n(E(n,1/2)+16)}} dn\,.
\label{surfe}
\end{equation}
Although $S_v^\prime$ is constrained by this relation, it cannot be
uniquely determined from it.

\section{Neutron star crusts and their constraints}

There are a number of observations that pertain to properties of the
neutron star crust, which specifically is the region between the
nuclear surface and the phase transition separating uniform nuclear
matter and matter containing nuclei.  The phase transition occurs at
about 1/3 to 1/2 $\rho_s$, or about $10^{14}$ g cm$^{-3}$, depending
on the nuclear compression modulus and the density dependence of the
nuclear symmetry energy.  For example, if pulsar glitches are due to
interactions between the neutron superfluid confined to the crust and
the bulk of the star, the rate of angular momentum transfer can be
related to the fraction of the moment of inertia of the star which
resides in the crust \cite{LEL}.  Quasi-periodic oscillations observed
in X-ray emission following X-ray bursts on neutron stars are likely
related to fundamental and overtone vibrational frequencies of neutron
star crusts \cite{Duncan98}.  Finally, the cooling observed over the
first several years following superbursts from neutron stars or giant
flares from magnetars is likely due to the cooling of the crust and
can be a sensitive indicator of its extent \cite{Rutledge01} .

\subsection{Theoretical considerations}

It is possible to relate both the relative thickness of the crust,
$\Delta/R$, and the relative fraction of the moment of inertia
contained in the crust $\Delta I/I$, to the mass, radius and a single
parameter of the core-crust interface which depends on the EOS.  To see this, first consider the crust thickness $\Delta=R-R_t$
where $R_t$ is the radius of the core-crust interface.
Begin with hydrostatic equilibrium in the crust, setting $M(r)=M$,
ignoring $P$ relative to both $\epsilon$ and to $Mc^2/4\pi R^3$, and
ignoring the internal energy per baryon compared to $m_bc^2$. Then, 
since $\rho\simeq m_bn$,
\begin{equation}
{dp(r)\over m_bn}={d\mu\over m_b}=-{GM\over r^2-2GMr/c^2}dr\,,
\label{dpdr}
\end{equation}
where $\mu=\mu_n$ is the baryon chemical potential in beta
equilibrium.  Integrating from the base of the crust (values of
physical variables here are denoted by the subscipt $t$) to the
surface (denoted by subscript $0$), Eq. (\ref{dpdr}) becomes
\begin{equation}
{\mu_t-\mu_0\over m_bc^2}={1\over2}\ln{r_t(R-2GM/c^2)\over
 R(r_t-2GM/c^2)}.
\label{mut}
\end{equation}
Note that $r_0=R$ and $r_t=R-\Delta$.  Approximately, $\mu_0\simeq-9$
MeV.  Upon defining
\begin{equation}
{\cal H}\equiv e^{2(\mu_t-\mu_0)/m_bc^2},
\label{calh}
\end{equation}
Eq. (\ref{mut}) can be manipulated to yield
\begin{equation}
{\Delta\over R}={{\cal H}-1\over{\cal H}(1-2\beta)^{-1}-1}.
\label{dmr}
\end{equation}
In general, ${\cal H}-1<<1$, whence Eq. (\ref{dmr}) implies that
$\Delta\propto(1-2\beta)R^2/M$.

In a similar way, one can determine the crustal moment of inertia
fraction.  One begins with the definition of the moment of inertia in
general relativity \cite{Hartle67}:
\begin{equation}
I=-{2c^2\over3G}\int_0^Rr^3\omega(r){dj(r)\over dr}dr
={8\pi\over3c^2}\int_0^Rr^4(\rho(r)+p(r)/c^2)e^{\lambda(r)}j(r)\omega(r)dr\,,
\label{mom}
\end{equation}
where $j(r)=e^{-(\nu(r)+\lambda(r))/2}$ and $\omega(r)$ is the
rotational drag.  The metric functions $\nu(r)$ and $\lambda(r)$
satisfy
\begin{equation}
{d\nu(r)\over dr}=2G{M(r)+p(r)/c^2\over r(r-2GM(r)/c^2)},
\qquad e^{-\lambda(r)}=1-{2GM(r)\over rc^2}\,,
\label{nulam}
\end{equation}
whereas the rotational drag satisfies
\begin{equation}
{d\over dr}\left(r^4j(r){d\omega(r)\over
  dr}\right)=-4r^3\omega(r){dj(r)\over dr}\,.
\label{omega}
\end{equation}
The relevant boundary conditions are
\begin{eqnarray}
e^{\nu(R)}=e^{-\lambda(R)}=1-2\beta,\qquad
j(R)&=&1,\qquad{dj(R)\over dr}=0,\qquad\omega(R)=1-{2GI\over
  R^3c^2}\,,\cr
{d\nu(0)\over dr}={d\lambda(0)\over dr}&=&{dj(0)\over
  dr}={d\omega(0)\over dr}=0\,.
\label{bound}
\end{eqnarray}
Substitution of Eqs. (\ref{omega}) and (\ref{bound}) into
Eq. (\ref{mom}) yields
\begin{equation}
I={c^2\over6G}R^4{d\omega(R)\over dr}\,.
\label{mom1}
\end{equation}

While these equations are straightforward to solve for a given
EOS, it is useful to employ a fit established by
Lattimer \& Schutz \cite{LS05} which is valid for realistic hadronic
EOSs that permit maximum masses greater than about 1.6
M$_\odot$:
\begin{equation}
I\simeq(0.237\pm0.008) MR^2(1+2.84\beta+
18.9\beta^4) {\rm~M}_\odot
{\rm~km}^2\equiv\alpha MR^2f(\beta)\,,
\label{momls}
\end{equation}
which defines $\alpha$ and $f(\beta)$ for future reference.

One can estimate the moment of inertia connected with the crust by
employing the right-hand form of Eq. (\ref{mom}) together with
hydrostatic equilibrium, Eq. (\ref{tov}), and the approximations
$4\pi r^3p(r)<<M(r)c^2, M(r)\simeq M,
j(r)\omega(r)\simeq\omega(R)$, to find
\begin{equation}
\Delta
I={8\pi\over3c^2}\int_{R-\Delta}^Rr^4(\rho+p/c^2)e^{-\lambda(r)}j(r)\omega(r)dr
\simeq{8\pi\omega(R)\over3Mc^2}\int_{R-\Delta}^Rr^6dp\,.
\label{di}
\end{equation}
If one assumes that $p\propto\rho^\gamma$ in the crust, and that
$p(R-\Delta)\equiv p_t$, the integral $\int r^6dp$ is approximately
\begin{equation}
\int_{R-\Delta}^Rr^6dp\simeq
R^6p_t\exp\left(-{6\gamma\over2\gamma-1}{\Delta\over R}\right)\,.
\label{di1}
\end{equation}
Combining Eqs. (\ref{dii}), (\ref{bound}) and (\ref{momls}), one finds
\begin{equation}
{\Delta I\over I}\simeq{8\pi\over3c^2}{R^4p_t\over M^2}
[\alpha^{-1}f(\beta)^{-1}-2\beta]
\exp\left(-{6\gamma\over2\gamma-1}{\Delta\over R}\right) \,.
\label{dii}
\end{equation}

Considering the importance of the location of the core-crust
interface, it is worthwhile analyzing how this depends upon EOS
assumptions.  A heuristic determination of ${\cal H}$ can be made as
follows.  The core-crust interface corresponds to the phase transition
between nuclei and uniform matter.  The uniform matter is nearly pure
neutron matter, with a proton fraction of just a few percent
determined by the condition of beta equilibrium.  Ignoring the finite
size effects due to surface and Coulomb energies of nuclei, one can
consider the critical density for which uniform $npe$ matter becomes
unstable to separation into two coexisting phases (one corresponding
to nuclei, the other to a nucleonic sea) \cite{Kubis06}.  This is a
first approximation to the actual case, in which surface and Coulomb
effects are not negligible and in addition the dense phase disappears
when its volume fraction is finite \cite{LLPR83}.  The weak
interactions conserve both baryon number and charge.  At zero
temperature, the first law of thermodynamics can be written as
\begin{equation}
du=-pdv-\hat\mu dq\,,
\label{first}
\end{equation}
where $u$ is the internal energy per baryon, $v=1/n$, $q=x-Y_e$ is the
charge density, $x$ is the proton fraction in the baryons, $Y_e$ is
the electron concentration, and $\hat\mu=\mu_n-\mu_p=\mu_e$ from beta
equilibrium.  Stability of the uniform phase requires that $e(v,q)$ is
convex \cite{Callen85}, or
\begin{equation}
-\left({\partial p\over\partial v}\right)_q-\left({\partial p\over\partial
 q}\right)_v\left({\partial q\over\partial v}\right)_\mu>0,\qquad
{\rm and}\quad -\left({\partial\hat\mu\over\partial q}\right)_v>0\,.
\label{phase1}
\end{equation}

We assume baryonic and electronic energies of the form
Eqs. (\ref{energy}) and (\ref{eenergy}), which we rewrite as
\begin{equation}
E_N\simeq V(n)+E_{sym}(n)(1-2x)^2 \,,\qquad E_e={3\over4}Y_e\hat\mu\,.
\label{e}
\end{equation}
The beta equilibrium condition is
\begin{equation}
\hat\mu=4E_{sym}(n)(1-2x)=\hbar c(3\pi^2 nx)^{1/3}\,,
\label{betaeq}
\end{equation}
where charge neutrality $x=Y_e$ is used. The conditions in
Eq. (\ref{phase1}) therefore become equivalent to
\begin{eqnarray}
n^2{d^2V\over dn^2}+2n{dV\over dn}+(1-2x)^2\left[n^2{d^2E_{sym}\over
    dn^2}+2n{dE_{sym}\over dn}-2E_{sym}^{-1}\left(n{dE_{sym}\over
    dn}\right)^2\right]&>&0\cr -\left({\partial
    q\over\partial\hat\mu}\right)_v={1\over8S_v}+{3Y_e\over\hat\mu}&>&0\,.
\label{phase2}
\end{eqnarray}
The second of these is always true, so the first one determines
stability.

\begin{figure}
\hspace*{-1.35cm}
  \includegraphics[angle=90,width=.8\textwidth]{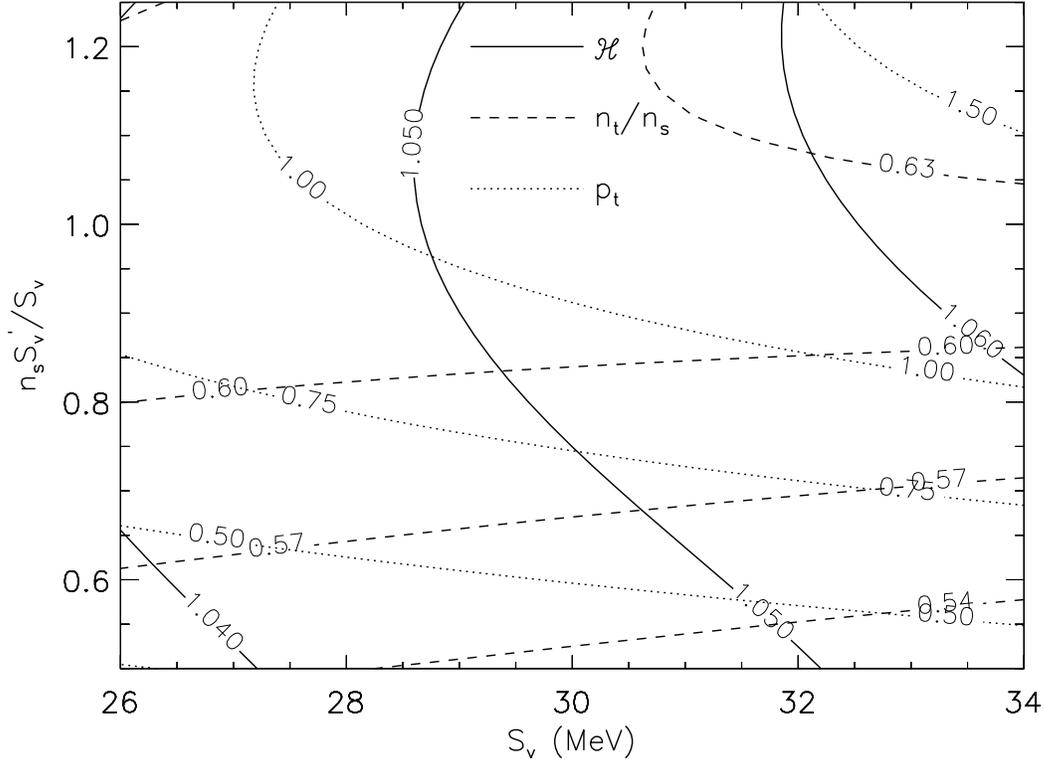}
\caption{Contours of enthalpy ${\cal H}_t$, transition
density $n_t/n_s$, and pressure $p_t$
are displayed as functions of $S_v$ and the logarithmic derivative of
$E_{sym}$ at $n_s$ for the simple model described by Eq. (\ref{vs}).}
\label{phasb}
\end{figure}
As an example, consider the following simple density
dependences for $V$ and $E_{sym}$:
\begin{equation}
V(n)=-16+{K\over18}\left(1-{n\over n_s}\right)^2\,,\qquad
E_{sym}(n)=T_k\left({n\over n_s}\right)^{2/3}+(S_v-T_k)\left({n\over 
n_s}\right)^i\,,
\label{vs}
\end{equation}
where the kinetic contribution to the symmetry energy is $T_k\simeq13$ MeV.
 A convenient parameter
describing the density dependence is $S_v^\prime$, defined by
\begin{equation}
n_s{S_v^\prime\over S_v}=\left({d\ln E_{sym}\over d\ln 
n}\right)_{n_s}={2T_k/3+i(S_v-T_k)\over S_v}\,.
\end{equation}
Applying the stability condition Eq. (\ref{phase2}) determines the
phase boundary density $n_t$.  For reasonable ranges of $S_v$ and
$S_v^\prime$, one finds $0.5<n_t/n_s<0.7$ as seen in Fig. \ref{phasb}.
The quantity
${\cal H}$ is related to the neutron chemical potential,
\begin{equation}
\mu_n(n_t)=-16+{K\over18}\left(1-{n_t\over n_s}\right)\left(1-3{n_t\over 
n_s}
\right)+E_{sym}(n_t)(1-4x_t^2)+\left(n{dE_{sym}(n)\over 
dn}\right)_{n_t}(1-2x_t)^2\,,
\label{phase4}
\end{equation}
by Eq. (\ref{calh}).
The pressure at the boundary, $p_t$, is given by
\begin{equation}
p_t={K\over9}{n_t^2\over n_s}\left({n_t\over n_s}-1\right)+
n_t(1-2x_t)\left[x_tE_{sym}(n_t)+\left(n{dE_{sym}(n)\over
dn}\right)_{n_t}(1-2x_t)\right]\,.
\label{pt}
\end{equation}
The quantities ${\cal H}$ and $p_t$ at the phase boundary are also
shown in Fig. \ref{phasb} for this simple model, using $n_s=0.16$
fm$^{-3}$ and $K=225$ MeV.  The boundary density $n_t$ and $p_t$ {\it
are much more sensitive to the density dependence of the symmetry
energy than to $S_v$ or $K$}.  The general traits found in this simple
model apply to realistic EOSs.  In particular, for realistic EOSs
\cite{LP01},
\begin{equation}
1.04\le{\cal H}\le1.07\,;\qquad
0.20{\rm~MeV~fm}^{-3}<p_t<0.65{\rm~MeV~fm}^{-3} \,.
\label{ulimit}
\end{equation}
These ranges are measures of the current uncertainty in the density
dependence of the symmetry energy.  The range of values for ${\cal
H}_t$ found for the simple model (and shown in Fig. \ref{phasb}) is
very similar.  However, the realistic values for $p_t$ are somewhat
smaller than predicted by the simple model, due to the neglect of
finite-size effects.  Finite-size effects are much less significant
for the neutron chemical potential.

\subsection{Observational application 1: Pulsar glitches}

\begin{figure}
\hspace*{-1.35cm}
  \includegraphics[angle=90,width=.75\textwidth]{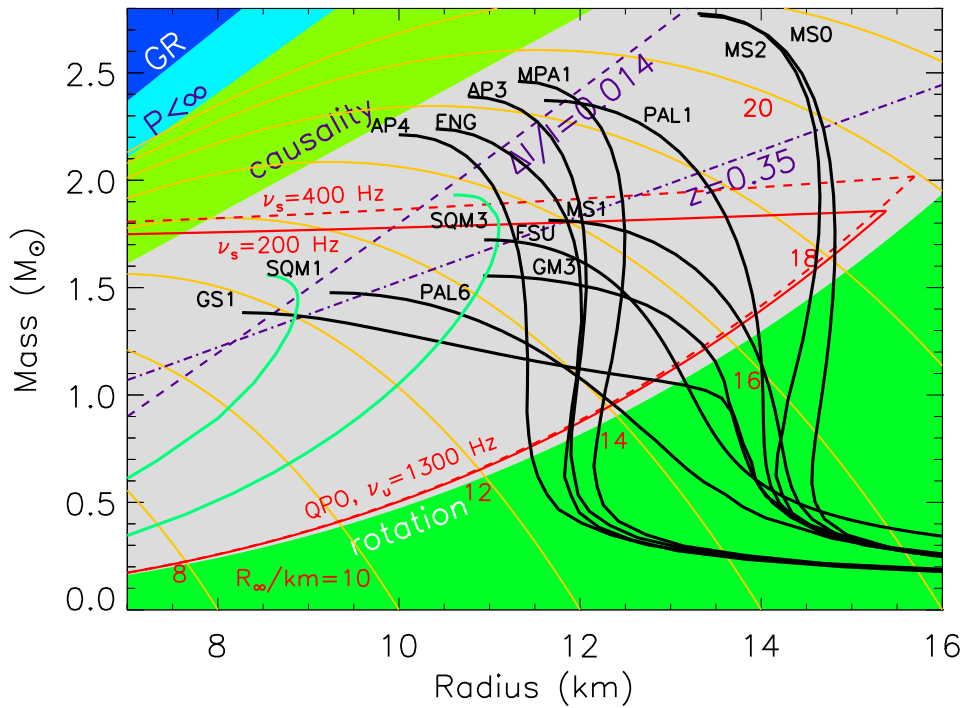}
\caption{The same as Fig. (\ref{mr}), but with additional contours of
fixed redshift $z=0.35$ (dot-dashed blue line), crustal fraction of
moment of inertia $\Delta I/I=0.14$ (dashed blue line), and QPO
constraints (red solid and dashed lines).  For the Vela pulsar, the
$\Delta I/I$ constraint and the assumption that the
pressure at the core-crust interface satisfies $p_t<0.65$ MeV
fm$^{-3}$ implies that allowed masses and radii lie to the right of
the dashed line \cite{LEL}.}
\label{mr3}
\end{figure}
Pulsar glitches, which are sudden discontinuties in the spin-down of
pulsars, seem to involve the transfer of angular momentum from an
isolated component to the entire star.  A leading model for glitches
supposes that the isolated component consists of superfluid neutrons
in the crust \cite{Anderson75}.  As the star spins down due to pulsar
electromagnetic emissions, the superfluid component's angular velocity
differs to a greater and greater extent from that of the
non-superfluid component.  When this differential becomes large
enough,
angular momentum is suddenly transferred from the superfluid component
to the non-superfluid part through vortex unpinning.  In the case of
the Vela pulsar, some 30 years of observations indicate a steady
overall angular momentum transfer rate that indicates at least 1.4 of
the total moment of inertia of the star is involved with the isolated
component \cite{LEL}.  Eq. (\ref{di}) indicats that the crustal
fraction of the moment of inertia is most sensitive to the pressure at
the transition from the crust to the core.  The location of the
transition and the pressure there are sensitive to, among other
nuclear parameters, the density dependence of the nuclear symmetry
energy.
Using the upper limit of $p_t$ from Eq. (\ref{ulimit}) in Eq. (\ref{dii}) allows one to set a
minimum $R$ for a given $M$ for Vela \cite{LEL} which is displayed in
Fig. \ref{mr3} and is approximately:
\begin{equation}
R\ge3.6 +3.9 {M\over{\rm M}_\odot}{\rm~km}\,.
\label{rm1}
\end{equation}
Better understanding of the nuclear symmetry energy would permit a
lower upper limit of $p_t$ to be determined, which could tighten this
constraint.

It should be pointed out that the standard picture of the outer core,
in which superfluid neutrons co-exist with type II superconducting
protons, has been challenged by Link \cite{Link06}.  He notes that
observations of the long-term precession in several pulsars indicates an
incompatibility with creeping neutron vortices, which predict too
large precession frequencies.  Furthermore, the observations seem to
require that the vortices of the inner crust must be able to move with
little dissipation with respect to the solid, which could be
incompatible with a neutron superfluid in the crust in precessing
neutron stars.  Nevertheless, since there is no evidence as yet that
the Vela pulsar is precessing, Link's argument may not apply to this
pulsar.  Why some stars show precession and others do not is an
unsolved problem, and until this is better understood, radius limits
based upon glitches remain tentative. 

\subsection{Observational application 2: Neutron star oscillations}

\begin{figure}
\hspace*{-1.35cm}
  \includegraphics[angle=90,width=.7\textwidth]{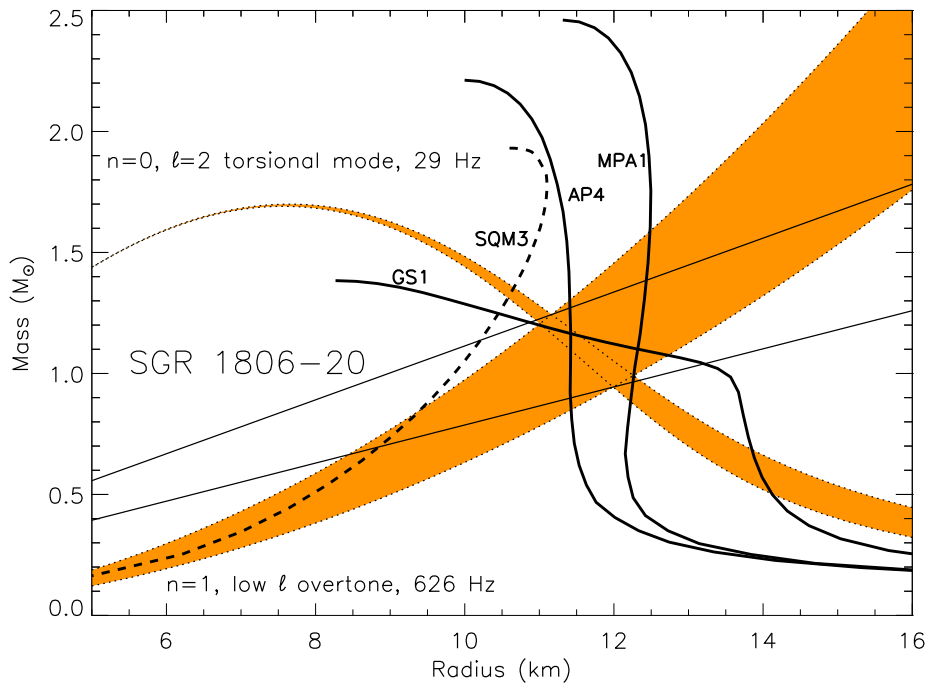}
\caption{Mass-radius diagram showing the constraints from neutron star
  seismology from the soft gamma-ray repeater SGR 1806-20.  ${\cal H}$
  values in the range 1.04 (lower boundaries of permitted regions) to
  1.07 (upper boundaries) were assumed.  A few representative $M-R$
  curves from Fig. \ref{mr} are shown for reference.  The straight
  lines follow the values of $\beta({\cal H}=1.04)=0.116$ and
  $\beta({\cal H}=1.07)=0.165$, determined from the observed
  frequencies, and show how the $(M,R)$ values change as the shear
  velocities are varied.}
\label{seisfig}
\end{figure}
Another kind of limitation on radii can be obtained from observations
of quasi-periodic oscillations in the X-ray emissions following giant
flares in three soft gamma-ray repeaters, SGR 0526-66 \cite{Barat83},
SGR 1806-20 \cite{Israel05} and SGT 1900+14 \cite{SW05}, presumably due
to torsional vibrations of the star's crust \cite{Duncan98}.
Neglecting magnetic fields and the effects of superfluids in the
crust, the frequencies of the fundamental mode and overtones have been
estimated \cite{SA06} to be
\begin{equation}
\omega_{n=0,\ell=2}=2v_t\sqrt{e^\nu\over RR_t}\,,\qquad
\omega_{n>0}=e^{(\nu-\lambda)/2}{n\pi v_r\over\Delta}\,,
\label{modes}
\end{equation}
where $\nu$ and $\lambda$ refer to the average values of the metric
functions, and $v_r$ and $v_t$ are the average radial and transverse
shear speeds, respectively.  Higher order modes ($\ell>2$ in the case
$n=0$ and $\ell>1$ in the case $n>0$) are ignored for a simple
analysis.  To a good approximation, one may assume
$e^\nu=e^{-\lambda}=1-2\beta$.  In the isotropic limit, $v_r=v_t$ and
we assume this; the shear velocities are of order $10^8$ cm s$^{-1}$.
Since the crust thickness $\Delta$ is small compared to $R$ one has,
roughly, $\omega_{n=0}\propto R_\infty^{-1}$, where
$R_\infty=R/\sqrt{1-2\beta}$ is the so-called radiation radius (see
\S \ref{sec:3}).  Also, 
$\omega_{n>0}\propto (1-2\beta)/\Delta\propto R^2/M$.  

Following the general approach of reference \cite{SA06}, in which this
simple analysis is compared to more sophisticated numerical models,
one can write, using Eq. (\ref{dmr}) for $\Delta/R$, the frequencies
($f=\omega/2\pi$)
\be
f_{n=0,\ell=2} &\simeq& 263.3\left({{\rm km}\over R}\right)
\sqrt{({\cal H}-1+2\beta)(1-2\beta)\over
\beta{\cal H}}{\rm~Hz}\,, \nonumber \\
f_{n>0} &\simeq& 1170n \left({{\rm km}\over R}\right)
{{\cal H}-1+2\beta\over{\cal H}-1}{\rm~Hz} \,.
\label{freq}
\ee
If more than one frequency can be measured, these relations
can be used to uniquely identify $R$ and $\beta$, that is, $R$ and
$M$, modulo uncertainties in ${\cal H}$. 

The chief model dependence of the above approach lies in the
velocities $v_r$ and $v_t$.  In the above, they are taken to be equal
and are fitted to numerical simulations ignoring magnetic fields and
superfluidity.  In a more realistic analysis, they could differ from
these values.  However, if $v_r\simeq v_t$, Eqs. (\ref{freq}) indicates
that $\beta({\cal H})$ can be found by eliminating $R$: it is the
solution of a quadratic equation.

The soft gamma-ray repeater SGR 1806-20 has quasi-periodic
oscillations which correspond to the $n=0,\ell=2,6,10$ and
$n=1,\ell=1$ modes \cite{SA06}.  The identification of the larger
$\ell$ modes is more speculative, but the frequencies
$f_{n=0,\ell=2}\simeq29$ Hz and $f_{n=1,\ell=1}\simeq626.5$ Hz are
claimed.  Fig. {\ref{seisfig} shows the allowed regions in the
mass-radius plane for the two modes, and their overlap, assuming
$1.04\le{\cal H}\le1.07$.  It is clear that the overtones are much
more sensitive to assumptions about ${\cal H}$ (an EOS attribute) than
is the fundamental mode.  The two values of $\beta$ associated with
these values of ${\cal H}$ are $\beta({\cal H}=1.04)=0.116$ and
$\beta({\cal H}=1.07)=0.165$, respectively.  Because lines of constant
$\beta$ are approximately parallel to the overtone constraint, but
perpendicular to the fundamental mode constraint, uncertainties in the
velocity would be reflected as relatively larger uncertainties in the
latter than the former.

\subsection{Observational application 3: Thermal relaxation of the crust}

After the formation of a proto-neutron star in the aftermath of a
gravitational collapse supernova event, cooling of the core proceeds
via neutrino emission, leading to temperatures of order $10^9$ K
within years (see \S \ref{sec:pns}).  During this period, the star is
not in complete thermal equilibrium due to the relatively long thermal
relaxation time of the crust, which is expected to be of order 10 to
100 years \cite{LvRPP}.  The excess heat of the crust is
preferentially transported by electron conduction into the star's
isothermal core, dominating radiation by thermal emission from the
surface.  However, after a time which we will denote by $t_w$, the
surface will finally come into thermal equilibrium with the core,
causing a reduction in the surface temperature.  Depending upon
whether or not the core neutrino emission processes are rapid or
conventional (see the terminology in \S \ref{sec:stand}), the drop in
surface temperature will be a factor of 2 (and relatively slow) to 10
(and relatively abrupt).  Following this reduction, the surface
temperature stabilizes at a value about 100 times the core
temperature \cite{vanriper88}.  The time $t_w$ is simply related to
the thermal conductivity and specific heat of the crust, its
thickness, and redshift \cite{LvRPP,Gnedin01}, so that its observation
could be used to constrain neutron star structure and composition.

In addition, some accreting neutron stars are observed to undergo
long-duration X-ray bursts, sometimes called superbursts, which appear
to be powered by thermonuclear He fusion.  During the burst, some of
the heat is transported into the interior, warming the crust.
Observations of quiescent thermal emission following the bursts
would therefore probe the thermal relaxation timescale of the crust as well
as the rate of core neutrino emission \cite{Rutledge01}.  It is
therefore useful to investigate the relation between the thermal
relaxation time and the structure of neutron stars.

To begin, it is necessary to supplement the hydrostatic equilibrium
equations, Eq. (\ref{tov}), with those of relativistic heat transport,
\begin{eqnarray}
L_\nu&=&-4\pi r^2Ke^{-(\lambda-\nu)/2}{\partial
    Te^{\nu/2}\over\partial r}\,,\cr
{\partial e^\nu L_\nu\over\partial r}&=&-4\pi r^2
e^{(\lambda+\nu)/2}C_V{\partial T\over\partial t}\,,
\label{trans}
\end{eqnarray}
where $L_\nu(r)$ is the local neutrino luminosity, $C_V$ is the
specific heat capacity at fixed volume, and $K$ is the thermal
conductivity.  Explicit neutrino emission is neglected in
Eq. (\ref{trans}).  Following thermal relaxation, the redshifted
temperature $T(r)e^{\nu(r)/2}$ becomes constant within the star.
Applying these equations to the crust, and assuming $\Delta/R<<1$ so
that $e^\nu\simeq e^{-\lambda}\simeq\sqrt{1-2GM/Rc^2}$ throughout the
crust, and defining $x=(R-r)/\Delta$, one finds
\begin{equation}
\left({R\over \Delta}\right)^2{\partial\over\partial x}
\left[K{\partial T\over\partial
    x}\right]=C_Ve^{3\lambda/2}{\partial T\over\partial t}\,.
\label{trans1}
\end{equation}
If it is assumed that the conductivity $K$ and specific heat $C_V$ can
each be written as separable functions of density and temperature,
then one can make the same assumption for $T$:  
$T(r,t)=T_0\psi(r)\phi(t)$.  Generally, $C_V\propto T$, and $K\propto
T^0$ or $T$ depending on density.  The temporal dependence $\phi$ in
these two cases is then either linear or quadratic: $\phi=1-t/t_w$ or
$\phi=\sqrt{1-t/t_w}$, where the thermal lifetime of the crust is
\begin{equation}
t_w\propto \Delta^2e^{3\lambda/2}\,,
\label{trans3}
\end{equation}
with the proportionality constant scaling as $C_V/K$.
This behavior is borne out in detailed numerical simulations of neutron star
cooling \cite{LvRPP} in which, for superfluid crusts,
\begin{equation}
t_w\sim5\left({\Delta\over{\rm km}}\right)^2\left(1-{2GM\over
  Rc^2}\right)^{-3/2}{\rm~yr}.
\label{trans4}
\end{equation}
For non-superfluid crusts, the cooling times are about a factor of 3.5
times larger.

Although no newly-formed neutron stars have ever been observed so
young as to show this surface temperature drop, it has been suggested
that the resaon magnetars (highly magnetized neutron stars) have
abnormally large surface temperatures is that they have heat sources
(presumably powered by magnetic field decay) located in a thin layer
in the outer crust \cite{Kaminker06}.  
If the heat source was located deeper, the energy
would be radiated efficiently by neutrinos and could not heat the
crust. This heat source warms up the crust and produces an
inhomogeneous temperature distribution within the star, with parts of
the crust much warmer than the deep crust and core.  The
thermalization times calculated above are consistent with
observations, but tell us more about the heat source than about
structural constraints.

This theory can be applied to superbursts or other cases in which the
crust is heated for a long-enough duration that the crust thermalizes
throughout.  For example, observations of the soft X-ray transient 1H
1905+000, which underwent a burst about 20 years ago, indicate that
the crust cooling times are somewhat shorter than Eq. (\ref{trans4})
predicts, suggesting that the crust's thermal conductivity is higher
than expected or that enhanced core neutrino cooling is taking place
\cite{Jonker06}.  Another source, KS 1731-260
\cite{Kuulkers02,Wijnands02}, which produced a superburst lasting
about 12 years, went into quiescence and has since shown a strong
decline of X-ray luminosity, also suggesting a high thermal
conductivity for the crust and rapid neutrino core cooling.  Further
work needs to be done, however, to test the assumption that the crusts
of these sources are being heated to complete thermal equilibrium as
otherwise shorter cooling times might be expected.

\section{Cooling neutron stars}
\label{sec:3}
\subsection{Observations of thermal emission}
The quantity inferred from thermal observations of a neutron star's
surface is the so-called radiation radius
\begin{eqnarray}
R_\infty=R/\sqrt{1-2GM/Rc^2}\,,
\end{eqnarray}
which results from a combination of flux and temperature measurements,
both redshifted at the Earth from the neutron star's surface.  For
example, the un-redshifted Kirchoff's law for a black body,
\begin{equation}
F=\left({R\over d}\right)^2\sigma T^4\,,
\label{flux}
\end{equation}
becomes 
\begin{equation}
F_\infty=\left({R_\infty\over d}\right)^2\sigma T_\infty^4 
\end{equation}
when redshifted to the Earth.
Contours of the quantity $R_\infty$ are displayed in Fig. \ref{mr}.  A
measured value of $R_\infty$ sets upper limits to both $R$ and $M$,
but without an independent estimate of mass or radius only limited
constraints are possible.  The major uncertainties involved in the
determination of $R_\infty$ include the distance, interstellar H
absorption that is important near the peak and at lower energies of
the spectrum, and details concerning the composition of the atmosphere
and its magnetic field strength and structure.  The most reliable
measurements will originate from sources for which one or more of these
uncertainties can be controlled.  For example, the nearby sources RX
J1856-3754 \cite{Walter02} and Geminga \cite{Walter06} have measured
parallaxes, and the quiescent X-ray binaries in globular clusters for
which reliable distances are (or will soon be) available.  In addition,
the X-ray binaries in globular clusters, having undergone recent
episodes of accretion, are expected to have low magnetic field H-rich
atmosperes, presumably, the simplest of all situations.
\begin{figure}
\hspace*{-1.35cm}
  \includegraphics[angle=90,width=.75\textwidth]{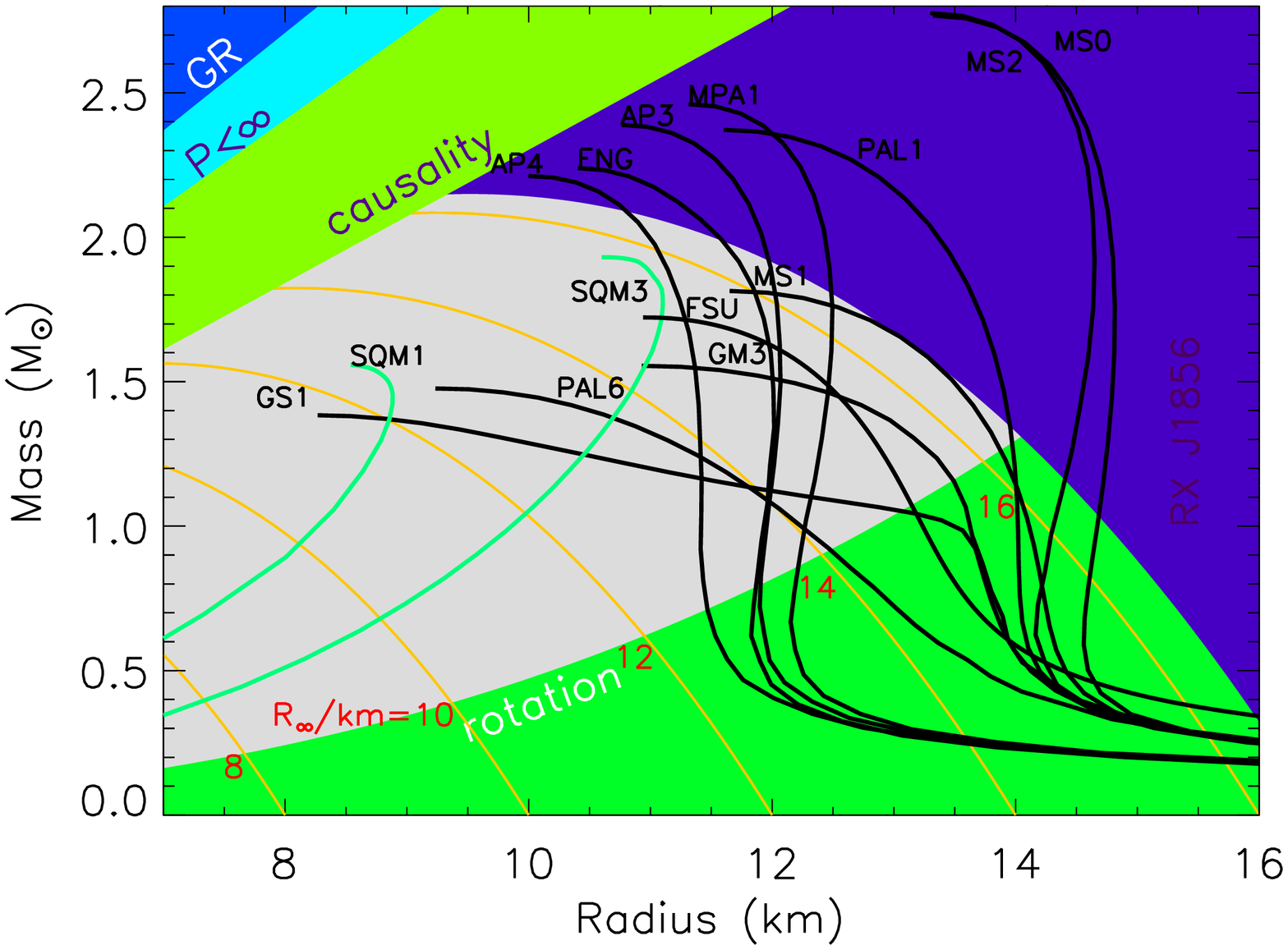}
\hspace*{-1.35cm}
\includegraphics[angle=90,width=0.75\textwidth]{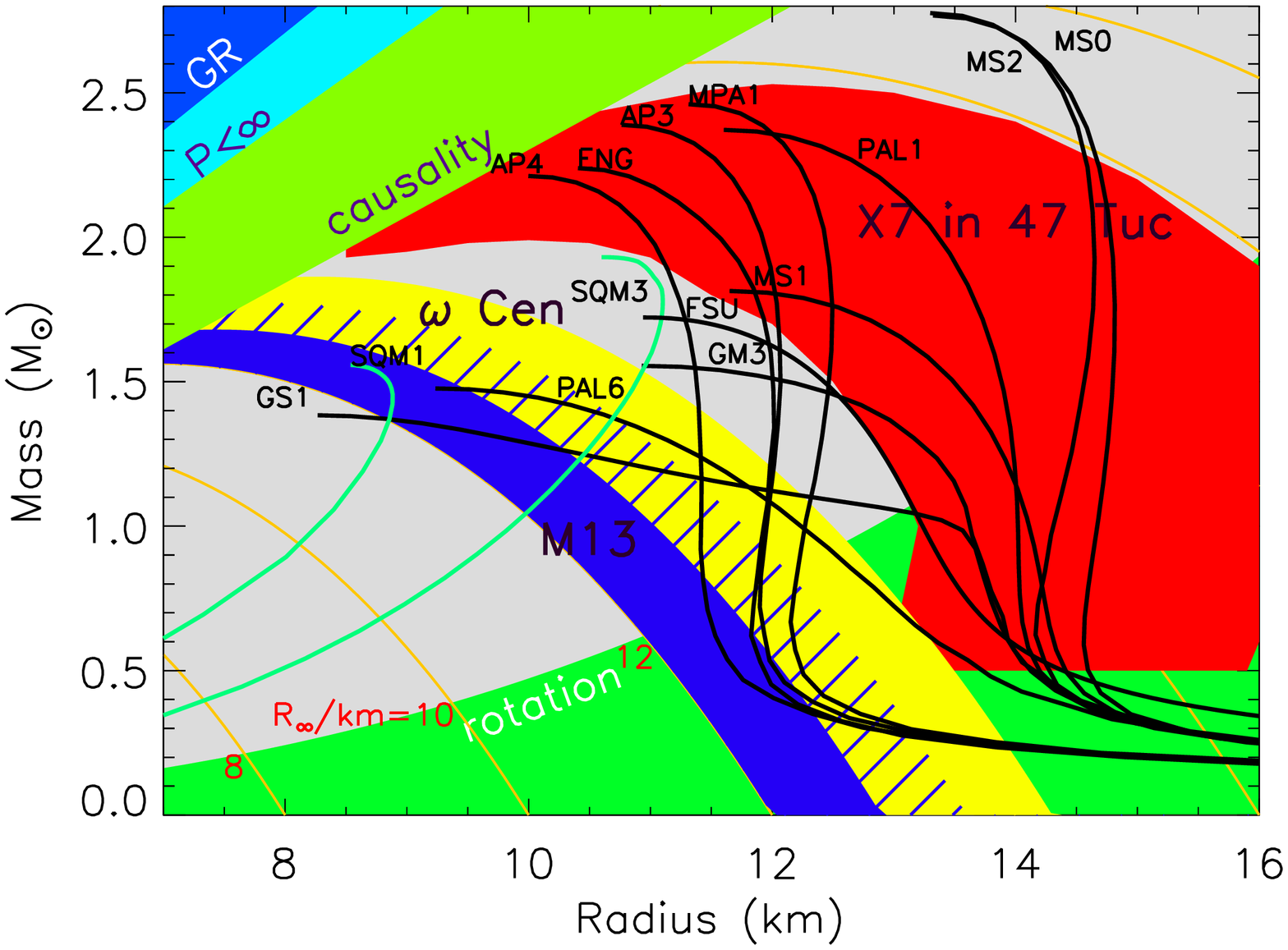}
\caption{The same as Fig. \ref{mr}, but displaying allowed mass and
radius regions for thermally-emitting neutron stars. Top panel is for
RX J1856-3754 (assuming a distance of 120 pc) \cite{Walter02}, while
the bottom panel shows results for globular cluster sources in M13
\cite{Gendre03a}, $\omega$ Cen \cite{Gendre03b} and 47 Tuc
\cite{Heinke06}.}
\label{mr1}
\end{figure}

A characteristic of these X-ray sources, in the cases in which the
distances are small enough to allow detection of optical emission, is
that the optical fluxes are a factor $f\simeq 5-7$ times the amount
expected from extrapolating the X-ray blackbody fits onto the
Rayleigh-Jeans tail.  This is one consequence of the neutron star
atmosphere and its redistribution of the flux from a simple blackbody.
The slope of the Rayleigh-Jeans tail is a measure of the optical
temperature, and typically $T_{opt}\sim T_X/2$.  Therefore, in order
to properly fit the overall flux distribution, a larger radius is
needed than the X-ray fit alone would imply.  The optical flux may be
written as
\begin{equation}
F_{opt}\propto 4\pi\left({R_{opt}\over d}\right)^2T_{opt}=
4\pi f \left({R_X\over d}\right)^2T_X \,,
\end{equation}
whereas the total stellar radius is a function of the effective optical
and X-ray radii:
\begin{equation}
R=\sqrt{R_{opt}^2+R_X^2}=R_X\sqrt{1+{T_X\over T_{opt}}f}
\simeq R_X\sqrt{1+2f}\,.
\end{equation}
This explicitly shows that a factor $f\sim 5-7$ results in an increase
in inferred radius of about a factor 3--4 over that inferred from X-rays
alone.  Radii inferred from X-ray data alone are therefore suspect,
and often much too small.

Results from atmospheric fitting of the data for RX J1856-3754 and
various globular cluster sources are displayed in Fig. \ref{mr1}.
An interesting feature of Fig. \ref{mr1} is that if one accepts the
results at face value, only EOS curves that pass through
all the permitted regions can be accepted.  This would eliminate
several relativistic field-theoretical EOSs such as GM3,
FSU, MS0, MS1 and MS2, while still permitting non-relativisitic
potential models like AP3 and AP4.  (It must be stressed, however,
that the approach of Muller \& Serot, of which the EOSs MS0, MS1 and
MS2 are typical examples, can be straightforwardly generalized to
yield results that can accomadate the mass and radius constraints
implied in Fig. \ref{mr1}.)  The Dirac-Brueckner relativisitic
field-theoretical models like ENG and MPA1 are allowed.  Note that the
surviving EOSs all support large masses, which could be
crucial if the 2.1 M$_\odot$ value for PSR J0751+1807 remains robust.

\subsection{Implications for neutron star cooling}
\label{sec:stand}

For many years, it was common to adopt a {\it standard cooling}
scenario, with relatively slow cooling, in which the dominant neutrino
processes were the relatively slow modified Urca processes involving
nucleons.  However, neutrino production would become significantly
more rapid if the symmetry energy and proton fraction is large or in
the presence exotic matter such as hyperons, kaons or deconfined
quarks, which would permit the direct Urca processes \cite{LPPH91,PPLP92}.

The direct Urca process involving baryons is the simple weak decays 
and the reverse reactions
\begin{eqnarray}
B_1 \rightarrow B_2 + \ell + {\overline \nu}_\ell \,; \qquad 
B_2 + \ell \rightarrow B_1 + \nu_\ell \,,
\label{bproc}
\end{eqnarray}
where $B_1$ and $B_2$ are baryons, and $\ell$ is a lepton, either an
electron or a muon. 
When the decaying baryon is a neutron, the occurence of this process
is contingent upon the presence of a sufficient number of protons
(determined entirely by the density dependent symmetry energy) so that
energy and momentum can be conserved simultaneously.

In contrast to the direct Urca process, the modified Urca process
includes a bystander baryon that facilitates the simultaneous
conservation of energy and momentum.  The additional particle slows
the neutrino energy emission rate by a factor $\sim(T/T_F)^2$, where
$T_F\sim30$ MeV is the Fermi energy at the equilibrium density $n_s$.
Since a typical temperature is of order $0.1-1.0$ MeV, the modified
Urca rate is millions of times slower than the direct Urca rate.
However, the direct Urca process becomes possible, albeit with a
smaller strength (matrix element), when a Bose condensate, deconfined
quarks, or hyperons are present.

The paradigm of slow standard (as opposed to rapid exotic) behavior
stemmed from models of the nuclear force which had relatively small
isospin dependences, such that the beta equilibrium proton and
electron fractions would be very small and the direct Urca process
among baryons would be forbidden.  With the realization that newer
versions of nuclear force models have greater isospin dependence and
predict larger proton fractions has come the possibility that the
direct Urca process can occur even in the absence of exotic matter 
\cite{LPPH91}.
Therefore, the detection of a rapidly cooling neutron star is not
sufficient to unambiguously prove the existence of exotic
matter in a neutron star interior.  Compounding the difficulty of
interpreting cooling observations is the role of superfluidity, which
can turn off the direct Urca process near the superfluid's critical
temperature.

While it is outside the intended scope of this review to develop the
details of neutron star cooling (see, {\it e.g.}, Ref. \cite{Page04}
and references therein), it should be clear that it
sensitively involves the internal composition.  Inasmuch as the
internal composition depends on the overall mass of the neutron star,
an indirect measure of the mass of the star is possible.  This is
especially pertinent for those supernova remnants which have yet to
yield detectable sources of thermal emission, such as 3C58
\cite{Slane04}, and supernova remnants G084.2-0.8, G093.3+6.9,
G127.1+0.5 and G315.4-2.3 \cite{Kaplan04}.  If these remnants in fact
contain neutron stars, the stars may have cooled rapidly due to one of
the several direct Urca processes, which is much more efficient than
standard cooling due to the modified Urca process.  Stars with high
proton fractions may cool via the direct Urca process, and this
depends upon the density dependence of the symmetry energy.  A
radiation radius determination for the same star would allow an
extraction of the radius itself.

\section{Other radius observables}
\label{sec:4}

Besides measurements of radiation radii from cooling neutron stars,
ther limits to the radius could be set by pulse profiles, redshifts,
explosions or accretion on neutron star surfaces with
Eddington-limited fluxes, quasi-periodic oscillations (QPO's) from
accreting neutron stars, and measurements of the moment of inertia of
binary pulsars.
We highlight these avenues in turn below.

\subsection{Pulse profiles}

Gravitational lensing strongly suppresses the amplitude of variations
from pulsing (rotating) neutron stars.  Essentially, gravitational
light-bending allows an observer to see much more than the facing
hemisphere.  Therefore, analysis of pulsations in the thermal emission
from pulsars allows limits to be placed on the ratio $M/R$.  The
locally emitted flux is proportional the the local temperature to the
fourth power, which varies since is controlled by the local magnetic
field strength.  Assuming dipole fields, it has been shown for many
pulsars that pulse fractions comparable to
the observed ones can be obtained only with stellar radii larger than
those which are predicted by current models of neutron star struture,
or with low stellar masses \cite{Page95}.  However, different field
geometries can alter this conclusion.  

The same physics influences other pulsed emissions from neutron stars,
such as Her X-1, which is an eclipsing X-ray emitting source with
X-ray pulsations.  The emission is observed to be occulted by the
inner-disk edge which implies that the neutron star emits a pencil
beam from its near pole and a fan beam from its far pole
\cite{Scott00}.  Gravitational light-bending strongly influences the
emission, and Leahy \cite{Leahy04} derived $M/R\simeq0.121 - 0.128$
M$_\odot$ km$^{-1}$, or $z\simeq0.247 - 0.268$, for this source.  For
the inferred mass of Her X-1, 1.29 -- 1.59 M$_\odot$ \cite{van95}, this
translates into a radius range $10.1 < R/{\rm km} < 13.1$. This
result may be model dependent, and more complex geometries and
emissivities should be considered, but it nevertheless indicates the
potential of the technique.

\subsection{Redshift}
\label{sec:4:1}
Another observable is the redshift
\begin{equation}
z=\left(1-{2GM\over Rc^2}\right)^{-1/2}-1\,.
\label{red}
\end{equation}
Possibly the best case is that of the active X-ray burster EXO
0748-676 \cite{Cottam02} for which a pair of resonance scattering
lines consistent with Fe XXV and XXVI imply $z\simeq0.345$.  Later, 45
Hz oscillations in the average power spectrum of 38 thermonuclear
X-ray bursts were found and interpreted as the neutron star spin
frequency \cite{Villarreal04}.  Furthermore, the widths of the lines
observed by Ref. \cite{Cottam02} are consistent with this spin
frequency as long as the star's radius is between about 10 and 15 km
\cite{Villarreal04}.  However, the spectral identification of
Ref. \cite{Cottam02} is not universally accepted \cite{Sidoli05}.
Importantly, as long as the star's spin frequency is less than about
600 Hz, the measured $z$, which refers to the equatorial radius of the
spinning star, is less than 2\% different to that of the non-spinning
star \cite{Bhattacharyya06}.  More observations and better modeling of
the line profiles could lead to additional parameters involving $M$
and $R$ and serve as useful constraint.  Other sources with lines have
been claimed, but never confirmed with more sensitive instruments.

Clearly, a simultaneous
measurement of $R_\infty$ and $z$ would determine both $M$ and $R$:
\begin{eqnarray}
R &=& R_\infty(1+z)^{-1},\cr
M &=& {c^2\over2G}R_\infty(1+z)^{-1}[1-(1+z)^{-2}]\,.
\end{eqnarray}
The contour $z=0.35$ is shown in comparison to typical mass-radius
curves in Fig. \ref{mr3}. However, it is often the case that it is the
quantity $R_\infty/d$ that is determined,  where $d$ is the distance to the
source, which is often uncertain.

  It is expected that planned X-ray missions, like Constellation-X
which has a resolution and effective area several times larger than
Chandra and XMM, will discover and precisely measure more spectral
lines from neutron star surfaces.
\subsection{Eddington limit}
\label{sec:4:2}
A major hurdle in determining radii from observations of thermal
radiation from cooling neutron stars is that the distances, $d$, to the
sources are generally not well known.  Even if a source's redshift can
be determined, an additional relation
involving $M$, $R$ and/or $d$ is needed to independently find $M$ and $R$. 
Burst sources, which are believed to involve thermonuclear X-ray bursts on
neutron star surfaces, have long been known to have peak fluxes that are to
order of magnitude comparable to the Eddington limited flux:
\begin{equation}
F_{edd,\infty}={cGM\over\kappa d^2}\sqrt{1-2GM/Rc^2}\,,
\label{fedd}
\end{equation}
where $\kappa$ is the opacity.  If the opacity is assumed to be
dominated by electron scattering, one has 
\begin{equation}
\kappa=0.2(1+X)~{\rm cm}^2~{\rm g}^{-1}\,, 
\end{equation}
where $X$ is the hydrogen mass fraction.
Following a burst, the source apparently relaxes to a quiescent state,
which is believed to involve the radiation of thermal emission.  Study
of the X-ray spectrum can then yield a cooling flux and temperature
$F_\infty$ and $T_\infty$ observed at Earth:
\begin{equation}
F_\infty/(\sigma
T_\infty^4)=(T_{eff}/T_{\infty})^4R^2d^{-2}(1-2GM/Rc^2)^{-1},
\label{flux1}
\end{equation}
where $T_{eff}$ is the effective temperature and $T_\infty$ is the
observed color temperature.  The ratio $T_{eff}/T_\infty$ is a color
correction factor that depends on the composition, gravity and $T_{eff}$
\cite{Madej04}.

If the observed peak flux is approximately the
Eddington flux, if the opacity and color correction factor are
well-understood, if the quiescent flux observed in-between bursts is
thermal emission with a measureable temperature $T_\infty$, and if spectral
features are observed which permit a determination of the object's
redshift, these multiple observations then contain the information
needed to identify uniquely the mass and the radius of a single star.
To make this clear, combine Eqs. (\ref{red}), (\ref{fedd}) and
(\ref{flux1}):
\begin{eqnarray}
M &=& {c^5\over4G\kappa}{\alpha\over F_{edd,\infty}}
~[1-(1+z)^{-2}]^2(1+z)^{-3},\cr
R &=& {c^3\over2\kappa}{\alpha\over 
F_{edd,\infty}}~[1-(1+z)^{-2}](1+z)^{-3},\cr
d &=& {c^3\over2\kappa}{\sqrt{\alpha}\over 
F_{edd,\infty}}~[1-(1+z)^{-2}](1+z)^{-2}\,.
\label{mrd}
\end{eqnarray}
In the above, the factor
$\alpha=(T_{\infty}/T_{eff})^4F_\infty/(\sigma
T_\infty^4)=(R_\infty/d)^2$.  The burst source EXO 0748-676 is
especially amenable to analysis \cite{Ozel06}, as this is one of the
few sources with an estimated redshift (see above), $z=0.345$.  Its
peak flux, to be identified with $F_{edd}$, is
$2.25\pm0.23\times10^{-8}$ erg cm$^{-2}$ s$^{-1}$, and the ratio
$F_\infty/(\sigma T_\infty^4)$ is observed to be $1.14\pm0.10$
(km/kpc)$^2$.  Using Eqs. (\ref{mrd}) and these observables, along
with standard assumptions for $\kappa$ and $T_\infty/T_{eff}$, Ozel
\cite{Ozel06} finds for this burst source $M=2.1\pm0.28$ M$_\odot$,
$R=13.8\pm1.8$ km, and $d=9.2\pm1.0$ kpc.  These errors do not include
uncertainties in the assumed opacity or color-corrections.

Note, however, that statistical studies \cite{Van Paradijs81} indicate
that the average maximum luminosity of X-ray bursts at infinity are
about 60\% larger than Eq. (\ref{fedd}) predicts.  Various suggestions
have been proposed to explain the super-Eddington luminosities in
these bursts.  Additionally, the value of $\kappa$ is as yet difficult
to ascertain to good precision because the composition is uncertain
and contributions from bound-free opacities are uncertain.
Nevertheless, if these problems can be addressed, observations of this
source could ultimately lead to the simultaneous determination of the
mass and radius of a neutron star.  In short, this is a promising
avenue to pursue as it holds the promise to establish the mass and radius
of the same neutron star.

An interesting variation of the theme of the Eddington flux concerns
the emission from neutron stars from accretion in low-mass X-ray
binaries (LMXRBs).  In the spreading layer model \cite{Inogamov99},
accreting matter spreads from the equator to the poles on the neutron
star surface, locally radiating at the Eddington flux.  The total
emission can approach the Eddington limit if the matter spreads all
the way from the equator to the poles.  Calculations indicate that the
overal emitted spectrum is that of a diluted blackbody.
Schematically, the effective temperature of the radiation ({\it
i.e.,}, the Eddington temperature $T_{Edd}$, is determined by the
balance between surface gravity and radiative acceleration:
\begin{equation}
{GM\over R\sqrt{1-2GM/Rc^2}}={\sigma T_{Edd}^4\over c}\kappa\,,
\end{equation}
where $\kappa=0.2(1+X)$ cm$^2$ g$^{-1}$ is the electron scattering
opacity and $X$ is the hydrogen mass fraction.  The observed color
temperature is
\begin{equation}
T_c=T_{Edd}f_c\ell^{1/4}\sqrt{1-2GM/Rc^2}\,,
\end{equation}
where $\ell\sim0.8$ is the correction for centrifugal reduction of
gravity and $f_c$ is a correction factor for spectral hardening.  As a
result, the neutron star radius can be found from
\begin{equation}
R^2={\ell f_c^4c^3\over2\sigma T_c^4\kappa}{2GM\over
c^2}\left(1-{2GM\over Rc^2}\right)^{3/2}\,.
\end{equation}
This is essentially equivalent to the middle formula in
Eq. (\ref{mrd}).  In practice, modeling of the radiative transfer
makes some corrections to the above, but it can be observed that the
contours in an $M-R$ diagram for fixed values of $T_c$ are nearly
orthogonal to contours of $R_\infty$, and are less sensitive to
variations in $M$.  For 1.4 M$_\odot$ neutron stars, and taking $X\sim1$ and
$2.3<T_c/{\rm keV}<2.5$ (the observed range in LMXRBs), this method
implies $12<R/{\rm km}<15$ \cite{Suleimanov06}.  It is important to
note that the spreading layer method only utilizes the color
temperature of the source.  If the same source could be observed after
accretion ceases, the information obtained during accretion could be
supplemented during quiescence to independently determine $M$ and $R$
as in the preceding analysis for EXO 0748-676.

\subsection{Quasi-periodic oscillations from accreting stars}
\label{sec:4:5}

Accreting matter in LMXRBs flows from the companion through
its inner Lagrangian point into an accretion disc.  Through viscous
dissipation, the matter slowly loses angular momentum and spirals
closer to the neutron star.  Eventually, the matter reaches the inner
edge of the accretion disc, which is thought to be close to the
innermost circular stable orbit (ISCO), which in the Schwarzschild
geometry is $R_{ISCO}=6GM/c^2$.  Thereafter, the matter quickly
spirals onto the neutron star's surface, emitting X-rays as
gravitational potential energy is converted into thermal energy.  In
some of these binaries, 25 so far \cite{vanderklis06}, Fourier
analysis of the X-ray emission shows quasi-periodic oscillations, or
QPOs.  Usually, a frequency of 200-400 Hz is observed, which is
thought to be either the spin frequency of the star or half of that.
In adddition, two high frequency kHz peaks are also seen, separated by
the presumed neutron star spin frequency.  Explanations of the QPOs
range from beat-frequency models connected with the ISCO frequency and
the star's spin \cite{Miller98} to vertical epicyclic frequencies or
resonant interactions \cite{Stella98}.  Observations indicate that
these QPOs rise slowly in frequency and suffer a drop in their quality
factor as these frequencies saturate at their largest values
\cite{Barret06}.  The quality factor is the frequency divided by the
peak's full width at half maximum.  This behavior is what is expected
if the drop is produced by the approach of accreting matter to the
ISCO.  It should be noted that all proposed models connect the upper
kHz QPO with the ISCO orbital frequency to within a percent for
neutron stars.

Since the orbital frequency at the ISCO is a function of neutron star
mass (if the star is rotating slowly), and the inner edge of the
accretion disc must lie at or outside the ISCO, the inferred frequency
thus provides a limiting mass for the neutron star.  Inferred ISCO
frequencies range from $\nu_{ISCO}\simeq1220$ Hz for 4U 1636-53,
$\nu_{ISCO}\simeq1230$ Hz for 4U 1608-52, and $\nu_{ISCO}\simeq1310$
Hz for 4U 1728-34 \cite{Barret06}.  The inferred upper limit to the
masses, including the lowest order correction for neutron star spin,
is \cite{Miller98}
\begin{equation}
M\le \left(2200{\rm~Hz}\over\nu_{ISCO}\right)
\left(1+0.75{cJ\over GM^2}\right){\rm~M}_\odot\,,
\label{qpo1}
\end{equation}
where $J=2\pi fI$ is the angular momentum of a star spinning with
frequency $f$.  $I$ is the moment of inertia (see \S \ref{sec:mom}).
Eq. (\ref{qpo1}) is a slowly varying function of $R$, and the upper
limit is shown by the nearly straight lines in the range 1.8-2.0
M$_\odot$ in Fig. \ref{mr3} for $\nu_{ISCO}=1300$ Hz and neutron star
spin frequencies $f=200$ and 400 Hz.  The closer the observed QPO
frequency is to that of the ISCO, the closer the neutron star mass
must be to this upper limit.

In addition, the neutron star radius must be smaller than the ISCO,
which implies \cite{Miller98}
\begin{equation}
R\le\left({1950{\rm~Hz}\over\nu_{ISCO}}\right)\left(1+0.20{cJ\over
GM^2}\right){\rm~km}\,.
\label{qpo2}
\end{equation}
Radii limits for masses less than the upper limit Eq. (\ref{qpo2})
scale with $M^{1/3}$ and are also shown in Fig. \ref{mr3}, for the
same $\nu_{ISCO}$ and $f$'s as for the mass limit.  These radii limits
are somewhat more restrictive than that implied by the 716
Hz pulsar J1748-2446ad.

As upper limits to mass and radius, QPOs are not yet restrictive.
However, if the measured frequencies are actually measurements of the
ISCO, the implied masses are near 2 M$_\odot$.  Interestingly, in the
case of 4U 1636-536, phase-resolved spectroscopy with the VLT suggest
the neutron star is in the range of 1.6--1. M$_\odot$
\cite{Casares06}, providing support for this interpretation and for
the existence of high-mass neutron stars.

\subsection{Moments of inertia} 
\label{sec:mom}

The discovery of the double-pulsar system PSR J0737-3039A \&
B~\cite{Burgay03,Lyne04} provides physicists with the most
relativistic system yet. Perhaps most intriguingly, it could provide a
measurement of spin-orbit coupling that could eventually lead to a
determination of the moment of inertia $I_A$ of star A \cite{Lyne04}.
Spin-orbit coupling advances the periastron of the orbit (apsidal
motion) beyond the standard post-Newtonian advance as well as causing
a precession of the orbital plane about the direction of the total
angular momentum of the system \cite{Damour88} (geodetic
precession). Since the masses of both stars are already accurately
determined, and the moment of inertia can be expressed as a tight
relation involving only $M$ and $R$ and no other EOS parameters, the
measurement of $I_A$ could have enormous importance.

To date, the best-determined moment of inertia is for the Crab pulsar
\cite{Bejger03}. This was based on an estimate \cite{Fesen97} for the
mass of the ionized portion of the Crab's remnant, $4.6\pm1.8$
M$_\odot$, which implies a lower limit to the Crab pulsar's moment of
inertia of $97\pm38$ M$_\odot$ km$^2$.  This limit would rule out only
very soft EOSs.

According to Barker \cite{Barker75}, the spin and orbital angular
momenta evolve satisfying
\begin{eqnarray}
\dot{\vec S_i}&=&{G(M_i+3M)\over2M_i a^3c^2(1-e^2)^{3/2}}
\vec L\times\vec S_i\,,
\label{eqn:evolve,a}\\
\dot{\vec L^{SO}}&=&\sum_i{G(4M_i+3M)\over2M_ia^3c^2(1-e^2)^{3/2}}
(\vec{S_i}-3{\vec L\cdot\vec{S_i}\over|\vec L|^2}\vec L)\,,
\label{eqn:evolve,b}
\end{eqnarray}
where the superscript $SO$ refers to the spin-coupling contribution
only (there are also first- and second-order post-Newtonian terms, 1PN
and 2PN, respectively, unrelated to the spins, that contribute to this
order).  Here $a$ is the semimajor axis of the effective one-body
orbital problem (sum of the semi-major axes of the two stellar
orbits), $e$ is the eccentricity, $M=M_A+M_B$, and $M_i$ refers to the
mass of either binary component.  To this order, one may employ the
Newtonian relation for the orbital angular momentum:
\begin{equation}
|\vec L|={2\pi\over P} {M_AM_Ba^2(1-e^2)^{1/2}\over 
M}=M_AM_B\sqrt{Ga(1-e^2)\over M}\,,
\end{equation}
where $P$ is the orbital period and $M=M_A+M_B$.  Then, from
Eq. (\ref{eqn:evolve,a}), the spin precession periods are
\begin{equation}
\label{eqn:kidder} 
P_{p,i} = \frac{2c^2aPM(1-e^2)}{G(M-M_i)(M_i+3M)}\,, 
\end{equation} 
which are not identical for the two components unless they are of
equal mass.  Note that the spin precession periods are independent of the
spins.  Also note that if the spins are parallel to $\vec L$, there is,
first, no spin precession, and second, the spin-orbit contribution
to the advance of the periastron is in the opposite sense to the
direction of motion.

The relevant observational parameters for PSR J0737-3039 extracted
from Ref. \cite{Lyne04} are:
\be
M_A &\simeq& 1.34{\rm~M}_\odot\,, \quad M_B\simeq1.25{\rm~M}_\odot\,, \quad
\quad a/c = 2.93~{\rm s}\,, \quad e\simeq0.088\,, \cr 
P_A &\simeq& 22.7\rm{~ms}\,, \quad P_B\simeq2.77~ {\rm s}\,, \quad
 \quad P\simeq 0.102~{\rm day}\,. 
\ee
From these values, we observe that $P_{pA}\simeq74.9$ yrs and
$P_{pB}\simeq70.6$ yrs.  With these parameters, we can form the useful
combinations
\begin{eqnarray}
{GM\over ac^2}=4.32\times 10^{-6}\,,\qquad 
{I_A\over Ma^2}=(7.74\times 10^{-11})~I_{A,80}\,,\qquad {\rm and} \quad
{P\over P_A}=3.88\times 10^5 \,,
\end{eqnarray}
where $I_{A,80}=I_A/(80 {\rm~M}_\odot{\rm~km}^2)$ is a typical value for
the moment of inertia. 

The spin precession leads to changes in the directions of the pulsar
beams.  In many cases, this will lead to the periodic appearance and
disappearance of the pulsar beam from the Earth.  Second, since the total
angular momentum is conserved (to this order), the orbital plane will
change orientation, which will be observed as a change in the
inclination angle $i$.  Damour \& Schaefer \cite{Damour88} have
considered the question of how these effects affect the timing of
binary pulsars.  For the change in $i$, 
\begin{eqnarray}\label{eqn:idot}
{di\over dt}={G\over ac^2}{\pi\over (1-e^2)^{3/2}}
\sum_i{I_i(4M_i+3M_{-i})\over M_ia^2P_i}\sin\theta_i\cos\phi_i\,,
\end{eqnarray}  
where $\theta_i$ is the angle between $\vec{S_i}$ and $\vec{L}$, and
$\phi_i$ is the angle between the line of sight to pulsar $i$ and the
projection of $\vec{S_i}$ on the orbital plane. 
The amplitude of the change in the orbital inclination angle $i$
due to A's precession will be
\begin{equation}\label{eqn:i}
\delta_i={|\vec{S_A}|\over|\vec L|}\sin\theta_A 
\simeq{I_AM\over a^2M_AM_B(1-e^2)^{1/2}}{P\over P_A}\sin\theta_A\,.
\end{equation}  
This will cause a periodic departure from the expected
time-of-arrival of pulses from pulsar A of amplitude  
\begin{equation}
\label{eqn:deltai}
\delta t_A={M_B\over M}{a\over c}\delta_i\cos{i}=
{a\over c}{I_A\over M_Aa^2}{P\over P_A}\sin\theta_A\cos i\,,
\end{equation}
if one can assume that the orbital eccentricity is small.
The facts that the
orbit of PSR J0737-3039 is seen nearly edge-on and that $\vec{S_A}$ is
only slightly misaligned from $\vec L$ makes this a special case in
which the amplitude of the timing change produced by the orbial plane
precession will be extremely small, $\delta
t_A\simeq(1.05\pm1.33)I_{A,80}~\mu$s \cite{LS05}. Not only is the
magnitude very small, but the large relative uncertainties in both
$\cos i$ ($|\cos i|\simeq0.028\pm0.028$) and in $\sin\theta_A$
($\sin\theta_A\simeq0.22\pm0.17$) combine to give a huge uncertainty
in $\delta t_A$.  We do not consider this effect further, but it could
be relevant for future  discoveries of binary pulsars with more favorable
inclinations.

\begin{figure}
\hspace*{-1.35cm}
  \includegraphics[angle=90,width=.75\textwidth]{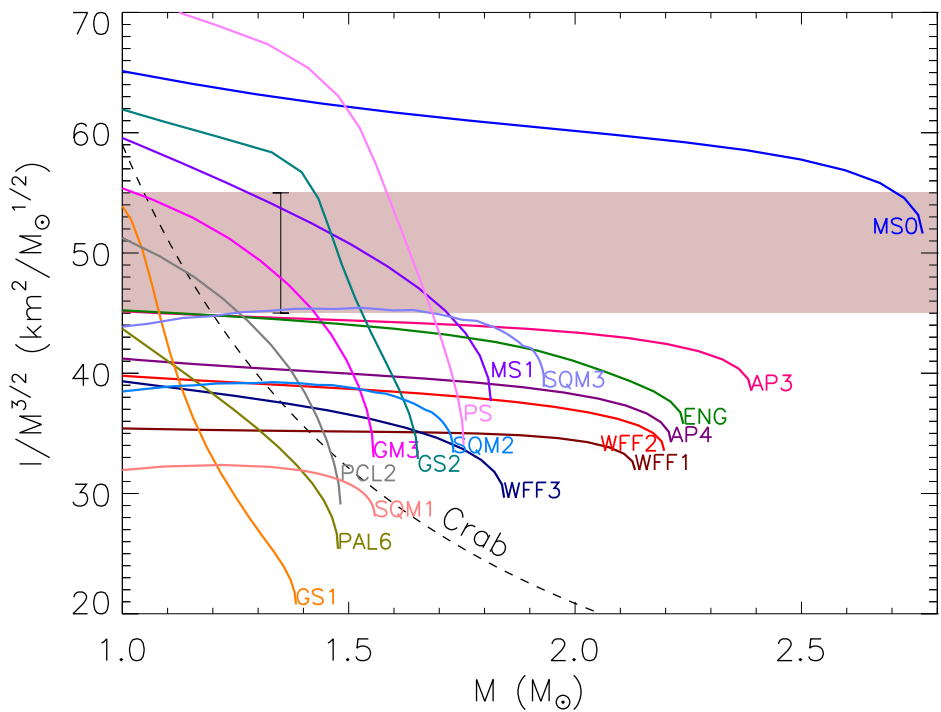}
\caption{The moment of inertia scaled by $M^{3/2}$ as a function of
stellar mass $M$ for EOSs described in \cite{LP01}.
The shaded band illustrates a $\pm10$\% error on a hypothetical $I/M^{3/2}$
measurement with centroid 50 km$^2$ M$_\odot^{-1/2}$; the error bar
shows the specific case in which the mass is 1.34 M$_\odot$ with
essentially no error.  The
dashed curve labelled "Crab" is the lower limit derived by
\cite{Bejger03} for the Crab pulsar.}
\label{mom:fig}
\end{figure}

For the advance of the periastron, the ratio of the spin-orbit
and 1PN contributions is \cite{Damour88}
\begin{eqnarray}
\label{eqn:periastron}
{A_p\over A_{1PN}}=-{P\over6(1-e^2)^{1/2}Ma^2}
\sum_i{I_i(4M_i+3M_{-i})\over M_iP_i}
(2\cos\theta_i+\cot i\sin\theta_i\sin\phi_i) \,.
\label{peri}
\end{eqnarray}
In the case that $|\vec{S_A}|>>|\vec{S_B}|$, only the $i=A$ term contributes
substantially. For comparison, the ratio of the 2PN to 1PN contributions is
\cite{Damour88}
\begin{eqnarray}
\label{eqn:a2pn}
{A_{2PN}\over A_{1PN}}&=&{GM\over 4ac^2}\sum_i\left
(\left[27+6{M_i\over M}+6\left({M_i\over M}\right)^2\right](1-e^2)^{-1}-1-
{46M_i\over3M}+{10\over3}\left({M_i\over M}\right)^2\right)\nonumber\\
&\simeq&{GM\over12ac^2}\left({189\over1-e^2}-47\right)\,,
\end{eqnarray}
where both $i=A$ and $i=B$ terms contribute. The second line of Eq.
(\ref{eqn:a2pn}) is valid in the case that $M_A=M_B$.  Observational
effects of the periastron advance are proportional to $\cos i$ and so
the effect is maximized in this system.  Furthermore, Eq. (\ref{peri})
illustrates that uncertainties in the angle $\phi_A$ are largely
irrelevant in the case $i\simeq90^\circ$.  The 1PN periastron advance
for PSR J0737-3039 is
\begin{eqnarray}
A_{1PN}={6\pi\over1-e^2}{GM\over ac^2} {\rm~radians~per~orbit}\,,
\end{eqnarray}
or 0.294 radians per year.  The periastron advance ratio works out to
be \cite{LS05}
\be
A_{pA}/A_{1PN} \simeq 6.6^{+0.2}_{-0.6}~10^{-5}I_{A,80}\,.
\ee
However, a practical difficulty remains.  The spin-orbit contribution
depends upon the individual masses, not just the total mass which is
known to extremely high precision, requiring for our purposes the
determination of essentially three post-Newtonian parameters, as
Damour \cite{Damour88} has emphasized, to an accuracy of a part in
$10^5$.  The accuracy of $a_B\sin i$ at present is a few parts in
$10^3$, and the determination of $I_A$ depends on its refinement by a
factor of 100 or more.  There is reason to be optimistic that this can
occur within a decade \cite{LS05}.

A sufficiently accurate measurement of $I$ say, to within 10\%, would
usefully discriminate among families of EOSs because the
mass will be known to great precision.  This is displayed in
Fig. \ref{mom:fig}, in which the quantity $I/M^{3/2}$ (in order to
reduce the vertical range) is plotted versus $M$.  The shaded band
illustrates a hypothetical measured value with error bars of
$\pm10$\%.  Only curves that pass through the line with the error bar
would be allowed.  In this hypothetical case, the vast majority of the
EOSs illustrated would not be adequate.  For comparison,
the lower limit derived for the Crab pulsar \cite{Bejger03}, which
rules out only the softest EOSs, is also shown in
Fig. \ref{mom:fig}.  This limit is not nearly so restrictive.

The fit in Eq. (\ref{momls}) for EOSs previously cited
as supporting masses greater than about 1.6 M$_\odot$ is reasonably
tight.  This relation may be inverted to yield a radius estimate given
$I$ and $M$.  Furthermore, since $I\propto R^2$, the inferred relative
error of $R$ using this procedure would be approximately half the
relative measured error of $I$ itself.

\section{Neutrino emission from a proto-neutron star}
\label{sec:pns}

A proto-neutron star (PNS) is born in the aftermath of a successful
supernova explosion resulting from the gravitational collapse of the
core of a massive star.  During the first tens of seconds of
evolution, nearly all ($\sim$ 99\%) of the remnant's binding energy is
radiated away in neutrinos of all flavors
\cite{Burrows86,KJ95,burr99a,pons99,Pon00a,Pon01b}.  The neutrino
luminosities and the emission timescale are controlled by several
factors including the total mass of the PNS and the opacity of
neutrinos at supranuclear density, which depend on the star's
composition and the EOS of strongly interacting dense matter.  One of
the chief objectives in modeling PNSs is to infer their internal
compositions through future supernova neutrino signals detected
in neutrino observatories like SuperK, SNO and others, including UNO
\cite{AIP}.

\subsection{The birth of proto-neutron stars}

The evolution of a PNS proceeds through several distinct stages
\cite{Burrows86,supernova} and with various outcomes~\cite{prak97a,
Ellis96}, as shown schematically in Fig. \ref{pict1}. Immediately
following core bounce and the passage of a shock through the PNSs
outer mantle, the star contains an unshocked, low entropy core of mass
$\simeq0.7$ M$_\odot$ in which neutrinos are trapped (stage I in the
figure). The core is surrounded by a low density, high entropy
($5<s<10$) mantle that is both accreting matter from the outer iron
core falling through the shock and also rapidly losing energy due to
electron captures and thermal neutrino emission. The mantle extends up
to the shock, which is temporarily stalled about 200 km from the
center prior to an eventual explosion.

\begin{figure}
\begin{center}
\includegraphics[angle=90,scale=0.6]{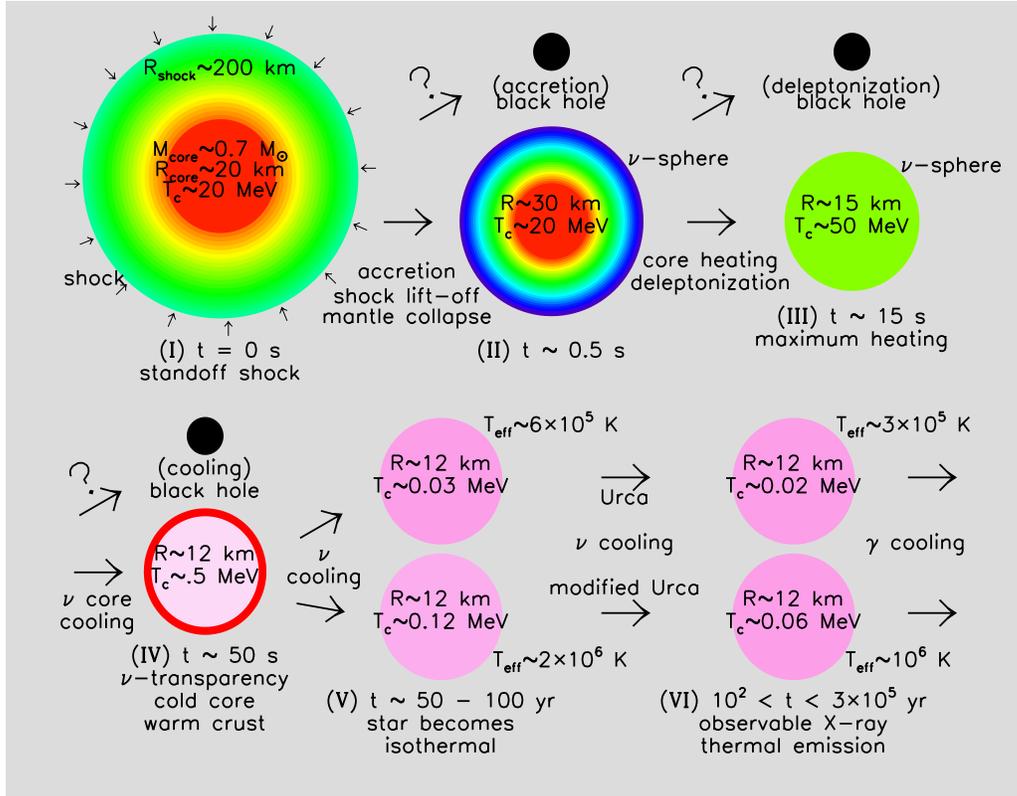}
\end{center}
\caption{The main stages of evolution of a neutron star, from
Ref.\cite{prak97a}.  Shading indicates approximate relative
temperatures.}
\label{pict1}
\end{figure}

After a few seconds (stage II), accretion becomes less important if the
supernova is successful and the shock has ejected the stellar envelope.
Extensive neutrino losses and deleptonization will have led to a loss
of lepton pressure and the collapse of the mantle.  If enough
accretion has occurred, however, the star's mass could increase beyond the
maximum mass capable of being supported by the hot, lepton-rich
matter.  If this occurs, the remnant collapses to form a black hole
and its neutrino emission is believed to quickly cease \cite{Burrows88}.

Neutrino diffusion deleptonizes the core on time scales of 10--15 s
(stage III).  Diffusion time scales are proportional to
$R^2(c\lambda_\nu)^{-1}$, where $R$ is the star's radius and
$\lambda_\nu$ is the effective neutrino mean free path.  This generic
relation illustrates how both the EOS and the composition influence
evolutionary time scales.  The diffusion of high-energy (200--300 MeV)
$\nu$'s from the core to the surface where they escape as low-energy
(10--20 MeV) $\nu$'s generates heat (a process akin to joule heating)
through neutrino-matter interactions. The core's entropy approximately
doubles, producing temperatures in the range of 30--60 MeV during this
time, even as neutrinos continue to be prodiguously emitted from the
star's effective surface, or neutrinosphere.

Strange matter, in the form of hyperons, a Bose condensate, or quark
matter, suppressed when neutrinos are trapped, could appear at the end
of the deleptonization.  Its appearance would lead to a decrease in
the maximum mass that the internal pressure of matter is capable of
supporting against gravity, implying metastability of the neutron star
and another chance for black hole formation~\cite{prak97a, Ellis96}.  This
would occur if the PNSs mass, which must be less than the maximum
mass of hot, lepton-rich matter (or else a black hole would already
have formed), is greater than the maximum mass of hot, lepton-poor
matter.  However, if strangeness does not appear, the maximum mass
instead increases during deleptonization and the appearance of a black
hole would be unlikely unless accretion in this stage remains
significant.

The PNS is now lepton-poor, but it is still hot.  While the star has
vanishing net neutrino number, thermally produced neutrino pairs of
all flavors dominate the emission.  The average neutrino energy slowly
decreases, and the neutrino mean free path increases.  After
approximately 50 seconds (stage IV), $\lambda\simeq R$, and the star
finally becomes transparent to neutrinos.  The neutrino luminosity
rapidly decreases beyond this time.  Since the threshold density for
the appearance of strange matter decreases with decreasing
temperature, a delayed collapse to a black hole is still possible
during this epoch.

Following the onset of neutrino transparency, the core continues to
cool by neutrino emission, but the star's crust remains warm and cools
less quickly. The crust serves as an insulating blanket which prevents the
star from coming to complete thermal equilibrium and keeps the surface
relatively warm ($T\approx3\times10^6$ K) for up to 100 years (stage
V).  The temperature of the surface after the interior of the
star becomes isothermal (stage VI) is determined by the rate of
neutrino emission in the star's core and the composition of the surface.

\subsection{Theoretical expectations}
\label{sec:PNS}

To understand what aspects of the EOS and structure can be probed by
neutrinos, we examine some analytic models for proto-neutron star
evolution \cite{prak97a,reddyt}.  For clarity and simplicity, we employ Newtonian
gravitation, as this does not affect the qualitative conclusions we
will draw.  We will assume that the neutrino distribution function is
well-approximated by a Fermi-Dirac distribution, so the neutrino
number density is $n_\nu=\int_0^\infty n_\nu(E_\nu)dE_\nu$, where
\begin{equation}
n_\nu(E_\nu)={E_\nu^2\over2\pi^3(\hbar c)^3}f_\nu(E_\nu),\qquad {\rm and~}
f_\nu(E_\nu)=[1+e^{(E_\nu-\mu_\nu)/T}]^{-1}\,.
\label{fd}
\end{equation}
We also make the diffusion approximation, in which both number and
energy fluxes are driven by density gradients:
\begin{eqnarray}
F_\nu&=&-{c\over3}\int_0^\infty
\left[\lambda_\nu{\partial n_\nu(E_\nu)\over\partial r}-
\lambda_{\bar\nu}{\partial n_{\bar\nu}(E_\nu)\over\partial r}\right] dE_\nu,\cr
L_\nu&=&-4\pi r^2\int_0^\infty\sum_i{c\lambda_E^i\over3}{\partial\epsilon_i(E_\nu)\over\partial r}dE_\nu\,,
\label{flux0}
\end{eqnarray}
where $F_\nu$ and $L_\nu$ are the net electron neutrino number flux
and total neutrino luminosity, respectively, $\lambda_\nu$
($\lambda_{\bar\nu}$)and $\lambda_E^i$ are the mean free paths for
number and energy transport, respectively, and $\epsilon_i$ is the
neutrino energy density.
$i=[\nu_e,\bar\nu_e,\nu_\mu,\bar\nu_\mu,\nu_\tau,\bar\nu_\tau]$ refers
to the neutrino species.  The net changes in electron lepton number
$Y_L=Y_e+Y_\nu-Y_{\bar\nu}$ and the total entropy per baryon
are
\begin{eqnarray}
n{\partial Y_L\over\partial t}&=&-{1\over r^2}{\partial\over\partial
r}(r^2F_\nu),\cr 
nT{\partial s\over\partial t}&=&-{1\over
4\pi r^2}{\partial L_\nu\over\partial r}-n\sum_j\mu_j{dY_j\over dt}\,,\label{change}
\label{transport}
\end{eqnarray}
where $j=n,p,e,\nu_e$.

The net neutrino mean free paths $\lambda_\nu$ and $\lambda_E^i$ are
due to both absorption and scattering and should also include inverse
processes.  In what follows, we will want to have an approximation for
$\lambda_\nu$ under degenerate conditions and when electron
neutrino absorption dominates.  Similarly, we will want to have an
approximation for $\lambda_E^i$ under non-degenerate conditions and
when it is dominated by scattering.  In that case, the transport is
dominated by all neutrino types other than $\nu_e$, and one can
replace the sum in Eq. (\ref{transport}) by a single term involving
the scattering mean free path.  Ref. \cite{reddyt} shows
that suitable approximations are then
\begin{eqnarray}
\lambda_\nu&\simeq&\lambda_0\left({T_0\over T}\right)
\left[1+\left({E_\nu-\mu_\nu\over\pi T}\right)\right]^{-1}\,,\cr
\lambda_E^i&\simeq&\lambda_C\left({T_0^3\over
TE_\nu^2}\right)\left({n_s\over n}\right)^{1/3}\,,
\label{lambda}
\end{eqnarray}
where $\lambda_0\simeq 50$ cm, $\lambda_C\simeq.2$ km and
$T_0\simeq10$ MeV.  Finally we note a property of the Fermi
distribution function under degenerate conditions:
\begin{equation}
{\partial f_\nu\over\partial r}\simeq{f_\nu(1-f_\nu)\over
  T}{\partial\mu_\nu\over\partial
  r}\simeq\delta(E_\nu-\mu_\nu){\partial\mu_\nu\over\partial r}\,.
\label{ferm}
\end{equation}

It is convenient to examine two distinct periods of proto-neutron star
evolution: the deleptonization era and the cooling era.  During the
first of these, the transport within the star is dominated by
degenerate electron neutrinos propagating through degenerate matter.
Only near the surface do these conditions break down.  We can use the
first of Eqs. (\ref{transport}) to examine the loss of leptons from
the star.  Using Eq. (\ref{lambda}) for $\lambda_\nu$,
Eq. (\ref{flux0}) for $F_\nu$, and
Eq. (\ref{ferm}), one has
\begin{equation}
{\partial nY_L\over\partial t}={c\lambda_0\over18\pi^2(\hbar c)^3}{1\over
  r^2}{\partial\over\partial r}\left[r^2\left({T_0\over
  T}\right)^2{\partial\mu_\nu^3\over\partial r}\right]\,.
\label{delep1}
\end{equation}
Using 
$\partial Y_L/\partial Y_\nu\simeq(\partial Y_L/\partial Y_\nu)_0\simeq5$ and
  $6\pi^2nY_\nu(\hbar c)^3=\mu_\nu^3$ for degenerate neutrinos, 
ignoring spatial variations of $T$ since the core is nearly
isothermal, and separating the spatial and temporal variations using
$\mu_\nu^3=\mu_{\nu,0}^3\psi(x)\phi(t)$, this can be
  rewritten as
\begin{equation}
{\tau_D\over\phi}{d\phi\over dt}={1\over x^2\psi}{\partial\over\partial
  x}\left[x^2{\partial\psi\over\partial x}\right]=-1,
\label{delep2}
\end{equation}
where
\begin{equation}
x=r{x_1\over R},\qquad \tau_D={3\over c\lambda_0}\left({R\over x_1}\right)^2
\left({T_c\over T_0}\right)^2\left({\partial Y_L\over\partial
  Y_\nu}\right)_0\,.
\label{delep3}
\end{equation}
$T_c$ is the assumed uniform temperature.  The quantity $x_1$ is the
value of $x$ for which $\psi$ first vanishes.  The second equality of
Eq. (\ref{delep2}) is the Lane-Emden equation of index 1 for which the
solution is $\psi(x)=\sin(x)/x$ and has the outer boundary $x_1=\pi$.
The first equality of Eq. (\ref{delep1}) is an exponential decay
$\phi(t)=\exp(-t/\tau_D)$ and $\tau_D$ is clearly the diffusion
timescale.  It is interesting that $\tau_D$ is independent of
$\mu_{\nu,0}$ but does depend sensitively on the ambient temperature
$T_c$.  Since one expects $T_c\propto n_{center}^{2/3}$ at fixed
entropy, $\tau_D$ will scale with the mass of the remnant like
$M^{4/3}$ since $R$ changes very slowly with $M$.

From Eq.(\ref{flux0}), one can determine the neutrino flux
\begin{equation}
F_\nu(x,t)=-{c\lambda_0\over18\pi^2}\left({\mu_{\nu,0}\over\hbar
  c}\right)^3 \left({T_0\over T_c}\right)^2{x_1\over
  R}{\partial\psi(x)\over\partial x}\phi(t)\,.
\label{flux2}
\end{equation}
At the outer boundary, where  $x\partial\psi/\partial x=-1$, 
the emerging flux is then
\begin{equation}
F_\nu(R,t)={c\lambda_0\over18\pi^2R}\left({\mu_{\nu,0}\over\hbar
  c}\right)^3 \left({T_0\over T_c}\right)^2\phi(t)\simeq{cF_2(0)\over8\pi^2}\left({T_e(t)\over\hbar c}\right)^3
\,,
\label{flux3}
\end{equation}
where $T_e(t)$ is thereby defined as the effective blackbody temperature, and
$F_i$ is the standard Fermi integral of index $i$.  We assume that the
effective value of the chemical potential when the neutrinos decouple
from the matter is $\mu_{\nu,e}\le T$.  The effective temperature is
therefore
\begin{equation}
T_e(t)\simeq\left({4\lambda_0\over9RF_2(0)}\right)^{1/3}
\left({T_0\over T_c}\right)^{2/3}\mu_{\nu,0}\phi(t)^{1/3}\,.
\end{equation}
The average emergent neutrino energy is then
\begin{equation}
<E_\nu>\simeq T_e(t)(F_3(0)/ F_2(0))\simeq3.15 T_e(t)\,,
\label{emerg}
\end{equation}
and the star's emergent luminosity will be $4\pi R^2F_\nu<E_\nu>$.
The effective value of the chemical potential at decoupling can be
estimated using the Eddington approximation, which gives
\begin{equation}
\psi_e=-{4\over3}{\lambda_e\over R}
\label{edd}
\end{equation}
where $\lambda_e$ is the neutrino mean free path at the decoupling
point.  From this, one can show that $\mu_e\simle T_e$ so we are justified
in using zero arguments in the above Fermi integrals.

During the beginning of the proto-neutron star phase, $R\simeq20$ km
and $T_c\sim20$ MeV.  One thus obtains $\tau_D=16$ s, $F_\nu(R,0)=
2.5\times10^{42}$ neutrinos cm$^{-2}$ s$^{-1}$, $T_e\simeq 2.9$ MeV
and $<E_\nu>\simeq9.2$ MeV.  The initial electron neutrino luminosity
due to deleptonization is $L_\nu(R,0)=2.1$ bethe s$^{-1}$.  We remark
that the initial luminosity burst predicted by detailed models is much
larger, around 100 bethe s$^{-1}$, but these are neutrinos originating
from the outer mantle of the proto-neutron star near the
neutrinosphere.  Within a few seconds, as the mantle deleptonizes and
collapses, the emergent luminosity becomes dominated by core emission.

Energy transport is also important during the deleptonization stage.
  Noting that
\begin{equation}
\sum_j\mu_jdY_j=(-\mu_n+\mu_p+\mu_e-\mu_\nu)dY_e+\mu_\nu dY_L\,,
\label{betaeq1}
\end{equation}
the first term on the right-hand side vanishes in beta
  equilibrium.  For electron neutrinos, which dominate in the
  interior, the mean free paths are so small that the first term on
  the right-hand side of Eq. (\ref{transport}) is negligible during
  deleptonization compared to the second term.  We therefore have
\begin{equation}
{s\over a}{ds\over dt}\simeq-\mu_\nu{\partial Y_L\over\partial
  t}\simeq-\left({\partial Y_L\over\partial Y_\nu}\right)_0
\left({Y_\nu\over Y_{\nu,0}}\right)^{1/3}\mu_{\nu,0}{\partial
  Y_\nu\over\partial t}\,
\label{delep4}
\end{equation}
This shows that continuous replenishment of diffusing neutrinos
enhances the growth of entropy in the star.  For degenerate matter,
$s\simeq aT$ where $a\sim0.05$ MeV$^{-1}$ is the supra-nuclear specific heat.
Integration leads to
\begin{equation}
s_f^2-s_i^2\simeq{3a\over2}\left({\partial Y_L\over\partial Y_\nu}\right)_0
\mu_{\nu,0}Y_{\nu,0}\simeq5\,,
\label{delep5}
\end{equation}
where $s_f$ and $s_i$ are the final and initial entropies.  For
$s_i\sim1$ one obtains $s_f\sim2.5$.  The final entropy scales roughly
as $M^{1/6}$.  Thus it is clear that the central portions of the star
heat significantly during deleptonization.

Following deleptonization, when the central temperature reaches a
maximum, core cooling begins.  By this time, $\mu_\nu\le T$ and its
gradient is also small.  The second term in the second of
Eq. (\ref{transport}) becomes negligible.  Assuming that the energy
transport mean free paths of all flavors of neutrinos are roughly
equal, we can replace the sum in the second of Eq. (\ref{flux0}) by a
factor 6, and the neutrino luminosity becomes
\begin{equation}
L_\nu=-4\pi r^2 {c\lambda_C\over6}\left({T_0\over\hbar c}\right)^3
{\partial T\over\partial r}\,.
\end{equation}
where we used the fact that $\int_0^\infty EfdE=\pi^2T^2/12$.
During the cooling phase, the density gradients in the star's center
are small, and one can remove the density dependence from the integral in
the energy transport equation, which beomes
\begin{equation}
T{\partial s\over\partial T}{dT\over
  dt}={c\lambda_C\over6n_s}\left({n_s\over n_*}\right)^{4/3}
\left({T_0\over\hbar c}\right)^3{1\over
  r^2}{\partial\over\partial r}\left[r^2{\partial T\over\partial r}\right]\,.
\label{cool1}
\end{equation}
We let $n_*$ be the density in the star's center.
The matter specific heat is dominated by baryons, so that
\begin{equation}
s\simeq a\left({n_s\over n_*}\right)^{2/3}T\,,
\end{equation}
where $a\simeq0.1$ is the level density of nucleons at $n_s$.
Once again, we can seek a separable solution with
$T=T_*\psi(x)\phi(t)$:
\begin{equation}
\tau_C{\partial\phi\over\partial
  t}={1\over\psi^2x^2}{\partial\over\partial
  x}\left[x^2{\partial\psi\over\partial x}\right]=-1\,,
\label{cool2}
\end{equation}
with
\begin{equation}
\tau_C={6an_sT_*R^2\over c\lambda_C x_1^2}
\left({\hbar c\over T_0}\right)^3\left({n_*\over
  n_s}\right)^{2/3}\,,
\label{cool3}
\end{equation}
where $x_1^2\simeq19$ is the eigenvalue for the index 2 Lane-Emden
equation.  The temporal solution is trivial, $\phi=1-t/\tau_C$.  $T_*$
is the core temperature at the end of deleptonization and the
beginning of cooling ($t=0$ in the immediate above).

\begin{figure}[hbt]
\includegraphics[scale=0.7, angle=90]{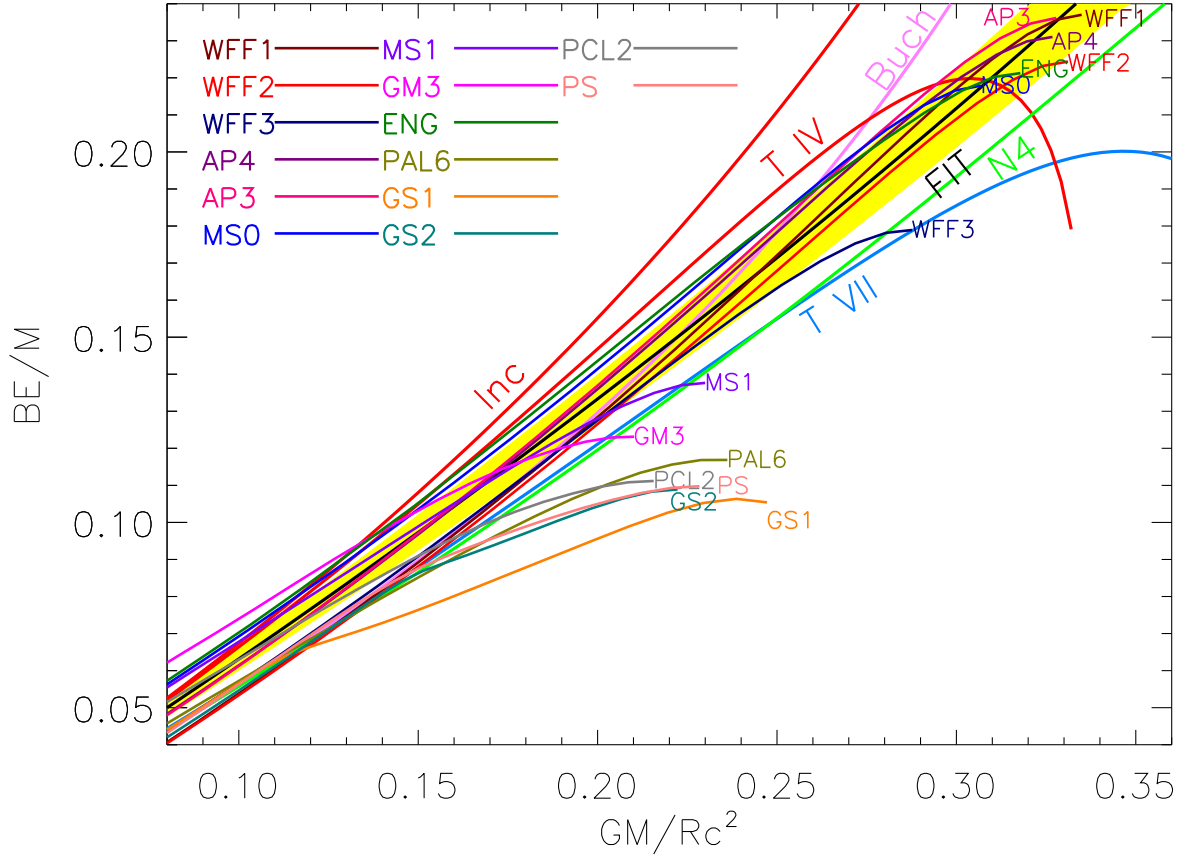}
\caption{The binding energy per unit mass of various EOSs as a
  function of $M/R$.  The lighter curves are EOSs for nucleonic stars
  as well as those containing hyperons, Bose condensates and
  deconfined quark matter (see Ref. \cite{LP01} for details).  The
  five heavier curves are for analytic solutions of GR structure
  equations.  The shaded band is the fit with error extents as
  given in Eq. (\ref{bind}).  Figure taken from Ref. \cite{LP01}.}
\label{bindc}
\end{figure}
We can establish the emergent luminosity and define an effective temperature
with
\begin{equation}
L_\nu(R,t)=-4\pi RT_*{c\lambda_C\over6}\left({T_0\over\hbar c}\right)^3
\left(x{\partial\psi\over\partial x}\right)_{x_1}\phi(t)
={cF_3(0)\over2(\hbar c)^3}RT_e(t)^4\,.
\label{cool4}
\end{equation}
For the index 2 Lane-Emden equation, $-(x\partial\psi/\partial
x)_{x_1}\simeq0.56$.  Inserting values for $T_*=50$ MeV and
$n_*=4n_s$, we find $\tau_C\simeq18$ s, $L_\nu(R,0)\simeq11$ bethe
s$^{-1}$, $T_e(0)\simeq5.5$ MeV, and $<E_\nu>\simeq17$ MeV.  Note that
$T_e$ at the beginning of the cooling era is nearly twice as large as
at the beginning of deleptonization.  This is attributable to core
heating which occurs during deleptonization and results in a steady
increase in the average energy of emitted neutrinos during the
deleptonization era.  Finally, note that like $\tau_D$, $\tau_C\propto
M^{4/3}$.  Also $T_e\propto T_*^{1/4}\propto M^{1/6}$.

In summary, there are a multitude of observables that can provide
details of the proto-neutron star: timescales, average energies, and
neutrino light curves.  Of course, the unknown details of the
proto-neutron star such as its mass, EOS and opacities are not related
to these variables in unambiguous ways.  However, one observable that
does not depend upon opacities can be considered: the total neutrino
emission which equals the binding energy released in the gravitational
collapse to form the neutron star.  In Newtonian gravity, the binding
energy of a uniform sphere is $BE=3GM^2/(5R)$.  Ref. \cite{LP01}
determined an approximate fit to the binding energies for EOSs that is
valid for EOSs which permit maximum masses larger than about 1.65
M$_\odot$:
\begin{equation}
BE/M \simeq (0.60\pm0.05)\beta(1-\beta/2)^{-1}\,,
\label{bind}
\end{equation}
where $\beta=GM/Rc^2$.
The binding energies of neutron stars formed from a variety of EOSs is
shown, together with the above fit, in Fig. \ref{bindc}.  Additionally
are shown binding energies for five analytic solutions of GR
applicable to neutron or self-bound stars \cite{LP01}.  With the
detection of thousands or millions of neutrinos, the binding energy
could be determined to better than a percent, and the tightness of
this fit would imply a correspondingly tight relation between $M$ and
$R$ to be established.  

\subsection{Model calculations}

\begin{figure}[hbt]
\includegraphics[scale=0.65]{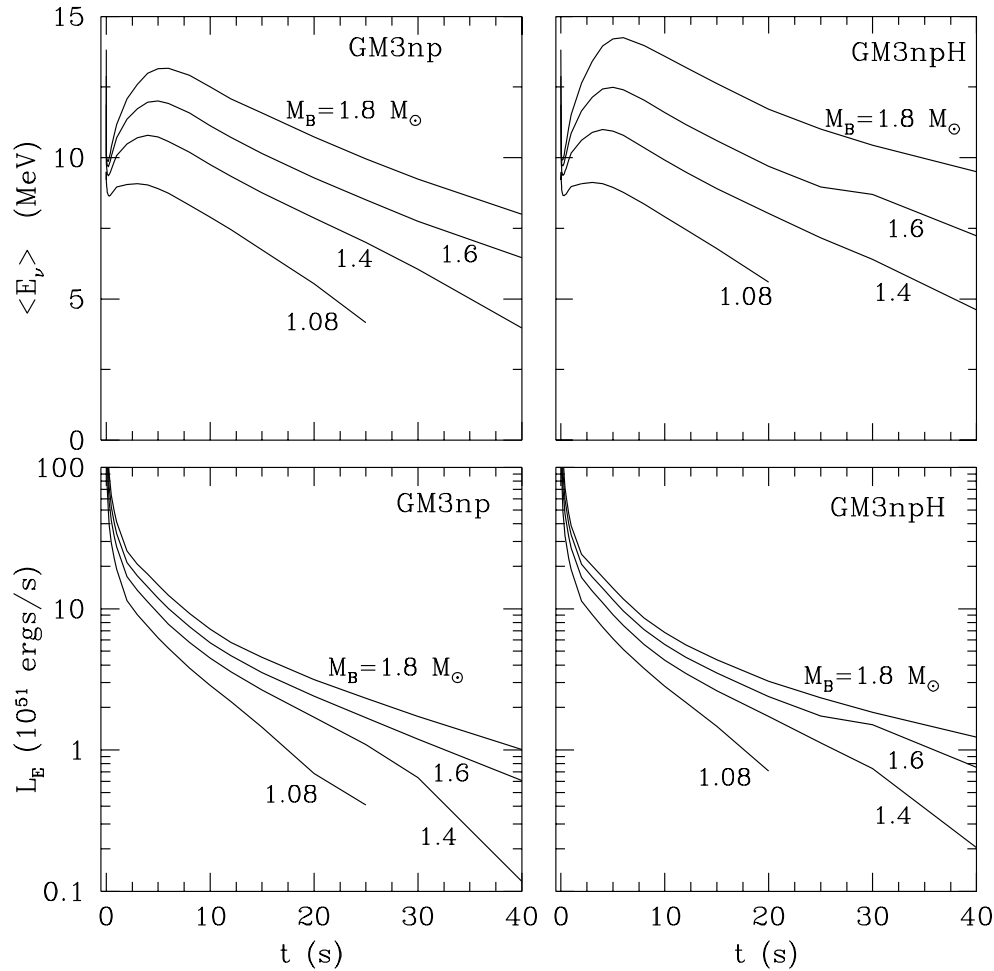}
\caption{The evolution of the average energy and total luminosity of
neutrinos in PNSs composed of baryons only (left panel)
and baryons and hyperons (right panel).  The figure is from
Ref. \cite{pons99}.}
\label{res_mass}
\end{figure}
Neutrino signals from PNSs depend on many stellar properties,
including the mass; initial entropy, lepton fraction and density
profiles; and neutrino opacities.  In Figs.~\ref{res_mass} --
\ref{fig:ttc}, the dependence of neutrino emission on PNS
characteristics are shown from the detailed study of Pons et
al. \cite{pons99,Pon00a,Pon01b}.  The generic results (see
Fig.~\ref{res_mass}) are that both $L_\nu$ and $<E_\nu>$ increase
with increasing mass~\cite{Burrows86,pons99}.  $<E_\nu>$ for all flavors
increases during the first 2-5 seconds of evolution, and then
decreases nearly linearly with time.  For times larger than about 10
seconds, and prior to the occurrence of neutrino transparency, the
$L_\nu$ decays exponentially with a time constant that is sensitive to
the high-density properties of matter. Significant variations in
neutrino emission occur beyond 10 seconds: $L_\nu$ is larger during
this time for stars with smaller radii and with the inclusion of
hyperons in the matter.  Finally, significant regions of the stars
become convectively unstable during the evolution, as several works
have found \cite{convec}.

\begin{figure}
\begin{center}
\includegraphics[scale=0.6]{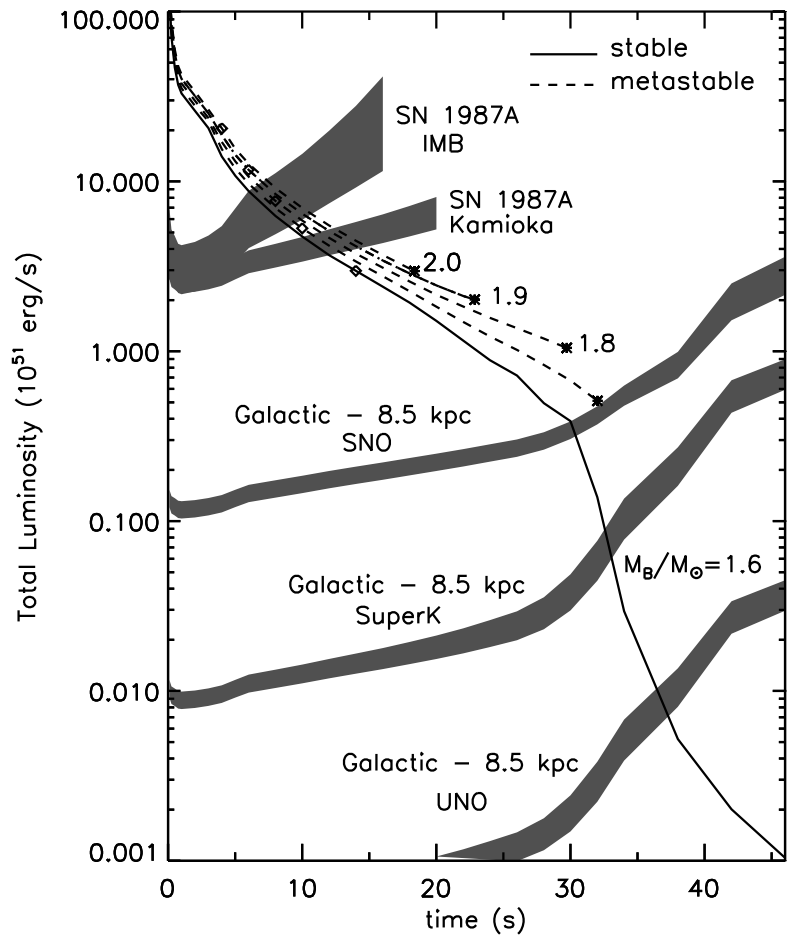}
\end{center}
\caption{The evolution of the total neutrino luminosity for $npQ$
PNSs.  Shaded bands illustrate the limiting
luminosities corresponding to a count rate of 0.2 Hz, assuming a
supernova distance of 50 kpc for IMB and Kamioka, and 8.5 kpc for SNO
and SuperK. The widths of the shaded regions represent uncertainties
in the average neutrino energy from the use of a diffusion scheme for
neutrino transport.  Figure taken from Ref. \cite{Pon01b}.}
\label{fig:lum1}
\end{figure}
The main effect of the larger mean free paths produced by in-medium
corrections \cite{burr98,burr99a} is that the inner core
deleptonizes more quickly.
In turn, the maxima in central temperature and entropy are reached on
shorter timescales. In addition, the faster increase in thermal
pressure in the core slows the compression associated with the
deleptonization stage, although after 10 s the net compressions of all
models converge.  The relatively large, early, changes in the central
thermodynamic variables do not, however, translate into similarly
large effects on observables such as $L_\nu$ and $<E_\nu>$, relative
to the baseline simulation.  It is especially important that at and
below nuclear density, the corrections due to correlations are
relatively small.  Since information from the inner core is
transmitted only by the neutrinos, the time scale to propagate any
high density effect to the neutrinosphere is the neutrino diffusion
time scale. Since the neutrinosphere is at a density approximately
$0.01n_s$, and large correlation corrections occur only above $n_s/3$
where nuclei disappear, correlation corrections have an effect at the
neutrinosphere only after 1.5 s.  However, the corrections are still
very important during the longer-term cooling stage,
and result in a more rapid onset of neutrino
transparency compared to the Hartree results \cite{burr98,burr99a}.

\begin{figure}[htb]
\begin{center}
\includegraphics[scale=0.6]{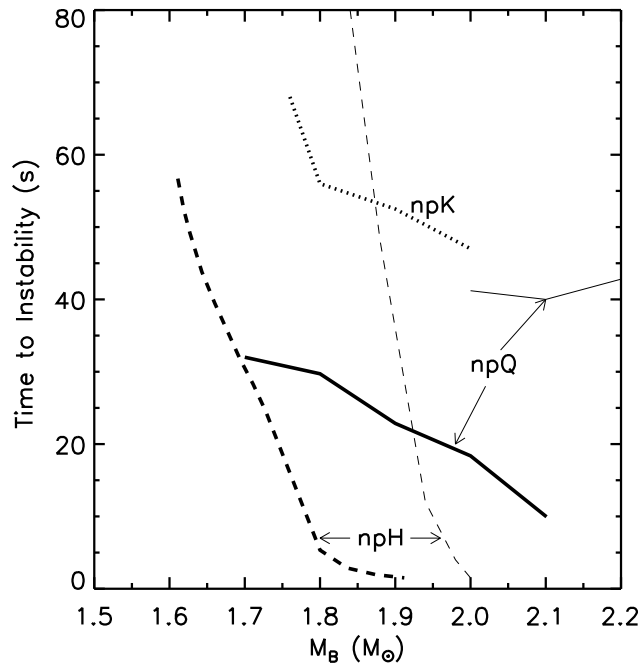}
\end{center}
\caption{Lifetimes of metastable stars versus the PNS
baryon mass $M_B$.  Thick lines denote cases in which the maximum
gravitational masses of cold, catalyzed stars are near 1.45 M$_\odot$,
which minimizes the metastability lifetimes.  The thin lines for the
$npQ$ and $npH$ cases are for EOSs with larger maximum gravitational
masses (1.85 and 1.55 M$_\odot$, respectively.)  Figure taken from
Ref. \cite{Pon01b}.}
\label{fig:ttc}
\end{figure}
A comparison of the signals observable with different detectors is
shown in Fig. \ref{fig:lum1}, which displays $L_\nu$ as a function
of baryon mass $M_B$ for stars containing quarks in their cores.  In
the absence of accretion, $M_B$ remains constant during the evolution,
while the gravitational mass $M_G$ decreases.  The two upper shaded
bands correspond to estimated SN 1987A (50 kpc distance) detection
limits with KII and IMB, and the lower bands correspond to estimated
detection limits in SNO, SuperK, and UNO, for a Galactic supernova
(8.5 kpc distance).  The detection limits have been set to a count
rate $dN/dt=0.2$ Hz \cite{Pon00a}.  It is possible that this limit is
too conservative and could be lowered with identifiable backgrounds
and knowledge of the direction of the signal.  The width of the bands
represents the uncertainty in $<E_{\bar\nu_e}>$ due to the diffusion
approximation \cite{pons99,Pon00a,Pon01b}.  It appears possible to
distinguish between stable and metastable stars, since the
luminosities when metastability is reached are always above
conservative detection limits.

\subsection{Metastable proto-neutron stars}
Proto-neutron stars in which strangeness appears following
deleptonization can be metastable if their masses are large enough
\cite{Brown94, pcl95}.
This stems from the interesting behavior that the expression of
strangeness is suppressed in hot, lepton-rich stars because of the
large value of $\mu_e$.  However, as stars deleptonize, $\mu_e$ falls
and it is possible that strangeness eventually appears in the star's
core.  Generally, this softens the equation of state, causing a
decrease in the effective maximum mass.  This leads to a possibly important
diagnostic that could shed light on the internal composition of
neutron stars: the abrupt cessation of the neutrino signal when a
deleptonizing strangeness-containing star suddenly finds that its mass
exceeds the maximum mass.  This would be in contrast to a normal star
of similar mass for which the signal continues to fall until it is
obscured by the background.  In Fig. \ref{fig:ttc} the lifetimes for
stars containing hyperons ($npH$), kaons ($npK$) and quarks ($npQ$)
are compared \cite{Pon00a}.  In all cases, the larger the mass, the
shorter the lifetime.  For the kaon and quark PNSs, however, the
collapse is delayed until the final stage of the Kelvin-Helmholtz
epoch, while this is not necessarily the case for hyperon-rich stars.
In addition, there is a much stronger mass dependence of the lifetimes
for the hyperon case.

Clearly, the observation of a single case of metastability, and the
determination of the metastability time alone, will not necessarily
permit one to distinguish among the various possibilities. Only if the
metastability time is less than 10--15 s, could one decide on this
basis that the star's composition was that of $npH$ matter.  However,
as in the case of SN 1987A, independent estimates of $M_B$ might be
available \cite{THM90BB95}.  In addition, the observation of two or
more metastable neutron stars might permit one to differentiate among
these models.  This highlights the need for breakthroughs in lattice
simulations of QCD at finite baryon density in order to unambiguously
determine the EOS of high density matter.  In the meantime, intriguing
possible extensions of supernova and PNS simulations with $npQ$ and
$npK$ matter include the consideration of heterogenoeus structures and
quark matter superfluidity \cite{CR00}.

\section{Magnetic matters}

Recent investigations of the effects of ultra-strong
magnetic fields ($B>10^{14}$ Gauss) on neutron stars are spurred by
several independent arguments that link the class of soft $\gamma-$ray
repeaters and perhaps certain anomalous X-ray pulsars with neutron
stars having ultra strong magnetic fields -- the so-called magnetars
\cite{pac,th0,th1,Mel}.
In addition, some soft $\gamma-$ray repeaters directly imply, from
their periods and spin-down rates, surface fields in the range
$2-8\times 10^{14}$ Gauss.  (See Table I from Cardall
et. al. \cite{Card} for a summary of observations.)  Kouveliotou et
al. \cite{k}
argue from the population statistics of soft
$\gamma-$ray repeaters that magnetars constitute about 10\% of the
neutron star population.  While some observed white dwarfs have large
enough fields to give ultra-strong neutron star magnetic fields
through flux conservaton, it does not appear likely that such isolated
examples could account for a significant fraction of ultra-strong
field neutron stars.  Therefore, an alternative mechanism seems
necessary for the creation ultra-strong magnetic fields in neutron
stars.  Duncan \& Thompson \cite{dn0,dn1}  
suggested that large fields
(up to $3\times10^{17}\times (1\rm{~ms}/P_i)$ Gauss, where $P_i$ is
the initial rotation period) can be generated in nascent neutron stars
through the smoothing of differential rotation and convection.

The effects of magnetic fields on the EOS at low densities,
relevant for neutron star crusts, has been extensively studied
(see for example, Refs. \cite{cv,fgp,ls,rfgpy}). 
Only a handful of works have considered the
effects of very large magnetic fields on the EOS of dense neutron star
matter \cite{Ch0,ch:dense,yz}.  
Lai and Shapiro \cite{ls} considered non-interacting, charge neutral,
beta-equilibrated matter at subsaturation densities, whereas
Chakrabarty and co-authors \cite{Ch0,ch:dense} studied dense matter
including interactions using a field-theoretical description.  These
authors found large compositional changes in matter induced by
ultra-strong magnetic fields due to the quantization of orbital
motion. Acting in concert with the nuclear symmetry energy, Landau
quantization substantially increases the concentration of protons
compared to the field-free case, which in turn leads to a softening of
the EOS. This lowers the maximum mass relative to the field-free
value.  In these works, however, the electromagnetic field energy
density and pressure, which tend to stiffen the EOS, were not
included. In addition, changes in the general relativistic structure
induced by the magnetic fields (studied in detail by Bocquet et
al. \cite{boc} who, however, omitted the compositional changes in the
EOS due to Landau quantization) were also ignored.  Calculations
including the combined effects of the magnetic fields on the EOS and
on the general relativistic structure have been reported by Cardall,
Prakash \& Lattimer \cite{Card}.

The most intriguing questions are:
\begin{enumerate}
\item What is the largest  frozen-in magnetic field
a stationary neutron star can sustain?,
and, 
\item What is the effect of such ultra-strong magnetic fields on the maximum
neutron star mass?
\end{enumerate}

\subsection{Magnetic effects on the EOS}

The answers to both of these questions hinge upon the effects strong
magnetic fields have both on the EOS of
neutron-star matter and on the structure of neutron stars.  
The magnitude of the magnetic field strength $B$ needed to dramatically
affect neutron star structure directly can be estimated with a
dimensional analysis \cite{ls} 
equating the magnetic field energy $ E_b \sim (4\pi R^3/3)(B^2/ 8\pi)$
with the gravitational binding energy $ E_{B.E.}  \sim 3GM^2/5R$,
yielding the so-called virial limit
\be
B \sim 1.4\times 10^{18} 
\left(\frac{M}{1.4 {\rm M}_\odot}\right) 
\left(\frac{R}{10\ {\rm
  km}}\right)^{-2}~{\rm Gauss}\,, 
\ee
where $M$ and $R$ are, respectively, the neutron star mass and
radius.

The magnitude of $B$ required to directly influence the EOS can be
estimated by considering its effects on charged particles.  Charge
neutral, beta-equilibrated, nucleonic matter contains both
negatively charged leptons (electrons and muons) and positively
charged protons.  Magnetic fields quantize the orbital motion (Landau
quantization) of these charged particles.  Relativistic effects become
important when the particle's cyclotron energy $e\hbar~B/(mc)$ is
comparable to it's rest mass energy. The magnitudes of the so-called
critical fields (although there is nothing critical about this) are
\be
B_c^e &=& (\hbar c/ e) \lbar^{-2} =
4.414\times10^{13}~{\rm Gauss}\,, \nonumber \\
B_c^\mu &=& (m_\mu/m_e)^2 B_c^e =
1.755\times10^{18}~{\rm Gauss} \,, \nonumber \\
B_c^p &=& (m_p/m_e)^2 B_c^e =
1.487\times10^{20}~{\rm Gauss} \,,
\ee
for the electron, muon and proton, respectively ($\lbar=\hbar/m_ec\simeq
386$ fm is the Compton wavelength of the electron).  When the Fermi
energy of the proton becomes significantly affected by the magnetic
field, the composition of matter in beta equilibrum is significantly
affected.  In turn, the pressure of matter is significantly affected.
In neutron star matter this situation occurs when $B^*\equiv
B/B^c_e\sim 10^5$, and will lead to a general softening of the EOS as
shown by 
Broderick, Prakash \& Lattimer \cite{bpl00}.

In neutron stars, magnetic fields may well vary in strength from the
core to the surface.  The scale lengths of such variations are,
however, usually much larger than the microscopic magnetic scale
$l_m$, which depends on the magnitude of $B$.  For low fields, for
which the quasi-classical approximation holds, 
\be
l_m \simeq
(\lbar^2/B^*)(3\pi^2n_e)^{1/3}\approx 10^5(n_e/n_s)^{1/3}/B^*~{\rm fm}\,,
\ee
where $n_e$ is the number density of electrons and $n_s$ is the normal nuclear
saturation density (about 0.16 fm$^{-3}$).  For high fields, when only
a few Landau levels are occupied, 
\be
l_m \simeq 2\pi^2n_e
(\lbar^2/B^*)^2\approx7\times10^9(n_e/n_s)/B^{*2}~{\rm fm}\,.  
\ee
In either case,
the requirement that $R >> l_m$ is amply satisfied; hence, the magnetic
field $B$ may be assumed to be locally constant and uniform as far as
effects on the EOS are concerned.

In non-magnetic neutron stars, the pressure of matter ranges from
$2-5~{\rm MeV~fm}^{-3}$ at nuclear density to $200-600~{\rm
MeV~fm}^{-3}$ at the central density of the maximum mass
configuration, depending on the EOS \cite{prak97a}.  
These
values may be contrasted with the energy density and pressure from the
electromagnetic field: 
\be
\varepsilon_f = P_f = \frac {B^2}{8\pi} = 4.814
\times 10^{-8} B^{*~2}~{\rm MeV~fm}^{-3}\,.  
\ee
Note that $\varepsilon = P_f = 1 ~{\rm MeV~fm}^{-3}$ for a $B^* = 4.56
\times 10^3$ or $B \simeq 2 \times 10^{14}$ Gauss. Thus, the field
contributions can dominate the matter pressure for $B^* > 10^4$ at
nuclear densities and for $B^* > 10^5$ at the central densities of
neutron stars, and must therefore be included whenever the field
dramatically influences the star's composition and matter pressure.

In strong magnetic fields, contributions from the anomalous magnetic
moments of the nucleons must also be considered.  Experimentally,
\be
\kappa_p = \mu_N \left(\frac{g_p}{2} - 1 \right) 
\quad {\rm for~the~proton}\quad {\rm and} \quad 
\kappa_n = \mu_N \frac {g_n}{2}\quad  {\rm for~the~neutron}\,, 
\ee
where $\mu_N$ is the nuclear magneton and $g_p=5.58$ and $g_n=-3.82$
are the Land\'{e} g-factors for the proton and neutron, respectively.
The energy
\be
|\kappa_n + \kappa_p| B
\simeq 1.67 \times 10^{-5} B^*~{\rm MeV} 
\ee
measures the changes in the beta equilibrium condition and to the
baryon Fermi energies.  Since the Fermi energies range from a few MeV
to tens of MeV for the densities of interest, the
contributions from the anomalous magnetic moments become
significant for $B^* > 10^5$.  In fact, complete spin
polarization of the neutrons occurs when 
\be
|\kappa_n| B \simeq \frac{(6\pi^2n_n)^{2/3}}{4m_n}\,.
\ee
At nuclear density, this leads to $B^* \simeq 1.6 \times 10^5$ or $B
\simeq 7.1 \times 10^{18}$ Gauss. Such
spin polarization results in an overall stiffening of the EOS 
(due to the increased degeneracy pressure of neutrons) that
overwhelms the softening induced by Landau quantization \cite{bpl00}. 

Inasmuch as the anomalous magnetic moments of strangeness-bearing
hyperons are similar in magnitude to those of nucleons (for a
compilation, see Table 1 of \cite{bpl02}, 
magnetic field interactions will be present if hyperons appear in
dense matter.  The influence of magnetic fields on the EOS of matter
containing strangeness-bearing hyperons is characterized by a
suppression of hyperons relative to the field-free case as both Landau
quantization and the anomalous magnetic moment interactions serve to
postpone the densities at which hyperons appear in dense matter \cite{bpl02}.

In magnetized matter, the stress energy tensor contains terms
proportional to $HB$, where $H = B+4\pi{\cal M}$ and ${\cal M}$ is the
magnetization \cite{llp}. 
Thus, extra
terms, in addition to the usual ones proportional to $B^2$, are
introduced into the structure equations \cite{Card}. 
The
magnetization in a single component electron gas has been studied
extensively by Blandford and Hernquist \cite{BH:magsus}
for neutron star crust
matter. The generalization of this formulation to the case of
interacting multicomponent matter with and without the effects of the
anomalous magnetic moments leads to the result that deviations of $H$
from $B$ occur only for field strengths $B^* \gtrsim 10^5$ \cite{bpl00}. 

In a magnetic field, the energy of an electron receives contributions
from vacuum polarization effects as well. Up to one loop level, these
contributions have been calculated by Schwinger \cite{Schw} with the
result $E^e_{vac} = m_ec^2\kappa_e$, where the dimensionless factor
$\kappa_e$ (arising from the anomalous magnetic moment of the electron)
takes on sign changing values in the weak and strong field limits:
\be
\kappa_e^{wf} &=& - m_e \frac {\alpha_e}{4\pi} B^*\,, 
\quad {\rm  weak~field~result}\,, \cr
\kappa_e^{sf} &=& \frac {\alpha_e}{2\pi} \left(\ln 2B^* \right) \,, 
\quad {\rm  strong~field~result}\,, 
\ee
where $\alpha_e \simeq 1/137$ is the electromagnetic fine structure constant.
One can set $|\kappa_e^{wf}|=1$ and $\kappa_e^{sf}=1$ in order to gauge
the fields required to induce a change in the electron Fermi energy
(or chemical potential at zero temperature) by an amount $m_ec^2$. The
results are:
\be
B^* &=& \frac {4\pi}{\alpha_e} \sim 1.7\times 10^3 \quad 
{\rm or~} B \sim  10^{17}~{\rm Gauss}\,, \quad 
{\rm for~weak~fields} \cr
&=& \frac {\alpha_e}{2\pi}~ e^{{\sqrt{2\pi/\alpha_e}}} \sim 2.8\times 10^{12} 
\quad {\rm or~} B \sim 10^{26}~{\rm Gauss}\,, \quad 
{\rm for~strong~fields}\,. 
\ee
Insofar as the effects of symmetry energy (with shifts of tens of
MeV's), Landau quantization and the anomalous magnetic moments of
baryons overwhelm the influence on proton fractions, vacuum
polariztion effects of the electron are not significant except for
extremely large fields.  
 
In addition, it must be borne in mind that for super-strong fields
exceeding the proton critical field of $\sim 1.5\times 10^{20}$ Gauss,
the energy density in the field would be significantly larger than the
proton's mass energy density.  At the proton critical field, the field
energy density $B^2/(8\pi) \sim 0.5~{\rm GeV~fm}^{-3}$ is comparable
to the energy density of the proton $\sim 1~{\rm GeV~fm}^{-3}$, which
calls for taking the compositeness of the proton seriously.

\begin{figure}
\begin{center}
\includegraphics[scale=0.7]{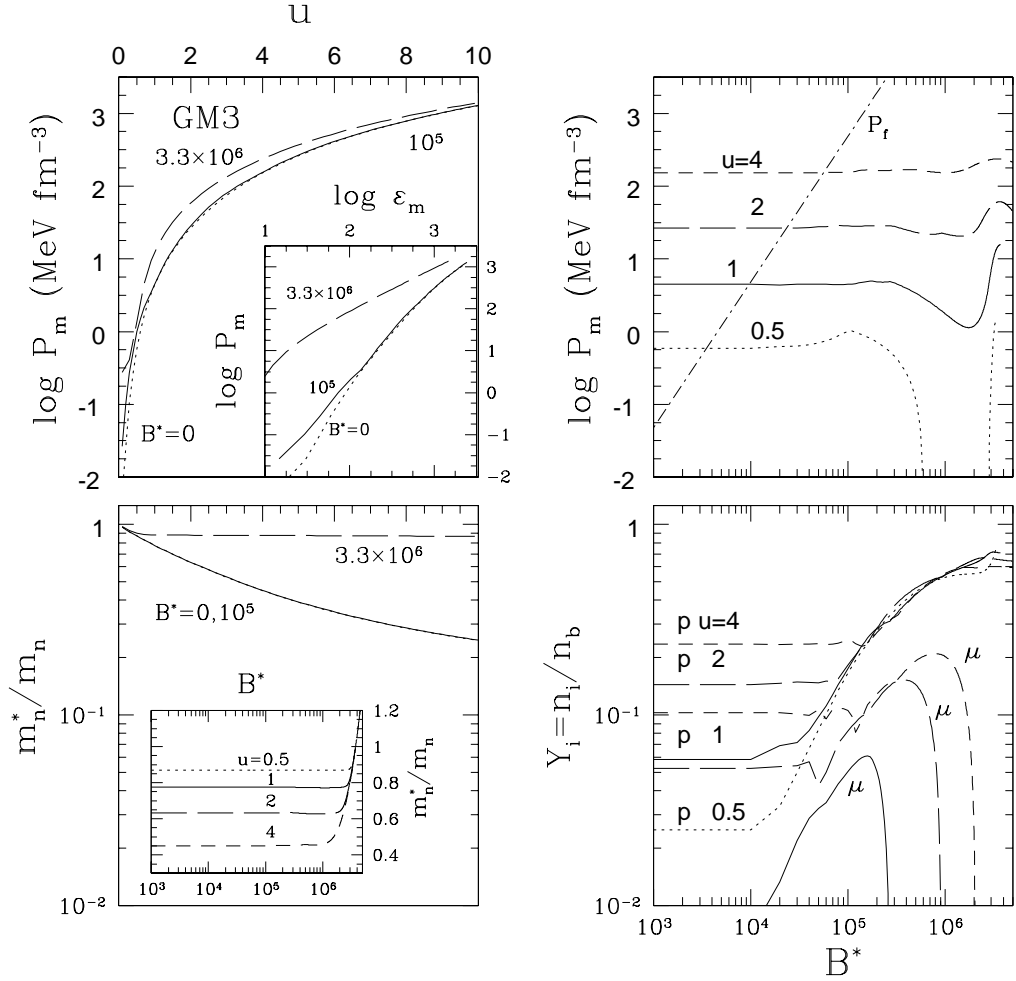}
\end{center}
\vspace*{-0.25in}
\caption{Matter pressure $P_m$, nucleon Dirac effective mass
$m_n^*/m_n$, and concentrations $Y_i=n_i/n_b$ as functions of the
density $u=n_b/n_s$ (left panels; $n_s=0.16~{\rm fm}^{-3}$ is the
fiducial nuclear saturation density) and magnetic field strength
$B^*=B/B_e^c$ (right panels; $B_e^c=4.414 \times 10^{13}$ Gauss is the
electron critical field), for the model GM3.  The inset in the upper
left panel shows $P_m$ as a function of the matter energy density
$\varepsilon_m$.  The curve labeled $P_f$ in the upper right panel
shows the $B^2/8\pi$ contribution to the total pressure.  The inset in
the lower left panel shows the effective mass as a function of $B^*$.
In the lower right panel, the electron and neutron concentrations have
been suppressed for clarity ($Y_e=Y_p-Y_\mu$ and $Y_n=1-Y_p$). These
results include the effcts of Landau quantization and the anomalous
magnetic moments in interacting nucleon matter. Figure taken
from Ref. \cite{bpl00}.}
{\label{fig5}}
\end{figure}

Figure \ref{fig5} shows typical results when the effects of Landau
quantization and anomalous magnetic moments are included in nucleonic
matter.  The main lesson to be learned from these results is the
softening of the EOS due to Landau quantization, which is, however,
overwhelmed by stiffening due to the incorporation of the anomalous
magnetic moments of the nucleons.  These effects become significant
only for fields in excess of $B^*\sim10^5$, for which neutrons become
completely spin polarized. Note that this field strength is
substantially less than the proton critical field.  In addition, the
inclusion of ultra-strong magnetic fields leads to a reduction in the
electron chemical potential and an increase in proton fraction.
Similar results obtain upon the inclusion of hyperons \cite{bpl02}.
These compositional changes have implications for neutrino emission
via the direct Urca process and, thus, for the cooling of neutron
stars \cite{ch:dense}.  The magnetization of the matter never appears
to become very large, as the value of $|H/B|$ never deviates from
unity by more than a few percent.

\subsection{Magnetic effects on neutron star structure}

We turn now to address whether or not stable stellar configurations
can exist in which the magnetic field is large enough
($B>5\times10^{18}$ G) that the properties of matter are significantly
affected by the magnetic field.  One step in this direction was
undertaken in Ref.~\cite{Card}, in which the limits of
hydrostatic equilibrium for axially-symmetric magnetic fields in
general relativistic configurations were analyzed, including the
effects of the magnetic field on the EOS.  As discussed in
detail in Ref. \cite{boc,Card}, in axially symmetric field
configurations with a constant current function, the magnetic field
contributes a centrifugal-like contribution to the total stress
tensor.  This can be understood by noting that for the field geometry
considered (namely axisymmetric), a superconducting fluid
can move along field lines but not across them.  Thus the ``pressure''
associated with the magnetic field will only act equatorially and not
vertically.  This flattens an otherwise spherical star, and for large
enough fields, decreases the central (energy) density even as the mass
is increased.  For large enough fields, the star's stability is
eventually compromised. 

\begin{figure}
\begin{center}
\includegraphics*[width=12cm]{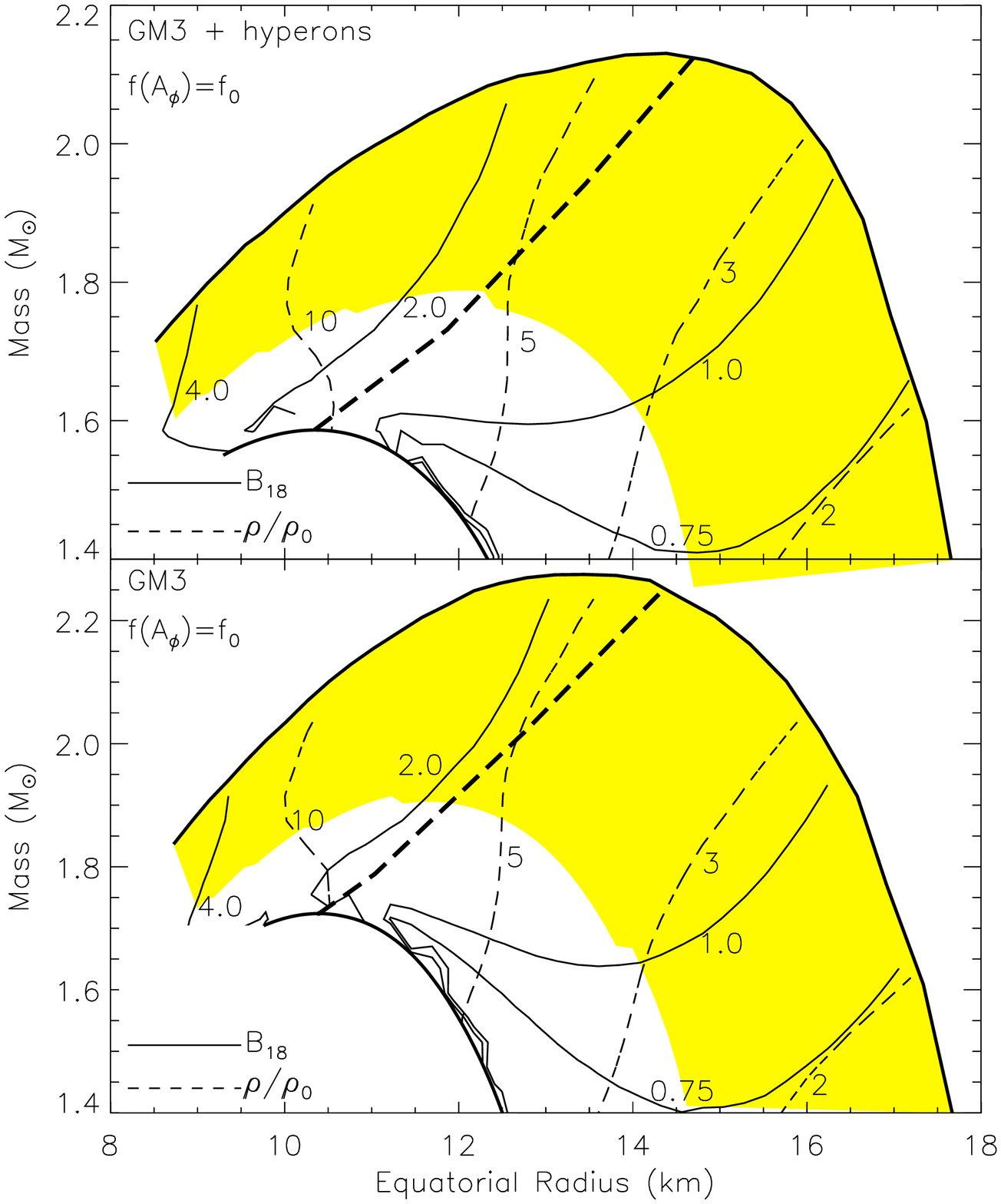}
\end{center}
\vspace*{-0.25in}
\caption{Limits to hydrostatic configurations for neutron stars
permeated by axially symmetric magnetic fields.  The upper (lower)
panel is for the EOS GM3 including (excluding) hyperons.  In each
panel, the lower heavy solid curve is the standard mass-radius
relation for field-free stars.  The upper heavy solid curve represents
the largest gravitational mass possible for a given equatorial radius
as the magnetic fields are increased, for the indicated current
function $f$.  The yellow shaded region contains configurations in
which the maximum density point is off-center.  The heavy dashed curve
is the locus of minima of the contours of fixed baryon mass
configurations: Ref.~\cite{Card} suggested that this could be the
limit to dynamical stability. Thin solid lines are contours of the
maximum magnetic field strength in units of $10^{18}$ G.  Similarly,
thin dashed lines are contours of maximum mass-energy density in units
of the energy density $\rho_0=\rho_s$ at the nuclear saturation
density. Figure taken from \cite{bpl02}.}
\label{jimc}
\end{figure}

The question of when is the magnetic field too large is answered to an
extent by the results shown in Fig. \ref{jimc}.  Results shown here
are for the GM3 EOS, both excluding (lower panel) and including (upper
panel) hyperons.  As with rotation, magnetic fields allow neutron
stars with a particular EOS and baryon number to have larger masses
and equatorial radii compared to the field-free case.  The maximum
mass attainable with a magnetic field governed by a constant current
function is noticeably larger than that attained by rotation.
Hydrostatically stable configurations (of which some may not be stable
to dynamical perturbations) are contained between the heavy solid
lines.  The lower heavy solid line in each panel is the usual
field-free, spherical result for the mass-radius relation.  The upper
heavy solid line represents the largest possible stable mass for a
given equatorial radius as the internal magnetic field strength is
increased.  Large axially-symmetric fields tend to yield flattened
configurations, and if large enough, shift the maximum densities
off-center (these configurations are contained in the yellow shaded
regions in the figure).  This results in toroidal shapes with
low-density centers.  As discussed in Ref.~\cite{Card}, regions to the
left of the heavy dashed line, which are the loci of minima of fixed
baryon mass configurations, are likely unstable to
small amplitude perturbations.

Superimposed on Fig. \ref{jimc} are contours of maximum mass
density and maximum magnetic field strength within a star.  For the
GM3 model, hyperons appear at zero field at about twice nuclear
density.  Therefore, the portions of the two panels in which the
maximum density is below about 3 times nuclear density are nearly
identical. In the case of the GM3 EOS, with or without hyperons, the
maximum field value for any presumably stable configuration is
$B\sim1.8\times10^{18}$ G.  For a constant current function, and for a
variety of EOSs, Ref. \cite{Card} found that the maximum value of the
magnetic field in stable stars never exceeds $3\times10^{18}$ G,
which corresponds to $B^*\sim 7\times10^4$.  {\it This field strength
is not nearly large enough to produce appreciable effects on the EOS.}
This includes changes to the hyperon or nucleon compositions, with or
without the inclusion of anomalous magnetic moments.  Additionally,
the magnetic fields with the assumed current function have relatively
small spatial gradients so that the ratio of the maximum field to the
average field within the star is not large.  Therefore, the dominant
effect of the field arises through the magnetic field stress
$B^2/8\pi$, which effectively dominates the matter pressure below a
few times nuclear saturation density, depending on the field's
orientation.  Whether other choices of current functions, or the
relaxation of the condition of axial symmetry, will alter these
conclusions has not been explored.

It might be that the shapes of the stars will significantly change
with a different field geometry.  It is even possible to imagine a
disordered field for which $\langle B^2\rangle$ is significantly larger
than $\langle \vec{B} \rangle^2$.  In this case the pressure will be
dominated by fluctuations in the field, but the stars will tend to be
spherical.  It is possible that strong magnetic fields may be held in
the core for periods much longer than the ohmic diffusion time due to
interactions between the magnetic flux tubes and the vortex tubes
expected to be present in a superconducting or superfluid rotating
neutron star~\cite{Rud95}.  Although the results to date
imply that average fields within a neutron star cannot exceed
$2-3\times10^{18}$ G before stability is compromised, too small to
have a significant impact on the EOS, it is possible that these
conclusions are not true in some magnetic
field geometries.  This is an important subject that deserves
additional investigation.

\section{Gravity wave emission from binary mergers}

Mergers of compact objects in binary systems, such as a pair of
neutron stars (NS-NS), a neutron star and a black hole (NS-BH), or two
black holes (BH-BH), are expected to be prominent sources of
gravitational radiation \cite{THORNE1}.  The gravitational-wave
signature of such systems is primarily determined by the chirp mass
$M_{chirp}=(M_1M_2)^{3/5}(M_1+M_2)^{-1/5}$, where $M_1$ and $M_2$ are
the masses of the coalescing objects. The radiation of gravitational
waves removes energy which causes the mutual orbits to decay. For
example, the binary pulsar PSR B1913+16 has a merger timescale of
about 250 million years, and the pulsar binary PSR J0737-3039 has a
merger timescale of about 85 million years \cite{Lyne04}, so there is
ample reason to expect that many such decaying compact binaries exist
in the Galaxy.  Besides emitting copious amounts of gravitational
radiation, binary mergers have been proposed as a source of the
r-process elements \cite{LSCH2} and the origin of the
shorter-duration gamma ray bursters \cite{EICHLER89}.

From the point of view of neutron star structure, observations of
gravity waves from merger events offer a unique opportunity to
simultaneously measure masses and radii, and could set extreme limits
on the neutron star maximum mass.  Current and detectors may not have
the ability to receive information expected at high frequencies from
NS-NS mergers, so our discussion focuses on BH-NS binaries.  From the
inspiral waves waves, masses of the binary components can be measured.
The inspiral terminates either due to tidal disruption or to the
stars' reaching their mutual ISCO.  In either case, measurement of the
gravity-wave frequency at this point can be used to infer the neutron
star radius \cite{FABER04}.  Furthermore, from the ringdown phase, the
mass of the final merged remnant can be found.  The binding energy
thus observed can be connected to $M/R$ and/or the EOS (see \S
\ref{sec:PNS}).  Moreover, a signal might be observable from a
super-sized neutron star formed from the pair if a black hole doesn't
form.  This would set a lower limit to the neutron star maximum mass
that is substantially larger than at present.  Importantly, as
demonstrated below, the wave pattern emitted if one of the components
is a strange quark star is substantially different than that of a
normal hadronic star.

\subsection{Tidal disruption and mass transfer} 

A neutron star in a binary merger with a larger mass companion will be
tidally disrupted.  This occurs when the radius of the less massive
component exceeds the innermost equipotential surface ({\it i.e.}, its
Roche lobe) surrounding the two stars that includes the first, or
inner, Lagrange point.  If the tidal disruption occurs far enough
outside the innermost stable circular orbit (ISCO) of the binary, it
is expected that an accretion disc will form or that mass transfer to
its companion will occur \cite{CLARK1,Jaranowski92,PORTEGIES98b}.  In
either case, the gravity wave signal is expected to be significantly
different than if tidal disruption occurs after the star penetrates
the ISCO.  

Tidal disruption or the onset of mass transfer depends upon
the EOS of dense matter and the mass ratio $q=M_1/M_2$ of the binary.
If mass can be transferred quickly enough from the lighter to the
heavier star, conservation of mass and orbital angular momentum can be
assumed, which generally leads to a reversal of inspiral.  The binary
separation widens.  Although this increases the Roche lobe volume, the
radius of the neutron star also increases in response to the mass
loss.  Mass transfer will continue in a stable fashion if the lighter
star can expand sufficiently fast such that it is able to continuously
fill its Roche surface.  Mass transfer under such conditions is termed
{\em stable mass transfer}.  Because the orbital separation now
increases, the gravity wave amplitude and frequency will decrease.
The signature of stable mass transfer in gravity waves should
therefore be strikingly different than for an amorphous tidal
disruption or a direct plunge.  

As we show below, important information about the radius of a neutron
star and the underlying EOS could be contained in the gravity wave
signal. For example, the expected gravity-wave signal from a merger
involving a neutron star should be quite different than from a
self-bound strange quark matter star.  In addition, the longer
timescale produced by stable mass transfer might also extend the
duration of the event from millisecond timescales, the orbital
timescale near the ISCO, to a few seconds, which then could explain
the duration of short-timescale gamma ray burst \cite{PORTEGIES98b}.

Another effect of stable mass transfer would be to modify the amount
of material potentially ejected from the system.  Matter from a
tidally disrupted neutron star, which could be accelerated to escape
velocities from the binary \cite{LSCH2}, undergoes decompression which
results in heavy nuclei and an intense neutron flux leading to the
copious production of r-process elements \cite{LATTIMER77,MEYER89}.
In the case in which stable mass transfer occurs, sudden disruption of
the neutron star near the last stable orbit is avoided, but mass could
be ejected more easily at later times and at larger separations when the
neutron star approaches its minimum stable mass \cite{COLPI93}.

\subsection{Merger evolution} 

Here we explore the qualitative character of a binary merger using a
relatively simple model that ignores relativistic corrections to the
Roche geometry and the orbital evolution but still reproduces the
qualitative features of more sophisticated models \cite{RPL05}.  This
model is similar in some respects to that of Ref. \cite{PORTEGIES98b}
but correctly includes the effects of the EOS.  To be definite, we
consider a low-mass BH-NS system, with $M_{NS}\simeq1.4$ M$_\odot$ and
$M_{BH}\simeq3-10 $M$_\odot$.  The original orbit decays by the
emission of gravitational radiation.  Denoting the total orbital mass
$M+M_{BH}+M_{NS}$, the reduced mass $\mu=M_{BH}M_{NS}/M$, and the mass
ratio $q=M_{NS}/M_{BH}$, the rate of change of orbital angular
momentum $J$ is \cite{Peters64}
\begin{equation}
\dot J_{GW}=-{32\over5}{G^{7/2}\over c^5}{\mu^2M^{5/2}\over a^{7/2}}
=-{32\over5}{G^{7/2}\over c^5}{q^2M^{9/2}\over(1+q)^4a^{7/2}}\,,
\label{GW}
\end{equation}
where $a$ is the orbital semi-major axis and
\begin{equation}
J^2=GM\mu^2a=GM^3aq^2(1+q)^{-4}\,.
\label{j}
\end{equation}
The orbital frequency is
\begin{equation}
\omega=\sqrt{GM\over a^3}\,.
\label{omegagw}
\end{equation}
These equations are valid for circular orbits, but it can be shown
that the timescale for decay of orbital eccentricity is much shorter
than the timescale for decay of semi-major axis $a$, which for
circular orbits is equivalent to the orbital separation.

The binary shrinks until tidal disruption ensues.  This occurs when
the neutron star fills its Roche radius, the gravitational equipotential
surface that passes through the inner Lagrange point, and mass begins
to flow to its companion.  In Newtonian
gravity, the Roche radius $R_\ell$ is well approximated by Kopal's formula
\cite{Kopal59}
\begin{equation}
R_\ell/a=0.46[q/(1+q)]^{1/3}\,,
\label{kopal}
\end{equation}
or a better fit by Eggleton \cite{Eggleton83}
\begin{equation}
R_\ell/a=0.49/[0.6+q^{-2/3}\ln(1+q^{1/3})]\,.
\label{eggleton}
\end{equation}
Thus, mass overflow begins at the moment that $R_\ell=R$, or
\begin{equation}
a=R(R_\ell/a)^{-1}\,,
\label{amo}
\end{equation}
where $R$ is the neutron star radius.  Note that both
$R_\ell/a$ and $a$ are decreasing functions of $q$ for fixed $M$ and
$J$.  Mass transfer will continue in a stable fashion if the star's
radius, after an increment of mass transfer, is less than $R_\ell$, so
that continued gravitational radiation will result in renewed mass
transfer, or
\begin{equation}
{d\ln R\over d\ln M_{NS}}\equiv\alpha\le{d\ln R_\ell\over d\ln
  M_{NS}}={d\ln a\over d\ln M_{NS}}+(1+q){d\ln (R_\ell/a)\over d\ln q}\,,
\label{drdm}
\end{equation}
where $M$ is assumed constant.  Eq. (\ref{drdm}) defines the parameter
$\alpha$, which is a function of the EOS and $M_2$.  It is $\alpha$
which determines both the onset of stable mass transfer and the
subsequent evolution.  If Kopal's formula is used, the second term on
the right-hand side of Eq. (\ref{drdm}) is just 1/3.

If the mass transfer is assumed to conserve orbital angular momentum,
the evolution of the system will be defined by
\begin{equation}
{d\ln J\over dt}={1\over2}{d\ln a\over dt}+{1-q\over 1+q}{d\ln q\over
  dt}=
{\dot J_{GW}\over J}=-{32\over5}{G^3\over c^5}{qM^3\over(1+q)^2a^4}\,.
\label{evolgw}
\end{equation}
Combining this with Eq. (\ref{drdm}), one finds
\begin{equation}
{\dot J_{GW}\over J}\ge\left({d\ln q\over dt}\right){1\over2(1+q)}
\left[\alpha+2(1-q)-(1+q){d\ln(R_\ell/a)\over d\ln q}\right]\,.
\label{evolgw1}
\end{equation}
Since both $\dot J_{GW}$ and $\dot q$ are negative, the condition for
stable mass transfer becomes
\begin{equation}
\alpha+2(1-q)-(1+q){d\ln(R_\ell/a)\over d\ln q}\ge0
\label{evolgw2}
\end{equation}
Using Kopal's formula, Eq. (\ref{kopal}), this is simply
$\alpha\ge2q-5/3$.  For hadronic stars near 1.4 M$_\odot$,
$\alpha\sim0$, so $q<5/6$ is the condition.  For strange quark stars
not near the maximum mass, $\alpha\sim1/3$, and $q<1$ is the
condition.  Thus, all NS-NS and BH-NS binaries containing strange
quark matter stars will likely have epochs of stable mass transfer.
However, some some NS-NS binaries with $q\sim1$ may not show this phenomenon.

During stable mass transfer, $\dot a>0$ and, hence, the binary spirals
apart.  However, if the condition Eq. (\ref{evolgw2}) is subsequently
violated, mass transfer becomes unstable at that point and the
remaining neutron star will quickly tidally disrupt.

Equations (\ref{drdm}) (with the equal sign) and (\ref{evolgw})
determine the time evolution of $a$ and $q$. From these quantities,
all other relevant quantities can be determined during mass transfer.
The details will depend upon the EOS. The two main observables will be
the gravity wave amplitude and the frequency of the gravity waves as a
function of time.  The scalar gravitational polarization amplitude (or
dimensionless strain) is
\begin{eqnarray}\label{eq:h}
h_+(t) &=& \frac{4}{r}\, \frac{4G^2M^2}{ac^4}\, \frac{q}{(1+q)^2}\, 
\cos 2\omega(t-r)\, ,  
\end{eqnarray}
where $r$ is the distance from the binary system to the observer. The
observed frequency of the emitted gravitational waves will be twice
the orbital frequency $\omega$ Eq. (\ref{omegagw}).  Typical values are
$|h_+r|\simeq2\times10^{-19}$ pc and $\nu=\omega/2!pi\simeq0.1-0.5$ kHz.


\subsection{The EOSs of normal versus self-bound stars}
\label{sec:EOS}

It is useful to examine the behavior of the function $\alpha(M_{NS}$
for both neutron stars and self-bound trange quark matter stars.  For the
discussion at hand, we will use the term normal star to refer to a
star with a surface of normal matter in which the pressure vanishes at
vanishing baryon density. The interior of the star, however, may
contain any or a combination of exotica such as (1)
strangeness-bearing matter in the form of hyperons, kaons, or quarks,
(2) Bose (pion or kaon) condensed matter, and (3) quark matter.

A self-bound star, as exemplified by Witten's conjecture that strange
quark matter might be more stable than normal matter at zero pressure 
\cite{Witten84, Alcock}, has a
bare quark matter surface in which the pressure vanishes at a finite
but supra-nuclear baryon density. In the context of the MIT bag model
with first order corrections due to gluon exhange, the baryon density
at which pressure vanishes is
\begin{equation}
n_b(P=0) =
(4B/3\pi^{2/3})^{3/4}(1-2\alpha_c/\pi)^{1/4} \,,
\end{equation}
where $B$ is the bag constant and $\alpha_c=g_c^2/(4\pi)$ is the
quark-gluon coupling constant. This density is not significantly
affected by the finite strange quark mass~\cite{PBP90} or by the
pairing phenomenon in quark matter ~\cite{JMad}.

\begin{figure}
\begin{center}
\includegraphics[width=.95\textwidth]{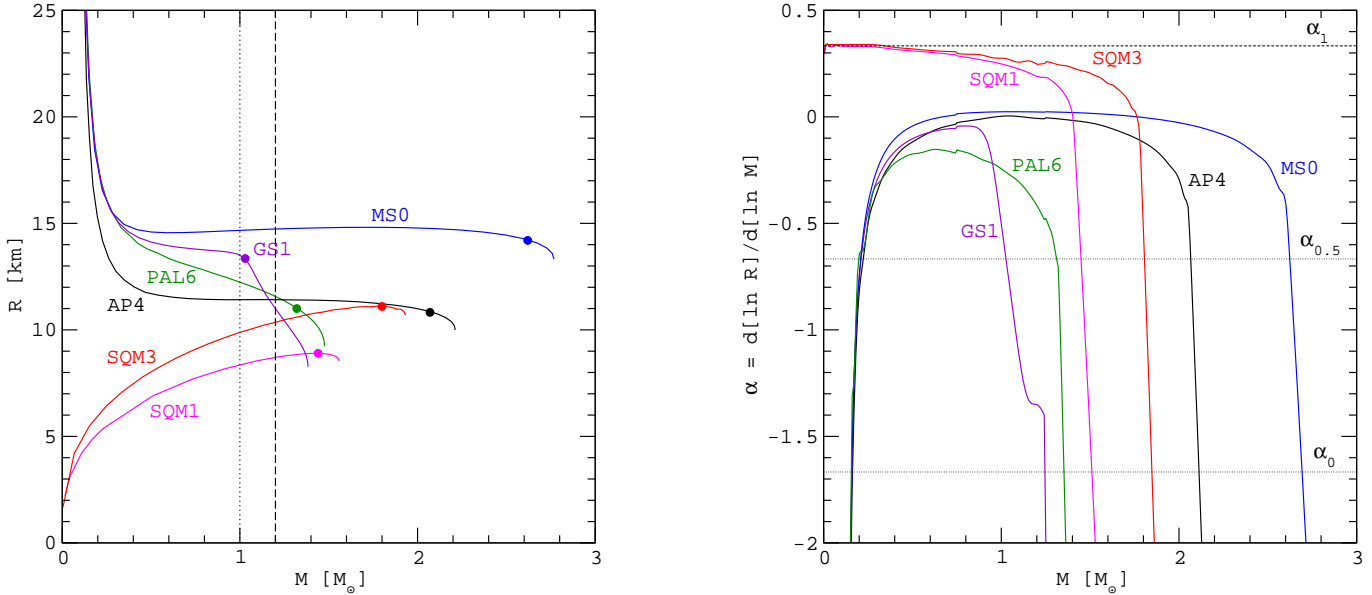}
\end{center}
\caption
{\label{fig:rmalpha}Radius versus mass (left panel) and its
logarithmic derivative (right panel) for prototype EOSs. The EOS
symbols are as in \cite{LP01}.  The vertical lines in the left panel
and the horizontal lines in the right panel are discussed in text. In
the left panel, the maximum masses for which stable mass transfer is
allowed in the case $q=M_{NS}/M_{BH}=0.5$ are denoted by filled
circles for each EOS. }
\end{figure}

Examples of radius versus mass for broadly differing EOSs
selected from \cite{LP01} are shown in Fig. 
\ref{fig:rmalpha}. 
While quantitative differences exist among normal stars, a rough
representation of $\alpha(M_{NS})$ for low mass stars is \cite{RPL05}
\begin{equation}
\alpha\approx-{0.09\over M_{NS}/{\rm M}_\odot-0.09}\,.
\label{alphaapp}
\end{equation}
This relation is useful in understanding detailed numerical
 simulations.
 Qualitative differences in the
outcomes of mergers with a black hole emerge, however, because of the
gross differences in the mass-radius diagram of normal and self-bound stars.
Thus, a normal star and a self-bound star represent two quite different
possibilities (see the right panel in Fig. \ref{fig:rmalpha}):
\begin{eqnarray}
\alpha \equiv  \frac {d\ln R}{d\ln M} 
\left\{ \begin{array}{ll} 
\leq 0 & \mbox{{\rm for~a~normal~neutron~star~(NS)}} \\
\ge 0 & \mbox{{\rm for~a~self-bound~SQM~star}} 
\end{array} 
\right. \,.
\label{lderiv}
\end{eqnarray}
We will explore the astrophysical consequences of these
distinctive behaviors in BH-NS mergers.  

Note that $\alpha$ is intimately connected with the dense matter EOS,
since there exists a one-to-one correspondence between $R(M)$ and
$P(n_B)$, where $P$ is the pressure and $n_B$ is the baryon density.
Gravitational mergers in which a compact star loses significant mass
during evolution is one of the rare examples in which the $R$ versus
$M$ (or equivalently, $P$ versus $n_B$) relationship of the same star
is sampled.  Although we focus here on the coalescence of a compact
star with a BH, the theoretical formalism and our principal findings
apply also to NS-NS mergers.

\subsection{Model evolutions}
\label{sec:evolution}

\begin{figure}
\begin{center}
\includegraphics[width=.65\textwidth, angle=90]{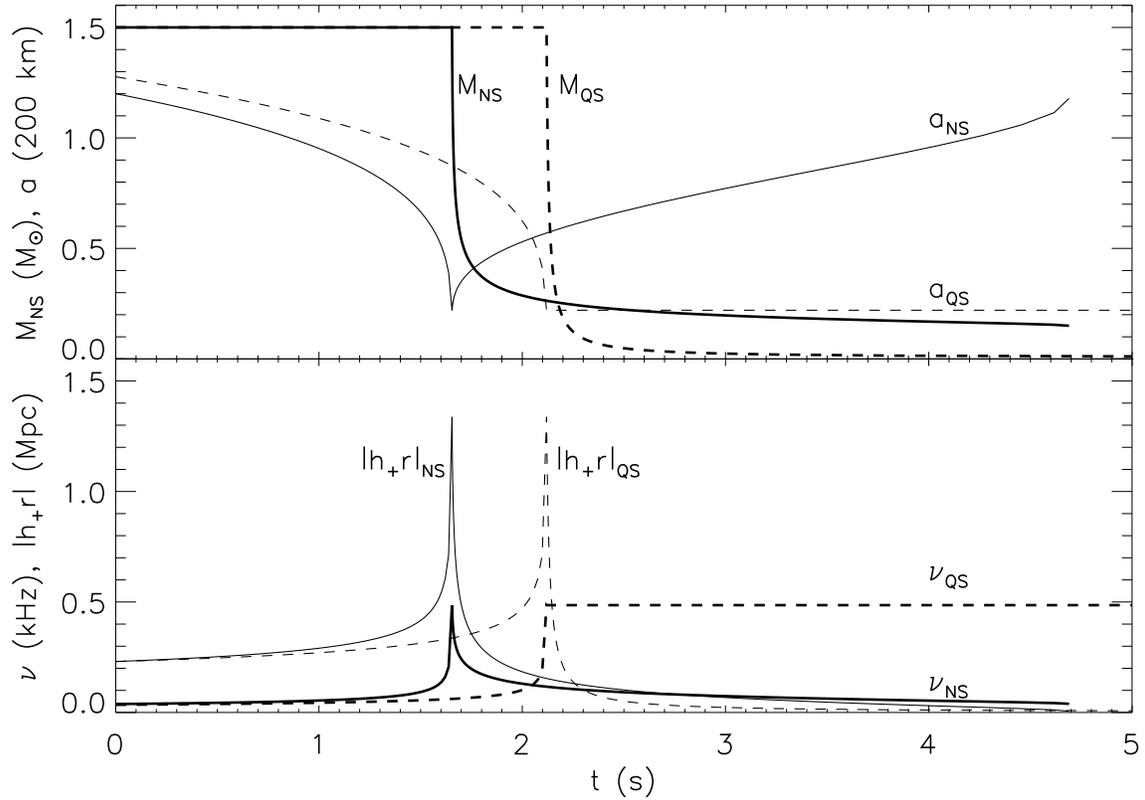}
\end{center}
\vspace*{-1cm}
\caption
{Schematic behavior of physical and observational
  variables in mergers between low-mass black holes and neutron stars
  or self-bound quark stars.  The total system mass is 6 M$_\odot$ and
  the initial mass ratio is $q=1/3$ in both cases.  The initial radii
  of the neutron star and quark star were assumed to be equal.  The
  time scales have arbitrary zero points.  Upper panel displays
  semi-major axis $a$ (thick lines) and component mass $M_{NS},
  M_{QS}$ (thin lines) evolution.  Lower panel displays orbital
  frequency $\nu$ (thick lines) and strain amplitude $|h_+r|$
  evolution.  In both panels, solid curves refer to the neutron star
  simulation and dashed curves to the quark star simulations.}
\label{gwevol} 
\end{figure}

Figure \ref{gwevol} shows the results of integrating Eqs. (\ref{drdm})
and (\ref{evolgw}) for simulations involving a 4.5 M$_\odot$ black
hole and a 1.5 M$_\odot$ normal or self-bound quark star.  Although
both simulations result in stable mass transfer, there is a pronounced
qualitative difference between the results after mass transfer begins.
Inspiral is characterized by increases in the orbital frequency
$\omega$ and scalar gravitational polarization amplitude $h_+$, a
decrease in orbital separation $a$, and a fixed $q$.  Stable mass
transfer ensues at the ``kinks'' visible in the evolution of these
quantities.    Within each class of EOS, i.e., normal and self-bound
stars, variations in the EOS only qualitatively alter the results.
During stable mass transfer, the decrease in orbital separation and
rise in frequency and waveform amplitudes are reversed.

In the neutron star case, the star spirals outwards and loses mass at
an approximately constant rate.  Mass loss continues for a few seconds
until the star approaches but does not decrease below its minimum mass
(about 0.1 M$_\odot$).  Mass transfer becomes unstable at this point,
and full tidal disruption including explosive decompression of the
remnant is to be expected in this case \cite{COLPI93}.  On the other
hand, the quark star loses mass exponentially and {\it remains at a
nearly fixed orbital separation} during this period.  Mass loss
continues for a virtually infinite time.  As a result, the temporal
behavior of the gravity wave emissions, both of the amplitude and the
frequency, from these two types of mergers significantly differ, as
seen in Fig. \ref{gwevol}.

The major effect of incorporating general relativistic corrections to
the potential is to speed up the evolution relative to the Newtonian
case \cite{RPL05}.  Stable mass transfer thus begins earlier in these
cases.  GR corrections also result in a somewhat larger value for the
orbital separations following the onset of mass transfer.

Using the relations established in the previous sections, one can
readily understand qualitatively and quantitatively the results shown
in Fig. \ref{gwevol}.  For example, the binary separation when mass
transfer begins is, from Eq. (\ref{amo}),
$a\simeq R(4/3)^{1/3}/0.46\simeq44$ km for the two cases shown (assuming
$R=12$ km).  The orbital frequency will be a maximum here,
$\nu=(\omega/2\pi)\simeq500$ Hz, from Eq. (\ref{omegagw}).  

In the neutron star case, mass loss continues until the condition
  Eq. (\ref{evolgw2}) is violated.  This occurs before the minimum
  mass is reached, since near the minimum mass
  $\alpha\rightarrow-\infty$.  The binary separation increases since,
  using Kopal's formula, Eq. (\ref{kopal}), $d\ln a/d\ln
  q=(\alpha-1/3)/(1+q)$ is negative for $\alpha<1/3$, which is always
  true for neutron stars.

However, in the quark star case, the binary separation remains nearly
fixed since $\alpha-1/3\simeq0$, especially for $q\rightarrow0$.  The
condition Eq. (\ref{evolgw2}) is never violated for quark stars, so
mass transfer is not terminated.  Using Kopal's formula and setting
$\alpha=1/3$, one arrives at the evolution equation
\begin{equation}
{dq\over dt}=-{32\over5}{G^3\over
  c^5}{M^3q^2\over1-q^2}\left({0.46\over R_Q}\right)^4\left({M_Q\over
  M}\right)^{4/3}\,,
\label{qsevol}
\end{equation}
where $R_Q$ and $M_Q$ are the quark star's initial radius and mass,
respectively.  This is trivially integrated to find the elapsed time
to go from an initial mass ratio $q_i$ to a final mass ratio $q_f$
\begin{equation}
\Delta t={5\over32M^3}{c^5\over
  G^3}\left({R_Q\over.46}\right)^4\left({M\over M_Q}\right)^{4/3}
\left[{q_f^2+1\over q_f}-{q_i^2+1\over q_i}\right]\simeq 0.002q_f^{-1}{\rm~s}\,,
\label{qsevol1}
\end{equation}
where we assumed in the final step that $q_f<<q_i$.  Thus, the mass of
  the self-bound star dwindles to extremely small values $<10^{-30}$
  M$_\odot$, the minimum mass limit being that of a strange quark
  nugget, determined in part by surface and Coulomb effects.  This
  takes, from Eq. (\ref{qsevol1}), more than $10^{27}$ s.

The envelope of the gravitational waveform amplitude $|h_+(t)|$
follows the behavior of $q$ and $a$.  In the normal neutron star case,
following the onset of stable mass transfer, the amplitude decreases
steadily until stable mass transfer terminates, but decreases much
more rapidly in the the quark star case.

\subsection{Observational parameters}

As discussed in Ref. \cite{CUTLER94}, careful analysis of the gravitational
waveform during inspiral yields values for not only the chirp mass
$M_{chirp}=(M_{BH}M_{NS})^{3/5}/M^{1/5}$, but for also the reduced mass
$M_{BH}M_{NS}/M$, so that both $M_{BH}$ and $M_{NS}$ can be found.  Thus,
observation of stable mass transfer effects in the gravitational wave
signal will allow several details about neutron star structure to be
discerned, including several features of the underlying EOS.  For
example, the onset of mass transfer can be determined by the peak in
$\omega$, and the value of $\omega$ there gives $a$.  The Roche limit
condition Eq. (\ref{amo}), as modified by general relativity
\cite{RPL05}, then allows the determination of the star's radius.
Thus a point on the mass-radius diagram can be estimated
\cite{FABER02}.

This information can be supplemented using the gravitational wave
amplitudes $h_+$ at its peak and at the end of stable mass transfer.
The combination $h_+\omega^{-1/3}$ depends only on a function of $q$,
so the ratio of that combination and knowledge of $q_i$ should allow
determination of $q_f$.  From the Roche condition and knowledge of
$a_f$ from $\omega_f$,  another mass-radius combination can be found.

The time elapsed between the onset of stable mass transfer and
its termination depends in a complicated way on $q, M$ and
$\alpha(M)$.  Nevertheless, measurement of this time, assuming that
$q$ and $M$ are already measured, implies that the function $\alpha$ can be
constrained.

Most importantly, the sharp contrast between the evolutions
during stable mass transfer of a normal neutron star and a strange
quark star should make these cases distinguishable even if other
information concerning the values of $q, M$ and $R$ is weak.  This
result is independent of the form of the gravitational potential or
the nuclear or quark matter EOS employed.

Finally, for the case of strange quark matter stars, the
differences in the height of the frequency peak and the plateau in the
frequency values at later times are related to the differences in
radii of the stars at these two epochs.  (In Fig. \ref{gwevol}, the
peak and the plateau are coincident because $\alpha=1/3$ was assumed
for all masses.)  It could be an indirect
indicator of the maximum mass of the star: the closer is the star's
mass before mass transfer to the maximum mass, the greater is the
difference between these frequency values, because the radius change
will be larger.  Together with radius information, the value of
the maximum mass remains the most important unknown that could reveal
the true equation of state at high densities.

\section{Constraints from laboratory data}

We turn now to opportunities afforded by laboratory experiments to constrain
the dense matter equation of state.  Supplements to our brief account
here are contained in several reviews, some of which are alluded to
below.

\subsection{Nuclear masses}

The most fundamental property of a nucleus is its binding energy, or
mass defect relative to the total individual masses of its $Z$ protons
and $N$ neutrons.  The first giant leap to understand the nuclear mass
systematics was taken by Bethe and von-Weiz{\" a}cker who developed
the semi-empirical mass formula using the liquid-drop model.
Incorporating realistic nuclear surfaces and effects stemming from
neutron-proton asymmetries, Myers and Swiatecki \cite{Myers69}
formulated the droplet model approach in which the nuclear energy can
be written as
\begin{eqnarray}\label{drop}
E(A,Z)&=&-BA+E_sA^{2/3}+S_vA{(1-2Z/A)^2\over1+S_s^*/(S_vA^{1/3})}+
 E_C{Z^2\over A^{1/3}}\cr
&&+E_{dif}{Z^2\over A}+E_{ex}{Z^{4/3}\over A^{1/3}}+a\Delta A^{-1/2}\,.
\end{eqnarray}
In this expression, $B\simeq16$ MeV is the binding energy 
per particle of bulk isospin
symmetric matter at saturation, $E_s, E_C, E_{dif}$ and $E_{ex}$ are
coefficients for the surface energy of symmetric matter, the Coulomb
energy of a uniformly charged sphere, the diffuseness correction and
the exchange correction to the Coulomb energy, respectively.  The last
term represents pairing corrections, where $\Delta$ is a constant and
$a=+1$ for odd-odd nuclei, 0 for odd-even nuclei, and $-1$ for
even-even nuclei.  For simplicity, the effects of curvature and 
higher-order terms are neglected.  
The quantity $S_v$ is the usual symmetry energy coefficient and
the quantity $S_s^*$ is related to the surface tension 
associated with the asymmetry parameter $\delta = (n_n-n_p)/(n_n+n_p)$ 
according to  
\begin{equation}
S_s^* = 4 \pi \left(\frac{3}{4 \pi n_s}\right)^{2/3} \sigma_{\delta} \,.
\end{equation}
Note that the surface energy 
$E_s=4\pi r_{0}^2\sigma(\delta=0)$ and the surface symmetry energy
$S_s=4\pi r_{0}^2\sigma_\delta$, where 
$(4\pi r_0^3/3) (0.16~{\rm fm}^{-3}) =1$, so that $S_s^*$ and $S_s$ differ 
only in the values of the equilibrium densities used.  

For the description of nuclei in the equation of state relevant for
astrophysical simulations of supernovae and neutron stars, the droplet
approach for isolated nuclei has been extended to the case in which
nuclei are immersed in a dense medium comprised of electrons,
positrons, protons, neutrons and alpha particles \cite{Lattimer85}.
The determination of the various parameters entering Eq.~(\ref{drop})
is afforded by the experimental nuclear masses.  Fits of
Eq.~(\ref{drop}) to currently available masses have revealed
interesting linear correlations, particularly for $S_s/S_v$ with
$S_v$.  Theoretically, $S_s/S_v$ is closely connected to the neutron
skin thickness $\delta R ={\langle r_n^2\rangle}^{1/2} - {\langle
r_p^2\rangle}^{1/2}$. (For recent accounts, see
\cite{Steiner05,Danielewicz03} and references therein.)  For example,
and depending on the precise way in which the nuclear surface and
Coulomb attributes are treated, correlations emerging from equivalent
fits to nuclear masses are
\begin{eqnarray}\label{eq:correl}
S_{s}/S_v = -5.253+0.254 S_v\qquad {\rm or} \quad 
S_s/S_v = -3.453+0.163 S_v\,.
\end{eqnarray}    
Tighter constraints will be provided by a larger number
of experimental masses for nuclei with $N>Z$, highlighting 
the importance of new mass measurements using rare isotope accelerators 
(RIA).  For new developments in techniques of measuring masses of 
short-lived nuclei off the stability line, see Refs. \cite{nmasses}.

\subsection{Nuclear matter compression modulus} 
The task of determining 
the isospin symmetric nuclear matter compression modulus 
\be
K = 9n_s^2 \left. \frac{d^2(E/A)}{dn^2}\right|_{n_s} \,,  
\ee 
where $E/A$ is the energy particle (determined from the empirical
nuclear masses) at the nuclear equilibrium density $n_s\simeq
0.16~{\rm MeV~fm^{-3}}$ (determined from electron and hadron
scattering experiments on nuclei), has been arduous on both
experimental and theoretical fronts. Thanks to vastly improved
experimental techniques, the identification and analysis of isoscalar
giant monopole and isoscalar giant dipole resonances have been accomplished
for light to heavy nuclei such as $^{70}$Zr, $^{116}$Sn and
$^{208}$Pb.  Equally challenging theoretical analyses of the data
based on fully self-consistent Hartree-Fock plus random phase
approximation (RPA) calculations have now been performed utilizing both
non-relativistic potential and relativistic field-theoretical models.
The conclusions that emerge from these combined experimental and
theoretical efforts are:
\begin{enumerate}
\item The isoscalar giant monopole resonance data yields $K =
  240\pm20$ MeV. The uncertainty of about 20 MeV is mainly due to the
  density dependence of the symmetry energy around $n_s$; and 
\item The isoscalar giant dipole resonance data tend to point toward
  lower values of $K$.  However, there is consensus that the
  extraction of $K$ in this case is largely experimental as the
  maximum cross section decreases very strongly at high excitation
  energy and likely drops below the current experimental sensitivity
  for excitation energies above 30 and 26 MeV for $^{116}$Sn and
$^{208}$Pb, respectively. With improved experimental techniques and
  analysis, these difficulties may be overcome. 
\end{enumerate}

An excellent account of these developments can be found in the
recent short review (and in other articles) by Shlomo et. al. \cite{shlomo06}.

\subsection{Symmetry energy and giant dipole resonances}

The analyses of giant resonances, particularly dipole resonances, have
long served to delineate the role of volume and surface effects in
nuclei; for a review see Ref.~\cite{Lipparini89}. In medium to heavy
nuclei in which magnetic contributions are small, the 
inverse-energy-squared weighted photoabsorption cross section $\sigma_{-2}$ 
can be related to the static polarizability $p$ as
\begin{equation}
\sigma_{-2} = \int \frac{\sigma(\omega)}{\omega^2}\,d\omega 
=2\pi^2(e^2/\hbar c)p \,.
\end{equation} 
In nuclei with mass number $A \geq 100$, $\sigma_{-2} =
(2.9\pm0.2)A^{5/3}~\mu{\rm b~MeV}^{-1}$ parameterizes the  
data~\cite{Bergere77,Laszewski79}. The dipole polarizability can be
evaluated as the response of a nucleus to an external dipole field
$\eta D$, where $\eta$ denotes the strength and $D=(1/2)\sum_{i=1}^A
z_i\tau_i^3$ is the dipole operator. Explicitly,
\begin{equation}
p = 2\sum_{n \neq 0} \frac {\left|\langle 0\left| D \right|
  n\rangle\right|^2}{\omega_n - \omega_0} 
\,, 
\end{equation}
where $|n\rangle$ and $\omega_n$ are the eigenstates and eigenenergies
of the nuclear Hamiltonian responsive to the dipole operator $D$.

Microscopic RPA calculations~\cite{Bohigas81} of $p$ employing
Skyrme-like interactions reproduce the general trends of the data. The
interplay between volume and surface effects are, however, difficult
to extract from RPA calculations. Semiclassical methods, in which the
energy density formalism is coupled with a hydrodynamic approach to
describe collective excitations, have thus been employed to explore
how the volume and surface symmetry energies, $S_{v0}$ and $S_s$,
determine the value of $p$ across the periodic table
\cite{Lipparini82,Krivine82,Lipparini89}.  Starting from a given Hamiltonian
density, and writing the transition density
as $\delta[\rho_n(r)-\rho_p(r)] = \eta \phi(r)\cos\theta$, the
polarizability is calculated from
\begin{equation}
p = \frac {2\pi}{3} \int \phi(r) r^3 \, dr \,,
\end{equation}
where $\phi(r)$ is the solution of the corresponding Euler-Lagrange
(integro-differential) equation.  
Investigations have found that similar results
for $p$ are obtained for correlated values of $S_v$ and $S_s$. 
Numerical results for
$^{40}$Ca, $^{120}$Sn, and $^{208}$Pb show that the choice of $S_v$
in the range 27--42 MeV requires $|S_s/S_v|$ = 1.2--2.2; lower values of
$S_v$ demand lower values of $S_s$ for good fits \cite{Krivine82}.

A qualitative understanding of this correlation can be gained by using
a schematic symmetry energy density functional and 
a leptodermous expansion of the density ~\cite{Lipparini89}, 
whence one obtains the result
\begin{equation}
p = \frac {A}{24} \frac {\langle r^2 \rangle}{S_v}
\left( 1 + \frac 53 \frac {S_s}{S_v} A^{-1/3} + \cdots  \right) \,,
\end{equation}
where 
$\langle r^2\rangle$ is the mean
square radius of the nucleus and 
higher order terms contain corrections from the diffuseness and
the skin thickness. Although values of $S_v=32.5$ MeV and $|S_s/S_v|=2.2$
describe the data adequately~\cite{Lipparini89}, correlated variations
in these numbers are allowed as pointed out in Ref.~\cite{Krivine82}.
                                                                             
Sum rules have been particularly useful to relate experiments and
theory in discussing the mean excitation energies, widths, and the
spreading of the excitation strengths \cite{Lipparini89}. The moments
$m_p$ of the strength function $S(\omega) = \sum_{n > 0} \left|
\langle n \left| F \right| 0 \rangle \right|^2 \delta (\omega -
\omega_n) $ defined by
\begin{equation}
m_p = \int_0^\infty S(\omega) \omega^p\,d\omega = \sum_{n > 0} 
\left| \langle n \left| F \right| 0 \rangle \right|^2 \omega_n^p \,, 
\end{equation}
where $F$ is the physical operator exciting the nucleus from its
ground state $| 0 \rangle$ to its eigenstate $| n \rangle$, are
especially helpful in this regard. For example, a good measure of the
mean excitation energy is provided by $E(D) = \sqrt
{m_1/m_{-1}}$, for which results from RPA and hydrodynamic
calculations have been compared with data \cite{Lipparini89}.  
Using a droplet model coupled with a hydrodynamic approach to excite
the dipole resonance, Ref.~\cite{Lipparini88} obtains
\begin{equation}
E(D) = \sqrt {
\frac {6\hbar^2(1+K_D)}{M\langle r^2 \rangle} 
\frac {S_v}{[1+5S_s/(S_vA^{1/3})] }} \,,
\end{equation} 
where $M$ is the nucleon mass.  The quantity $K_D$ is a model
dependent enhancement factor characterizing the relative contribution
of the nuclear interaction to the $m_1$ sum rule and depends
critically on the value of the energy up to which the energy
integration is carried out in analyzing the experimental data. Using
values of $S_v=32.5$ MeV and $S_s/S_v\simeq 2$ from a droplet model
fit to nuclear energies, and $K_D=0.2$ corresponding to $E_{max}=30$
MeV, Ref.~\cite{Lipparini88} finds that a hydrodynamic approach is
able to reproduce the experimental mean excitation energy of nuclei
ranging from $^{40}$Ca to $^{208}$Pb reasonably well.  Several Skyrme
interactions with different values of $S_v$ and $S_s/S_v$ are also able to
account for the data to the same level of accuracy.  However, since
the values of $S_v$ and $S_s$ for these forces are correlated because
they are fit to experimental masses, the additional constraint on the
permitted ranges of $S_v$ and $S_s$ resulting from fitting dipole 
resonances does not seem to be significant.

It must be emphasized that while a hydrodynamic approach is able to
account for the gross features of nuclear mass dependence, several 
detailed features of the data such as strength fractionation,
spreading widths, etc., are not naturally incorporated in its
scope. For such details one must adopt a more microscopic
approach that includes, for example, contributions from 2p--2h
excitations, etc. For a detailed account, see, for example,
Ref.~\cite{Khamerdzhiev97}.  

\subsection{Neutron skin thickness in nuclei}

Unlike proton distributions, neutron distributions in nuclei have
remained uncertain to this date.  Studies of neutron densities from a
global analysis of medium-energy proton scattering on $^{208}$Pb
indicate that $0.07 < \delta R < 0.16$ fm \cite{Clark03}. Related
information is also available from an analysis of antiprotonic atom
data that gives $\delta R = 0.15\pm 0.02$ fm~\cite{Trzcinska01}.  In
the latter work, nucleon density distributions are parameterized by
Fermi functions and it is found that the half-density radii for
neutrons and protons in heavy nuclei are the same, but the diffuseness
parameter for the neutrons is larger than that for the protons. Skin
thicknesses as large as 0.2 fm were obtained in earlier
analyses~\cite{Karataglidis02}.  Since these studies involve strongly
interacting probes, even to this date the value of $\delta R$ for a
nucleus such as $^{208}$Pb is not accurately known.  This situation
should improve as it is expected that the neutron rms radius will be
determined to about 1\% accuracy by measuring the parity-violating
electroweak asymmetry in the elastic scattering of polarized electrons
from $^{208}$Pb \cite{Horowitz01}.  This experiment, named PREX, is
planned at the Jefferson Laboratory \cite{Michaels00} for the summer
of 2008.

The neutron skin thickness is of interest as the
pressure of neutron star matter below and above the saturation density
$n_s$ depends on the density dependence of the isospin asymmetric part of the
nuclear interaction.  
Typel and Brown \cite{Brown00,Typel01} have noted that model calculations of the
difference between neutron and proton rms radii $\delta R ={\langle
r_n^2\rangle}^{1/2} - {\langle r_p^2\rangle}^{1/2}$ are linearly
correlated with the pressure of pure neutron matter at a density below
$n_s$ characteristic of the mean density in the nuclear surface (e.g.,
0.1 fm$^{-3})$.  The density dependence of the symmetry energy
controls $\delta R$ (we will call this the   
neutron skin thickness) in a
heavy, neutron-rich nucleus.  Explicitly, $\delta R$ is proportional
to a specific average of $[1-S_v/E_{sym}(n)]$ in the nuclear
surface, see Refs.~\cite{Krivine84,Lattimer96}.

Horowitz and Piekarewicz
\cite{Horowitz01b} have pointed out that models that yield smaller
neutron skins in heavy nuclei tend to yield smaller neutron star
radii. These authors, along with others \cite{Horowitz01,Michaels00},
have also pointed out the need for an accurate measurement of the
neutron skin.

For the connection between isospin asymmetry in nuclei and properties of 
neutron stars, see the recent review by Steiner et al. \cite{Steiner05}. 

\subsection{Heavy-Ion Collisions}

\subsubsection{Collective Flow}

Nuclear collisions in the range $E_{lab}/A = 0.5-2$ GeV offer the
possibility of pinning down the equation of state of matter above
normal nuclear density (up to $\sim 2 ~{\rm to}~ 3n_s$) from a study
of matter, momentum, and energy flow of nucleons \cite{Gutbrod89}.
The observables confronted with theoretical analyses include (i) the
mean transverse momentum per nucleon $\langle p_x \rangle /A$ versus
rapidity $y/y_{proj}$ \cite{Danielewicz85}, (ii) flow angle from a
sphericity analysis \cite{Gustafsson84}, (iii) azimuthal distributions
\cite{Welke88}, and (iv) radial flow \cite{Siemens79}.  Flow data
gathered to date are largely for protons (as detection of neutrons is 
more difficult) and for collisions of laboratory nuclei in which the
isospin asymmetry is not large.  Theoretical calculations have
generally been performed using Boltzmann-type kinetic equations. One
such equation for the time evolution of the phase space distribution
function $f({\vec r},{\vec p},t)$ of a nucleon that incorporates both
the mean field $U$ and a collision term with Pauli blocking of final
states is (see, for example, Ref. \cite{Bertsch88})
\def\sst{\scriptscriptstyle}
\begin{eqnarray}
\frac {\partial f}{\partial t} + {\vec \nabla}_p U \cdot {\vec \nabla}_r f
- {\vec \nabla}_r U \cdot {\vec \nabla}_p f =
&-& \frac {1}{(2\pi)^6} \int d^3p_2\,d^3p_{2^{\prime}}\,d\Omega 
\frac {d\sigma_{\sst NN}}{d\Omega} \, v_{12} 
(2\pi)^3\,\delta^3( {\vec p} + {\vec p}_2 - {\vec p}_{1^{\prime}} 
- {\vec p}_{2^{\prime}} ) \nonumber \\ 
&&  \times               
\left[ ff_2 (1-  f_{1^{\prime}}) (1-  f_{2^{\prime}})
- f_{1^{\prime}} f_{2^{\prime}} (1- f) (1- f_2) \right] \,.
\label{BUU}
\end{eqnarray}
Above, ${d\sigma_{\sst NN}}/{d\Omega}$ is the differential
nucleon--nucleon cross--section and $v_{12}$ is the relative velocity.
In general, the mean field $U$ depends on both the density $n$ and the
momentum ${\vec p}$ of the nucleon.  Equation (\ref{BUU}) contains
effects due to both hard collisions and soft interactions, albeit at a
semiclassical level.  Theoretical studies that confronted data have
thus far used isospin averaged nucleon-nucleon cross sections and mean
fields of symmetric nuclear matter.  It is now well established that
much of the collective behavior observed in experiments stems from
momentum dependent forces at play during the early stages of the
collision \cite{Gale90}. The conclusion that has emerged from several
studies is that as long as momentum dependent forces are employed in
models that analyze the data, a symmetric matter compression modulus
of $\sim 220-240$ MeV, as suggested by the analysis of the giant monopole
resonance data \cite{Youngblood99,shlomo06}, fits the heavy-ion data as well
(see the recent review by Danielewicz et. al.~\cite{Danielewicz02}).
 
The prospects of rare isotope accelerators (RIA's) that can collide
highly neutron-rich nuclei has spurred further work to study a system
of neutrons and protons at high neutron excess
\cite{Das03,Li04,Li04b}. Generalizing Eq.~(\ref{BUU}) to a mixture,
the kinetic equation for neutrons is
\begin{equation}
\frac {\partial f_n}{\partial t} + {\vec \nabla}_p U \cdot {\vec \nabla}_r f_n
- {\vec \nabla}_r U \cdot {\vec \nabla}_p f_n = J_n = \sum_{i=n,p} J_{ni}\,, 
\end{equation}
where $J_n$ describes collisions of a neutron with all other neutrons
and protons.  A similar equation can be written down for protons with
appropriate modifications.  On the left hand side of each coupled
equation the mean field $U\equiv U(n_n,n_p;{\vec p}\,)$ depends
explicitly on the neutron-proton asymmetry. The connection to the
symmetry energy arises from the fact that $U$ is obtained from a
functional differentiation of the Hamiltonian density.
Examples of such mean fields may be found in
Refs.~\cite{prak97a,Li04,Li04b}. Observables that are expected to
shed light on the influence of isospin asymmetry include
neutron-proton differential flow and the ratio of free neutron to
proton multiplicity as a function of transverse momentum at
midrapidity.  Experimental investigations of these signatures await
the development of RIA's at GeV energies.  In this connection, it will
be important to detect neutrons in addition to protons.

\subsubsection{Multi-fragmentation}

The breakup of excited nuclei into several smaller fragments during an
intermediate-energy heavy-ion collision probes the phase diagram of
nucleonic matter at sub-saturation density and moderate ($\sim 10-20$
MeV) temperatures. In this region of the phase diagram the system is
mechanically unstable if $(dP/dn)_{T,x}<0$, and/or chemically
unstable if $(d\mu_p/d x)_{T,P}<0$. (A pedagogical account of 
such instabilities can be found in
Ref.~\cite{Chomaz04}). These instabilities, which are directly related
to the symmetry energy at sub-saturation densities~\cite{Li97}, are
believed to trigger the onset  of multifragmentation. Because of the
instabilities, matter separates into coexisting liquid and gas phases,
which each have different proton fractions, {\it i.e.} ``isospin
fractionation''~\cite{Xu00}.  This fractionation is observed in the
isotopic yields which can potentially reveal information about the symmetry
energy.  Also, the scaling behavior of ratios of isotope yields  
measured in separate nuclear reactions, ``isoscaling'', is sensitive
to the symmetry energy~\cite{Tsang01,Ono03}. This scaling is expressed
in the empirically observed ratio of fragment yields from two similar
systems with different neutron-to-proton ratios:
\begin{equation}
Y_2(N,Z)/Y_1(N,Z) \sim \exp^{\alpha N+\beta Z} \,,
\end{equation}
where the constants $\alpha$ and $\beta$ can be related to the
neutron and proton chemical potentials in a canonical ensemble 
description, and thus to the proton fraction of the source.
To date, there are many suggestions of
how the symmetry energy may affect multifragmentation~\cite{Das04}. Ongoing
research is concerned with an extraction of reliable constraints on
the symmetry energy from the presently available experimental
information.

\subsubsection{Isospin diffusion}

Isospin diffusion is the
process in which the symmetry energy drives the exchange of neutrons
and protons between nuclei in a heavy-ion collision 
(see Fig.~\ref{fig:idiff}). This diffusion
process tends to force the isospin asymmetry, i.e. the charge to
baryon ratio, of the post-collision target and projectile nuclei to be
equal.  The isospin content of the post-collision
projectile-like fragment have been recently measured in 
experiments~\cite{Tsang04}. 
Starting from a parameterization of the symmetry energy near the nuclear
equilibrium density as in Eq. (\ref{vs},
information from multifragmentation~\cite{Shetty04} and isospin
diffusion~\cite{Chen05} observables have been combined to yield
the constraint $0.64
< i < 1.05$ and values of 31-33 MeV for $S_{v0}$, the symmetry
energy at saturation density. This constraint is consistent with
the Monte Carlo evaluation of the equation of state in Akmal et
al.~\cite{akmal98}, computed from two- and three-body interactions
which are matched to nucleon scattering phase shifts and the energy
levels of light nuclei. 
\begin{figure}[ht!]
\begin{center}
\includegraphics[width=0.8\textwidth, angle=0]{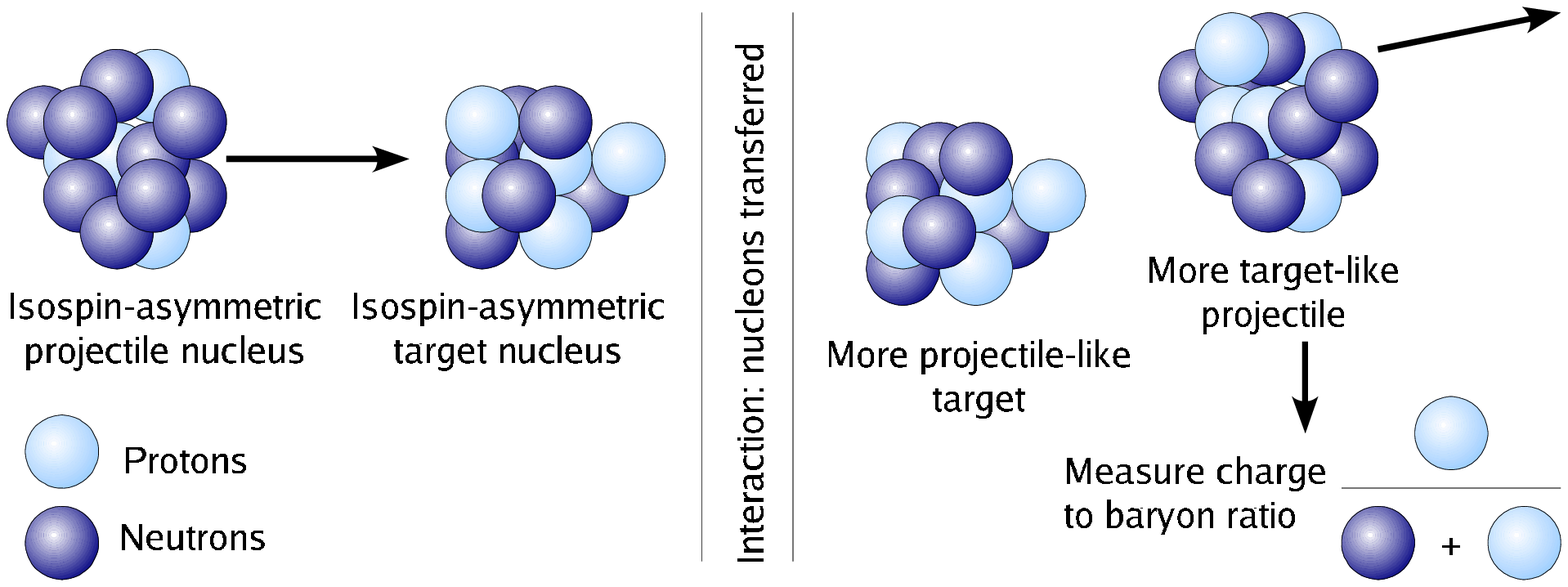}
\caption
[Isospin diffusion]
{Isospin diffusion process during a heavy-ion collision
\label{fig:idiff}. (Figure courtesy of A. W. Steiner.) }
\end{center}
\end{figure}

In the case of isospin diffusion, the constraints obtained
on the symmetry energy have also led to constraints in nuclear structure
physics. This simple picture is, however, modified by fragment 
emission during and
after the collision. This complication is alleviated~\cite{Rami00}
by considering the ratio
\begin{equation}
R_{\delta}=\frac{2{\delta}^{A+B}-{\delta}^{A+A}-{\delta}^{B+B}}
{{\delta}^{A+A}-{\delta}^{B+B}}\label{Ri}\,,
\end{equation}
where $A$ and $B$ are nuclei with different isospin asymmetries and
$\delta$ is the isospin asymmetry of the projectile-like fragment.
Using experimental data from the collision of $^{112}$Sn and
$^{124}$Sn obtained at the National Superconducting Cyclotron Laboratory 
(MSU) ~\cite{Tsang04}, a value $R_{\delta}
\sim 0.46$ was obtained. Isospin-dependent transport
models~\cite{Li04} of the same collision suggest that only symmetry
energies with $0.64 < \gamma < 1.05$ as discussed above are compatible
with the observed value of $R_{\delta}$. Because the symmetry energy
is tightly correlated with the neutron skin thickness 
(the difference between the neutron and
proton RMS radii) in
lead~\cite{Brown00},  the heavy-ion collision data also offer a
constraint on the neutron-skin thickness,
$R_n-R_p>0.15~\mathrm{fm}$~\cite{Steiner05a}.

Although the connection between the heavy-ion data and neutron star
radii is more tenuous, owing to the fact that neutron star radii are
sensitive to the symmetry energy at higher densities, isospin
diffusion also offers guidance on the radii of 1.4 solar mass neutron
stars. The symmetry energy from Eq.~(\ref{vs}) suggests that the
radius of a 1.4 solar mass star is likely between 11.5 and 13.6
km~\cite{Li05} - a range consistent with presently available
astrophysical observations.

\section{Outlook}

Even relative to a decade ago, the quantity of data and the variety of
approaches using light waves of all wavelengths to estimate neutron
star masses, radii, compactness, crustal thicknesses, etc. have
mushroomed.  Traditional techniques such as pulsar timing measurements
have produced more than two dozen masses accurate to the 10\% or better level,
enough that their statistical analysis are beginning to yield information about
their origin and evolution.  More than a dozen thermally emitting
neutron stars have had radiation radii estimated.  A few stars have
had redshifts determined either via direct measurements of atmospheric
absorption lines or from estimates obtained from light-bending
arguments.  Regions in the $M-R$ plane have been staked out from
observations of accreting neutron stars including QPOs and
Eddington-flux-limited sources.  Other observations, including pulsar
glitches, seismological studies of neutron star oscillations, and the
cooling behavior following bursts or accretion, have constrained the
extent of neutron star crusts.  With the exception of masses, none of
these other observations are yet sufficiently precise or
model-independent to significantly impact dense matter theories.

In fact, some apparently contradictory conclusions have been reported
over the last few years.  Some neutron stars are observed to be
spinning rapidly enough to exclude extremely stiff density-dependent
symmetry energies.  But the best-studied cooling, neutron star, the
nearby source RX J1856-3754, indicates a relatively large radius.
Rapidly cooling neutron star crusts following superbursts and the
non-detection of neutron stars in several young supernova remnants
both indicate high rates of neutrino emission in their interiors.
Yet the majority of thermally-emitting neutron stars are consistent
with relatively slow, or standard, core neutrino rates.

There are indications that new observations and techniques will mount
even more rapidly than in the last decade, which gives hope that some
of these contradictory results will be resolved.  The number of known
pulsars is growing rapidly, and new mass measurements implying that
the neutron star maximum mass is greater than 1.6 M$_\odot$ and
perhaps as large as 2 M$_\odot$ now exist.  This is large enough to
perhaps
delimit the amounts of exotic matter
({\it i.e.}, hyperons, Bose condensates, or deconfined quarks)
in neutron star cores.
Several new pulsars have been found in highly relativistic binaries,
including, in some cases, other neutron stars or pulsars.  The most
highly relativistic system discovered to date, PSR 0737-3039, might
reveal the moment of inertia of one of its stars, yielding perhaps the
most accurate measurement of a neutron star radius possible.
Furthermore, its relative closness and dimness strongly suggests that
several even more relativistic binaries will soon be found.

At the opposite extreme, there are indications that neutron stars
smaller than 1.2 M$_\odot$ exist.  This value is uncomfortably close
to the empirical minimum found in supernova models stemming from the
evolved cores of the smallest stars thought to be capable of the
event, approximately 8 M$_\odot$ initial mass.  Theoretically,
proto-neutron stars as small as 0.9 M$_\odot$ could be stable, and it
will be interesting to see if masses approaching this value are observed.

Less traditional methods, involving neutrinos and gravity waves, that
could be utilized to infer information about neutron stars should soon
become available.  The 20 neutrinos detected from SN 1987A twenty
years ago will be dwarfed by the thousands or even tens of thousands
that could be observed today from a galactic supernova.  Statistical
uncertainties in the total binding energy, the deleptonization and
cooling timescales, the average neutrino energies, etc. should be
quite small, permitting important conclusion to be drawn about the
proto-neutron star's mass and opacities.  In addition, tens or more
$\nu_e$'s could be detected during the collapse prior to bounce,
their number and average energy crucially depending on the density
dependence of the symmetry energy below nuclear density. The
observation of gravity waves from compact star binary mergers is
eagerly anticipated, even more so now that the pulsar binary PSR
0737-3039 has been discovered.  In fact, the gravity waves observed
from a binary merger might be the least unambiguous way to learn of
the existence of quark matter stars.

Complementing the increase in observational data has been the
development of new techniques, supporting theory and laboratory data
relevant to dense matter physics.  It is now known that the underlying
physics behind the neutron star maximum mass and typical neutron star
radii are quite different.  The maximum mass is set by the dense
matter equation of state well beyond $2n_s$, but neutron star radii
are determined by $dE_{sym}/dn$ in the range $1-2n_s$.  It is
similarly expected that better models of emission beams from accreting
neutron stars, pulsar glitches, superfluidity and superconductivity
involving nucleons and hyperons, neutron star atmospheres, and the
seismology of crustal oscillations will make better use of
accumulating data.

Laboratory experiments relevant to nuclear astrophysics in general,
and to neutron stars in particular, have accelerated significantly in
recent years.  With impressive efficiency,
masses of most nuclei between the path of beta stability and the
neutron drip line have now been measured.  With the Rare Isotope
Accelerator and Jefferson Lab experiments, many more mass measurements
of extremely neutron-rich nuclei and flow measurements of neutron-rich
heavy ion collisions will be forthcoming.  More knowledge of
hyperon-nucleon interactions through studies of hypernuclei should be
available.  Perhaps one of the more intriguing experiments is the PREX
experiment at the Jefferson Laboratory to measure the neutron radius
of Pb$^{208}$ to 1\% accuracy, which should yield a neutron skin
thickness estimate sufficiently precise to offer significant
constraints on the neutron star radius.  Such experiments should be
extended to other neutron-rich nuclei.  Multifragmentation and isospin
diffusion studies from heavy ion collisions should yield more accurate
information concerning the symmetry energy of nucleonic matter and the
critical density for phase separation in dense matter, which is
particulary relevant to the determination of the core-crust interface
density, pressure and enthalpy.  New studies of giant monopole, dipole
and other nuclear vibrational modes should be undertaken to study the
nuclear incompressibility and symmetry energy properties.  In
addition, new neutrino laboratories are planned, including UNO, that
will refine measurements of neutrino oscillations which could be
relevant for interpreting neutrino signals from supernovae.  These new
neutrino laboratories will also improve by an order of magnitude or
more, the number of neutrinos observed from a supernova, and will
increase the rate at which supernovae are observed as well.  New
experiments involving neutrino interactions with heavy nuclei could
shed light on neutrino absorption and scattering rates, quantifying
the effects of density, spin and isospin correlations for opacities of
dense matter.

Although we've not commented on recent developments in microscopic
theory, a topic to which Hans Bethe devoted much of his early career,
there have been many of interest for neutron stars.  Much effort has
gone into determining the EOS of neutron matter at subnuclear
densities which is consistent with phase shift data.  The effects of
Pauli blocking and density, spin and isospin correlations on neutrino
opacities have been explored.  New explorations of the pairing gaps of
nucleons and hyperons, the hyperon-nucleon interaction and the
threshold densities of hyperons have appeared.  Finally, new studies
 are targeting the properties
of pion and kaon condensates and delineating quark matter properties
including color compositions, gaps and critical densities.

\section*{Acknowledgements}

This work was supported in part by the U.S. Department of Energy under
grant DOE/DE-FG02-87ER-40317 (for JML) and DOE/DE-FG02-93ER40756 (for
MP).
                       


\begin{thebibliography}{3}
%

\bibitem[\protect\citeauthoryear{ (Witten 1984)}{}]{Witten84}
E. Witten, Phys. Rev. {\bf D30}, 272 (1984); 
E. Fahri \& R. L. Jaffe,
Phys. Rev. {\bf D30}, 2379 (1984)

\bibitem[\protect\citeauthoryear{ (Alford et al. 2006)}{}]{Alford06}
M. G. Alford, K. Rajagopal, S. Reddy, \& A. W. Steiner,
Phys. Rev. {\bf D73}, 4016 (2006)

\bibitem{Burrows86} A. Burrows \& J. M. Lattimer, 
ApJ {\bf 307}, 178 (1986)

\bibitem{Prakash01}M. Prakash et al., {\it Physics of Neutron Star
Interiors}, eds. D. Blaaschke, N. K. Glendenning and A. Sedrakian,
Springer-Verlag, New York, Lect. Notes. Phys. {\bf 578}, 364 (2001)

\bibitem[\protect\citeauthoryear{ (Rhoades \& Ruffini 1974)}{}]{RR74}
C. E. Rhoades Jr. \& R. Ruffini, 
Phys. Rev. Lett. {\bf 32}, 324 (1974)

\bibitem[\protect\citeauthoryear{ (Lattimer \& Prakash 2001)}{}]{LP01}
J. M. Lattimer \& M. Prakash, 
ApJ, {\bf 550}, 426 (2001)

\bibitem[\protect\citeauthoryear{ (Kalogera et al. 2004)}{}]{Kalogera04}
V. Kalogera et al., 
ApJL {\bf 614}, L137 (2004)

\bibitem[\protect\citeauthoryear{ (Lattimer \& Ravenhall 1978)}{}]{LR78}
J. M. Lattimer \& D. G. Ravenhall,  
ApJL {\bf 223}, L314 (1978)

\bibitem[\protect\citeauthoryear{ (Bethe et al. 1979)}{}]{BBAL} 
H. A. Bethe, G. E. Brown, J. Applegate \& J. M. Lattimer,
Nucl. Phys {\bf A324}, 487 (1979)

\bibitem[\protect\citeauthoryear{ (Lamb et al. 1978)}{}]{LLPR78}
D. Q. Lamb, J. M. Lattimer, C. J. Pethick \& D. G. Ravenhall, 
Phys. Rev. Lett. {\bf41}, 1623 (1978)

\bibitem{pabw} M. Prakash, T. L. Ainsworth, J. P. Blaizot and
H. Wolter, in {\em Windsurfing the Fermi Sea}, Proceedings of the
International Conference and Symposium on ``Unified Concepts of
Many-Body Problems'', Stony Brook, eds. T. T. S. Kuo
and J. Speth, 357 (1986)

\bibitem[\protect\citeauthoryear{ (Hartle \& Sabbadini 1977)}{}]{Hartle77}
J. B. Hartle  \& A. G. Sabbadini, 
ApJ {\bf 213}, 831 (1977)

\bibitem[\protect\citeauthoryear{ (Lindblom 1984)}{}]{Lindblom84}
L. Lindblom, 
ApJ {\bf278}, 364 (1984)

\bibitem[\protect\citeauthoryear{ (Glendenning 1992)}{}]{Glendenning92}
N. K. Glendenning, 
Phys. Rev. {\bf D46}, 4161 (1992)

\bibitem[\protect\citeauthoryear{ (Koranda, Stergioulas \& Friedman
1997)}{}]{Koranda97} 
S. Koranda, N. Stergioulas \& J. L. Friedman, ApJ {\bf 488}, 799 (1997); 
N. K. Glendenning, Phys. Rev. {\bf D46}, 4161 (1992)

\bibitem[\protect\citeauthoryear{ (Pandharipande \& Ravenhall
1989)}{}]{PR89} V. R. Pandharipande \& D. G. Ravenhall, in {\it
Proc. NATO Advanced Research Workshop on Nuclear Matter and Heavy Ion
Collisions}, ed. M. Soyeur et al., Plenum, New York, 103 (1989)

\bibitem[\protect\citeauthoryear{ (Lattimer \& Prakash 2005)}{}]{LP05}
J. M. Lattimer \& M. Prakash, 
Phys. Rev. Lett. {\bf94}, 1101 (2005)

\bibitem[\protect\citeauthoryear{ (Tolman 1939)}{}]{Tolman39}
R. C. Tolman, 
Phys. Rev. {\bf55}, 364 (1939)

\bibitem[\protect\citeauthoryear{ (Haensel \& Zdunik 1989)}{}]{Haensel89}
P. Haensel \& J.L. Zdunik, 
Nature {\bf340}, 617 (1989)

\bibitem[\protect\citeauthoryear{ (Friedman, Parker \& Ipser 1986)}{}]{FPI}
J. L. Friedman, L. Parker \& J. R. Ipser, 
ApJ {\bf 304}, 115 (1986);
J. M. Lattimer, M. Prakash, D. Masak \& A. Yahil, 
ApJL {\bf 355}, L241 (1990)


\bibitem[\protect\citeauthoryear{ (Lattimer \& Prakash 2004)}{}]{LP04}
J. M. Lattimer \& M. Prakash, 
Science {\bf 304}, 536 (2004)

\bibitem[\protect\citeauthoryear{ (Hessels et al. 2006}){}]{Hessels06}
J. W. T. Hessels et al., 
Science {\bf 311}, 1901 (2006)

\bibitem[\protect\citeauthoryear{ (Kaaret et al. 2006}){}]{Kaaret06}
P. Kaaret et al., astro-ph/0611716 (2006)

\bibitem[\protect\citeauthoryear{Clark et al.}{2002}]{Clark02}
J. S. Clark et al., 
A\&A {\bf 392}, 909 (2002) (a)

\bibitem[\protect\citeauthoryear{Barziv et al.}{2001}]{Barziv01}
O. Barziv et al.,
A\&A {\bf 377}, 925 (2001) (b)
\bibitem[\protect\citeauthoryear{ (Quantrel et al. 2003)}{}]{Quaintrell03}
H. Quaintrell et al.,
A\&A {\bf 401}, 303 (2003) (c)
%
\bibitem[\protect\citeauthoryear{Orosz \& Kuulkers}{1999}]{Orosz99}
J. A. Orosz \&  E. Kuulkers,
MNRAS {\bf 305}, 132 (1999) (d)
%

\bibitem[\protect\citeauthoryear{Thorsett \& Chakrabarty}{2001}]{Thorsett99}
S. E. Thorsett \&
D. Chakrabarty, ApJ {\bf 512}, 288 (1999) (e)

\bibitem[\protect\citeauthoryear{Nice et al.}{2004}]{Nice04}
D. J. Nice, E. M. Splaver, \& I. H. Stairs, IAU Symposium 218, 
eds. F. Camilo and B.-M. Gaensler, 49 (2004); astro-ph/0311296; 
also private communication (f)

\bibitem[\protect\citeauthoryear{Nice et al.}{2003}]{Nice03a}
D. J. Nice, E. M. Splaver
\& I. H. Stairs, ASP Conf. Ser. {\bf302}, {\it Radio Pulsars}, eds. M. Bailes,
D. J.  Nice and S. E. Thorsett, Ast. Soc. Pac., San Francisco, 75 (2003); astro-ph/0210637 (g)
%
\bibitem[\protect\citeauthoryear{Nice et al.}{2001}]{Nice01}
D. J. Nice, E. M. Splaver
\& I. H. Stairs, ApJ {\bf 549}, 516 (2001) (h)

\bibitem[\protect\citeauthoryear{Lyne et al.}{2004}]{Lyne04}
A. G. Lyne et al.,
Science {\bf 303}, 1153 (2004) (i)
%
\bibitem[\protect\citeauthoryear{Bailes et al.}{2003}]{Bailes03}
M. Bailes et al.,
ApJL {\bf 595}, L49 (2003) (j)

%
\bibitem[\protect\citeauthoryear{van Kerkwijk et al.}{1995}]{van95}
M. H. van Kerkwijk, J. van Paradijs \& E. J. Zuiderwijk,
A\&A {\bf 303}, 497 (1995) (k)
%

\bibitem[\protect\citeauthoryear{Gelino et al.}{2003}]{Gelino03}
D. M. Gelino, J. A. Tomsick \& W. A. Heindl, BAAS {\bf
  34}, 1199 (2003); J.A. Tomsick, private communication (l)

\bibitem[\protect\citeauthoryear{Langer et al.}{2001}]{Lange01}
Ch. Lange et al., MNRAS {\bf 326}, 274 (2001) (m)

\bibitem[\protect\citeauthoryear{Munoz-Darias et al.}{2003}]{Munoz05}
T. Mu$\tilde{\rm n}$oz-Darias, J. Casares \& I. G. Mart\'inez-Pais, ApJ {\bf
635}, 502 (2005) (n)
%
\bibitem[\protect\citeauthoryear{Jonker et al.}{2005}]{Jonker05}
P. G. Jonker, D. Steeghs, G. Nelemans \& M. van der Klis,
MNRAS {\bf356}, 621 (2005) (o)
%
\bibitem[\protect\citeauthoryear{van Stratan et al.}{2001}]{van01}
W. van Straten et al., Nature {\bf 412}, 158 (2001) (p)
%
\bibitem[\protect\citeauthoryear{ (Weisberg \& Taylor 2005)}{}]{Weisberg05}
J. M. Weisberg \& J. H. Taylor, ASP Conf. Ser. {\bf328}, {\it Binary
Radio Pulsars}, eds. F. A. Rasio and I. H. Stairs, Ast. Soc. Pac., San
Francisco, 25 (2005); astro-ph/0407149 (q)

\bibitem[\protect\citeauthoryear{Splaver et al.}{2005}]{Splaver05}
E. M. Splaver et al.,
ApJ {\bf620}, 405 (2005) (r)

\bibitem[\protect\citeauthoryear{Faulkner et al.}{2004}]{Faulkner04}
A. J. Faulkner et al., ApJL {\bf 618}, L119 (2004) (s)

\bibitem[\protect\citeauthoryear{Ransom et al.}{2005}]{Ransom05}
S. M. Ransom et al.,
Science {\bf307}, 892 (2005) (t)
%
\bibitem[\protect\citeauthoryear{Jacoby et al.}{2005}]{Jacoby05a}
B. A. Jacoby et al.,
ApJ {\bf629}, 113 (2005)  (u)
%
\bibitem[\protect\citeauthoryear{ (Nice et al. 2005)}{}]{Nice05}
D. J. Nice, et al.
ApJ {\bf634}, 1242 (2005) (v)
%
\bibitem[\protect\citeauthoryear{Freire et al.}{2003}]{Freire03}
P. C. Freire et al.,
MNRAS {\bf340}, 1359 (2003) (w)
%
\bibitem[\protect\citeauthoryear{Jacoby}{2005}]{Jacoby05b}
B. A. Jacoby, PhD dissertation, CalTech (x)
%
\bibitem[\protect\citeauthoryear{Splaver et al.}{2002}]{Splaver02}
E. M. Splaver et al., 
ApJ {\bf581}, 509 (2002) (y) 

\bibitem[\protect\citeauthoryear{Champion et al.}{2005}]{Champion05}
D. J. Champion  et al.,
MNRAS {\bf363}, 929 (2005) (z) 

\bibitem[\protect\citeauthoryear{Corongiu et al.}{2004}]{Corongiu04}
A. Corongiu  et al., 
Mem. S. A. It Suppl. {\bf5}, 188 (2004) (A) 

\bibitem[\protect\citeauthoryear{Lorimer et al.}{2006}]{Lorimer06} 
D. R. Lorimer et al., 
ApJ {\bf 640}, 428 (2006) (B) 

\bibitem[\protect\citeauthoryear{Nice et al.}{1999}]{Nice99} 
D. J. Nice, J. H. Taylor \& R. W. Sayer,  
{\em Pulsar Timing,
General Relativity, and the Internal Structure of Neutron Stars},
eds. E. Van den Heuvel, J. Van Paradijs and Z. Arzoumanian, Elsevier
(1999); http://pulsar.princeton.edu/ftp/pub/papers/ams-96.ps (C) 

\bibitem[\protect\citeauthoryear{Bethe \& Brown}{1998}]{Bethe98}
H. A. Bethe \& G. E. Brown, ApJ {\bf 506}, 780 (1998)

\bibitem{glenhyp} N. K. Glendenning, 
ApJ, {\bf 293}, 470 (1985); \\
{\em Compact Stars, Nuclear Physics, Particle Physics and Cosmology},
Springer, New York (1997)

\bibitem{kaphyp} J. Ellis, J. I. Kapusta \& K. A. Olive, 
Nucl. Phys. {\bf B348}, 345 (1991)

\bibitem{kpe} R. Knorren. M. Prakash \& P. J. Ellis, 
Phys. Rev. {\bf C52}, 3470 (1995)

\bibitem{schaffner} J. Schaffner \& I. N. Mishustin, 
Phys. Rev. {\bf C53}, 1416 (1996)

\bibitem{kapnel} D. B. Kaplan \& A. E. Nelson, 
Phys. Lett. {\bf B175}, 57 (1986);
{\bf B179}, 409 (1986)

\bibitem[\protect\citeauthoryear{ (Page 1995)}{}]{Page95}
D. Page, ApJ {\bf442}, 273 (1995)

\bibitem[\protect\citeauthoryear{ (Scott et al. 2000)}{}]{Scott00}
D. Scott, D. Leahy \& R. Wilson, ApJ {\bf539}, 392 (2000)

\bibitem[\protect\citeauthoryear{ (Leahy 2004)}{}]{Leahy04}
D. Leahy, ApJ {\bf613}, 517 (2004)

\bibitem{bro2} G. E. Brown, K. Kubodera, M. Rho \& V. Thorsson, 
Phys. Lett. {\bf B291}, 355 (1992)

\bibitem{pol} H. D. Politzer \& M. B. Wise, 
Phys. Lett. {\bf B273}, 156 (1991)

\bibitem{muto} T. Muto \& T. Tatsumi, 
Phys. Lett. {\bf B283}, 165 (1992);
T. Muto, Prog. Theor. Phys. {\bf89}, 415 (1993);
T. Muto et al, 
Prog. Theor. Phys. Suppl. {\bf112}, 221 (1993);
H. Fujii et al., Nucl. Phys, {\bf A571}, 758 (1994)

\bibitem{maruy} T. Maruyama, H. Fujii, T. Muto \& T. Tatsumi, 
Phys. Lett. {\bf B337}, 19 (1994)

\bibitem{bro3} G. E. Brown, C-H. Lee, M. Rho \& V. Thorsson, 
Nucl. Phys. {\bf A567}, 937 (1994)

\bibitem{bro4} G. E. Brown, C-H. Lee, H-J, Park \& M. Rho, 
Phys. Rev. Lett., {\bf 96}, 062303 (2006)

\bibitem{tpl} V. Thorsson, M. Prakash \& J. M. Lattimer, 
Nucl. Phys. {\bf A572}, 693 (1994)

\bibitem[\protect\citeauthoryear{ (Alford et al. 2005)}{}]{Alford05}
M. Alford, M. Braby, M. Paris \& S. Reddy, 
ApJ {\bf629}, 969 (2005)

\bibitem[\protect\citeauthoryear{ (Gondek et al. 1998)}{}]{Gondek98}
D. Gondek, P. Haensel \& J. L. Zdunik, Pacific Rim Conference
on Stellar Astrophyics, ed. K. L. Chan, K.S. Cheng and H.P. Singh, ASP
Conf. Ser. {\bf138}, 131 (1998)

\bibitem[\protect\citeauthoryear{ (Buchdahl 1967)}{}]{Buchdahl67}
H.-A. Buchdahl, ApJ {\bf147}, 310 (1967)

\bibitem[\protect\citeauthoryear{ (Blaizot 1980)}{}]{Blaizot80}
J. P. Blaizot, 
Phys. Rep. {\bf65}, 171 (1980)

\bibitem[\protect\citeauthoryear{ (Pearson 1991)}{}]{Pearson91}
J. M. Pearson, 
Phys. Lett. {\bf B271}, 12 (1991)

\bibitem{Krivine84}H. Krivine, 
J. de Phys. Supp.{\bf C6}, 153 (1984)

\bibitem{Lattimer96} J. Lattimer, {\it Nuclear Equation of State}, eds.
A. Ansari and L. Satpathy, World Scientific, Signapore, 83 (1996)

\bibitem[\protect\citeauthoryear{ (Link et al. 1999)}{}]{LEL} 
B. Link, R. I. Epstein  \& J. M. Lattimer, 
Phys. Rev. Lett. {\bf83}, 3362(1999)

\bibitem[\protect\citeauthoryear{ (Duncan 1998)}{}]{Duncan98}
R. C. Duncan, 
ApJL {\bf498}, L45 (1998)

\bibitem[\protect\citeauthoryear{ (Rutledge et al. 2001)}{}]{Rutledge01}
R. E. Rutledge et al., 
ApJ {\bf580}, 413 (2006)

\bibitem[\protect\citeauthoryear{ (Hartle 1967)}{}]{Hartle67}
J. B. Hartle, 
ApJ {\bf150}, 1005 (1967)

\bibitem[\protect\citeauthoryear{ (Lattimer \& Schutz 2005)}{}]{LS05}
J. M. Lattimer \& B. F. Schutz, 
ApJ {\bf629}, 979 (2005)

\bibitem[\protect\citeauthoryear{ (Kubis 2006)}{}]{Kubis06}
S. Kubis, 
astro-ph/0611740 (2006)

\bibitem[\protect\citeauthoryear{ (Lamb et al. 1983)}{}]{LLPR83}
D. Q. Lamb, J. M. Lattimer, C. J. Pethick \& D. G. Ravenhall, 
Nucl. Phys. {\bf A411}, 449 (1983)

\bibitem[\protect\citeauthoryear{ (Callen 1985)}{}]{Callen85}
H. B. Callen,  
{\it Thermodynamics}, John Wiley \& Sons, New York (1985)

\bibitem[\protect\citeauthoryear{ (Anderson \& Itoh
1975)}{}]{Anderson75} P. W. Anderson  \& N. Itoh,  
Nature {\bf256}, 25 (1975); 
M. Ruderman, ApJ {\bf203}, 213 (1976); 
D. Pines \& M. A. Alpar, Nature {\bf316}, 27 (1985)

\bibitem[\protect\citeauthoryear{ (Link 2006)}{}]{Link06}
B. Link,  
A\&A {\bf458}, 881 (2006)

\bibitem[\protect\citeauthoryear{ (Barat et al. 1983)}{}]{Barat83}
C. Barat et al., 
A\&A {\bf126}, 400 (1983)

\bibitem[\protect\citeauthoryear{ (Israel et al. 2005)}{}]{Israel05}
G. L. Israel et al., 
ApJL {\bf628}, L53 (2005)

\bibitem[\protect\citeauthoryear{ (Strohmayer \& Watts 2005)}{}]{SW05}
T. E. Strohmayer \& A. L. Watts, 
ApJL {\bf632}, L111 (2005)

\bibitem[\protect\citeauthoryear{ (Samuelsson \& Andersson 2006)}{}]{SA06}
L. Samuelsson  \& N. Andersson, 
submitted to MNRAS, astro/ph-0609265 (2006)

\bibitem[\protect\citeauthoryear{ (Lattimer et al. 1994)}{}]{LvRPP}
J. M. Lattimer, K. A. van Riper, M. Prakash \& Manju Prakash, 
ApJ {\bf 425}, 802 (1994)

\bibitem[\protect\citeauthoryear{ (van Riper 1988)}{}]{vanriper88}
K. A. van Riper, 
ApJ {\bf329}, 339 (1988)

\bibitem[\protect\citeauthoryear{ (Gnedin et al. 2001)}{}]{Gnedin01}
O. Y. Gnedin, D. G. Yakovlev  \& A. Y. Potekhin, 
MNRAS {\bf324}, 725 (2001)

\bibitem[\protect\citeauthoryear{ (Kaminker et al. 2006)}{}]{Kaminker06}
A. D. Kaminker et al., MNRAS {\bf371}, 477 (2006)

\bibitem[\protect\citeauthoryear{ (Jonker et al. 2006)}{}]{Jonker06}
P. G. Jonker, P.G. et al., 
MNRAS {\bf368}, 1803 (2006)

\bibitem[\protect\citeauthoryear{ (Kuulers et al. 2002)}{}]{Kuulkers02}
E. Kuulkers et al., 
A\&A {\bf382}, 503 (2002)

\bibitem[\protect\citeauthoryear{ (Wijnands et al. 2002)}{}]{Wijnands02}
R. Wijnands et al., 
ApJL {\bf560}, L159 (2002)

\bibitem[\protect\citeauthoryear{ (Walter \& Lattimer 2002)}{}]{Walter02}
F. M. Walter  \& J. M. Lattimer, 
ApJ {\bf576}, 145 (2002)

\bibitem[\protect\citeauthoryear{ (Walter \& Faherty
2006)}{}]{Walter06} F. M. Walter \& J. K. Faherty, 
 in {\it Isolated Neutron Stars}, 
ed. D. Page, R. Turolla \& S. Zane, ApSS, Springer (2006)

\bibitem[\protect\citeauthoryear{ (Gendre et al. 2003a)}{}]{Gendre03a}
B. Gendre, B. Barret \& N. A. Webb, 
A\&A {\bf403}, L11 (2003)

\bibitem[\protect\citeauthoryear{ (Gendre et al. 2003b)}{}]{Gendre03b}
B. Gendre, B. Barret \& N. A. Webb, 
 A\&A {\bf400}, 521 (2003)

\bibitem[\protect\citeauthoryear{ (Heinke et al. 2006)}{}]{Heinke06}
C. O. Heinke, G. B. , Rybicki, R. Narayan \& J. E. Grindlay, 
ApJ {\bf644}, 1090 (2006)

\bibitem[\protect\citeauthoryear{ (Lattimer et al. 1991)}{}]{LPPH91}
J. M. Lattimer, C. J. Pethick, M. Prakash \& P. Haensel, 
Phys. Rev. Lett. {\bf 66}, 2701 (1991)

\bibitem[\protect\citeauthoryear{ (Prakash et al. 1992)}{}]{PPLP92}
M. Prakash, M. Prakash, J. M. Lattimer, \& C. J. Pethick,  
ApJL {\bf390}, L77 (1992)

\bibitem[\protect\citeauthoryear{ (Page et al. 2004)}{}]{Page04}
D. Page, J. M. Lattimer, M. Prakash, \& A. W. Steiner, 
ApJS {\bf 155}, 623 (2004)

\bibitem[\protect\citeauthoryear{ (Slane et al. 2004)}{}]{Slane04}
P. Slane, D. J. Helfand, E. Van der Swaluw \& S. S. Murray, 
ApJ {\bf616}, 403 (2004)

\bibitem[\protect\citeauthoryear{ (Kaplan et al. 2004)}{}]{Kaplan04}
D. L. Kaplan et al., 
ApJS {\bf153}, 269 (2004)

\bibitem[\protect\citeauthoryear{ (Cottam, Paerls \& Mendez 2006)}{}]
{Cottam02} 
J. Cottam, F. Paerls \& M. Mendez, 
Nature {\bf 420}, 51 (2002)

\bibitem[\protect\citeauthoryear{ (Villarreal \& Strohmayer 2004)}{}]{Villarreal04}
A. R. Villarreal  \& T. E. Strohmayer, 
ApJL {\bf 614}, L121 (2004)

\bibitem[\protect\citeauthoryear{ (Sidoli et al. 2005)}{}]{Sidoli05}
L. Sidoli, A. N.  Parmar  \& T. Oosterbroek, 
A\&A {\bf 429}, 291 (2005)

\bibitem[\protect\citeauthoryear{ (Bhattacharyya et al. 2006)}{}]{Bhattacharyya06}
S. Bhattacharyya, M. C. Miller  \& F. K. Lamb, 
ApJ {\bf 644}, 1085 (2006)

\bibitem[\protect\citeauthoryear{ (Madej et al. 2004)}{}]{Madej04}
J. Madej, P. C. Joss \& A. Rozanska, 
ApJ {\bf602}, 904 (2004)

\bibitem[\protect\citeauthoryear{ (Ozel 2006)}{}]{Ozel06}
F. Ozel, Nature {\bf441}, 1115 (2006)

\bibitem[\protect\citeauthoryear{ (Van Paradijs 1981)}{}]{Van Paradijs81}
J. A. Van Paradijs, 
A\&A {\bf101}, 174 (1981)

\bibitem[\protect\citeauthoryear{ (Inogamov \& Sunyaev 1999)}{}]{Inogamov99}
N. A. Inogamov \& R. A. Sunyaev, Astron. Lett. {\bf25}, 269 (1999)

\bibitem[\protect\citeauthoryear{ (Suleimanov \& Poutanen 2006)}{}]{Suleimanov06}
V. Suleimanov \& J. Poutanen, MNRAS {\bf369}m 2036 (2006)

\bibitem[\protect\citeauthoryear{ (van der Klis 2006)}{}]{vanderklis06}
M. van der Klis,  in {\it Compact Stellar X-ray Sources},
eds. W.H.G. Lewin \& M. van der Klis (Cambridge: Cambridge
Univ. Press), in press (2006); astro-ph/0410551

\bibitem[\protect\citeauthoryear{ (Miller et al. 1998)}{}]{Miller98}
M. C. Miller, F. K. Lamb  \& D. Psaltis, 
ApJ {\bf508}, 791 (1998)

\bibitem[\protect\citeauthoryear{ (Stella \& Vietri
1998)}{}]{Stella98} 
L. Stella \& M. Vietri, 
 ApJL {\bf492}, L59 (1998); 
M. A. Abramowicz et al., PASJ {\bf55}, 567 (2003)

\bibitem[\protect\citeauthoryear{ (Barret et al. 2006)}{}]{Barret06}
D. Barret, J.-F. Olive \& M. C. Miller, 
MNRAS {\bf370}, 1140 (2006)

\bibitem[\protect\citeauthoryear{ (Casares et al. 2006)}{}]{Casares06}
J. Casares, 
MNRAS, in press (2006); astro-ph/0610086

\bibitem[\protect\citeauthoryear{ (Burgay et al. 2003)}{}]{Burgay03}  
M. Burgay et al., 
Nature {\bf426}, 531 (2003) 

\bibitem[\protect\citeauthoryear{ (Damour \& Schaefer 1988)}{}]{Damour88}
T. Damour \& G. Schaefer, 
Nuovo Cimento {\bf B101}, 127 (1988)

\bibitem[\protect\citeauthoryear{ (Bejger \& Haensel 2003)}{}]{Bejger03} 
M. Bejger \& P. Haensel, 
A\& A {\bf405}, 747 (2003) 

\bibitem[\protect\citeauthoryear{ (Fesen, Shull \& Hurford 1997)}{}]{Fesen97} 
R. A. Fesen, J. M. Shull  \& A. P. Hurford, 
ApJ {\bf113}, 354 (1997) 

\bibitem[\protect\citeauthoryear{ (Barker \& O'Connell 1975)}{}]{Barker75} 
B. M. Barker \& R. F. O'Connell, 
Phys. Rev. {\bf D12}, 329 (1975)


\bibitem{KJ95}W. Keil \& H-T Janka, 
A\&A {\bf 29}, 145 (1995)

\bibitem{burr99a}A. Burrows \& R. F. Sawyer, 
{Phys. Rev.} {\bf C59}, 510 (1999)

\bibitem{pons99}J. A. Pons, S. Reddy, M. Prakash \& J. M. Lattimer, 
J. A. Miralles, 
ApJ {\bf 513}, 780 (1999)

\bibitem{Pon00a} 
J. A. Pons, J. A. Miralles, M. Prakash \& J. M. Lattimer,  
ApJ {\bf 553} (2001) 382. 

\bibitem{Pon01b}
J. A. Pons, A. W. Steiner, M. Prakash \& J. M. Lattimer, 
{Phys. Rev. Lett.} {\bf 86}, 5223 (2001)

\bibitem{AIP}C. K. Jung, {\it Next Generation Nucleon Decay and Neutrino
Dectector}, ed. N. Diwan and C.K. Jung, AIP, New York, 29 (2000)

\bibitem{supernova} A. Burrows, 
{Ann. Rev. Nucl. Sci.} {\bf 40}, 181 (1990)

\bibitem{prak97a}M. Prakash et al.,
{Phys. Rep.} {\bf 280}, 1 (1997)

\bibitem{Ellis96}P. J. Ellis, J. M. Lattimer \& M. Prakash,  
{Comm. Nucl. Part. Phys.} {\bf 22}, 63 (1996);
M. Prakash, J. R. Cooke \& J. M. Lattimer, 
{Phys. Rev. D } {\bf 52}, 661 (1995);
A. W. Steiner, M. Prakash \& J. M. Lattimer, 
{Phys. Lett.} {\bf B486}, 239 (2000)

\bibitem{Burrows88}A. Burrows, ApJ {\bf 334}, 891 (1988)

\bibitem{reddyt}S. Reddy, PhD thesis, Stony Brook University (1998)

\bibitem{convec}
A. Burrows \& B. Fryxell, ApJL {\bf 418}, L33 (1993);
M. Herant et al., ApJ {\bf 435}, 339 (1994);
W. Keil, H.-T. Janka \& E. M\"uller, ApJL {\bf 473}, L111 (1995);
A. Mezzacappa et al., ApJ {\bf 495}, 911 (1998);
J. A. Miralles, J. A. Pons \& V. A. Urpin, ApJ {\bf 543}, 1001 (2000)

\bibitem{burr98}A. Burrows \&  R. F. Sawyer, 
{Phys. Rev. C} {\bf 58}, 554 (1998);
S. Reddy, M. Prakash, J. M. Lattimer \& J. A. Pons, 
{Phys. Rev. C } {\bf 59}, 2888 (1999)

\bibitem{Brown94} G. E. Brown \& H. A. Bethe, 
ApJ {\bf 423}, 659 (1994)

\bibitem{pcl95} M. Prakash, J. R. Cooke \& J. M. Lattimer, 
Phys. Rev. {\bf D52}, 661 (1995)
%

\bibitem{THM90BB95}
F. K. Thielemann, M. Hashimoto \& K. Nomoto, 
ApJ {\bf 349}, 222 (1990);
H. A. Bethe \& G. E. Brown, ApJL {\bf 445}, L129 (1995)
 
\bibitem{CR00}  G. W. Carter \& S. Reddy, 
{Phys. Rev.} {\bf D62}, 103002 (2000)


\bibitem{pac}B. Paczynski, 
Acta Astron. {\bf 42}, 145 (1992)

\bibitem{th0} C. Thompson \& R. C. Duncan, 
MNRAS {\bf 275}, 255 (1995)

\bibitem{th1} C. Thompson \& R. C. Duncan, 
ApJ {\bf 473}, 322 (1996)

\bibitem{Mel}A. Melatos, ApJL {\bf 519}, L77 (1999)

\bibitem{Card} C. Y. Cardall, M. Prakash \& J. M. Lattimer, 
ApJ {\bf 554}, 322 (2001)

\bibitem{k} C. Kouveliotou et al., Nature {\bf 393}, 235 (1998)

\bibitem{dn0} R. C. Duncan \& C. Thompson, ApJL {\bf 392}, L9 (1992)

\bibitem{dn1} R. C. Duncan \& C. Thompson, ApJ {\bf 469}, 764 (1996)

\bibitem{ls} D. Lai \& S. L. Shapiro, ApJ {\bf 383}, 745 (1991)

\bibitem{cv}V. Canuto \& J. Ventura, 
Fund. Cosmic Phys. {\bf 2}, 203 (1977)

\bibitem{fgp}I. Fushiki, E. H. Gudmundsson \& C. J. Pethick, 
ApJ {\bf 342}, 958 (1989); 
I. Fushiki, E. H. Gudmundsson, C. J. Pethick \&
J. Yngvason, Ann. Phys. {\bf 216}, 29 (1992);
A. M. Abrahams \& S. L. Shapiro, 
ApJ {\bf 374}, 652 (1991)

\bibitem{rfgpy} \"O. E. R\"{o}gnvaldsson, I. Fushiki, 
E. H. Gudmundsson, C. J. Pethick \& J. Yngvason, J.,
ApJ {\bf 416}, 276 (1993);
A. Thorlofsson, \"O. E. R\"{o}gnvaldsson, J. Yngvason \& E. H. Gudmundsson,
ApJ {\bf 502}, 847 (1998)

\bibitem{Ch0} S. Chakrabarty, 
Phys. Rev. {\bf D54}, 1306 (1996)

\bibitem{ch:dense} S. Chakrabarty, D. Bandyopadhyay \& S. Pal, 
Phys. Rev. Lett. {\bf 78}, 2898 (1997)

\bibitem{yz} Y. F. Yuan \& J. L. Zhang, 
ApJ {\bf 525}, 920 (1999)

\bibitem{boc} M. Boucquet, S. Bonozzola, F. Gourgoulhon \& J. Novak, 
A\&A {\bf 301}, 757 (1995)

\bibitem{bpl00} A. Broderick, M. Prakash \& J. M. Lattimer, 
ApJ {\bf 537}, 351 (2000)

\bibitem{bpl02} A. Broderick, M. Prakash \& J. M. Lattimer, 
Phys. Lett {\bf B531}, 167 (2002)

\bibitem{llp} L. D. Landau, L.M. Lifshitz \& L. P. Pitaevski\u{i}, 
{\it Electrodynamics of Continuous Media, 2nd ed.}, Pergamon, New York (1984)

\bibitem{BH:magsus} R. D. Blandford \& L. Hernquist, 
J. Phys. C: Solid State Phys. {\bf 15}, 6233 (1982)

\bibitem{Schw} J. Schwinger, {\em Particles, Sources and Fields},
  Vol. 3, Addison-Wesley, Redwood
  City, 164 (1988)

\bibitem{Rud95} M. Ruderman, {\it Millisecond Pulsars: A Decade of
Surprise}, eds. A. S. Fruchter, M. Tavani, and D. C. Backer, 
ASP Conf. Ser. {\bf72}, Ast. Soc. Pac., San
Francisco, 277 (1995)


\bibitem{THORNE1} K. S. Thorne, {\it Three Hundred Years of
  Gravitation}, ed. {S. W. Hawking and W. Israel}, Cambridge
  Univ. Press, Cambridge, Ch. 9 (1973).

\bibitem{LSCH2} J. M. Lattimer \& D. N. Schramm, 
ApJ {\bf 210}, 549 (1976)

\bibitem{EICHLER89} {D. Eichler, M. Livio, T. Piran \& D. N. Schramm,}
Nature {\bf 340}, 126 (1989)

\bibitem{FABER04} J. A. Faber, P. Grandclement \& F. A. Rasio,
Phys. Rev. {\bf D69}, 14036 (2004)

\bibitem{CLARK1} J. P. A. Clark, 
Astrophys. Lett. {\bf 18}, 73 (1977)

\bibitem{Jaranowski92} P. Jaranowski \& A. Krolak, 
ApJ {\bf 394}, 586 (1992)

\bibitem{PORTEGIES98b}S. F. Portegies Zwart, 
ApJ {\bf 503}, 53 (1998)

\bibitem{LATTIMER77} J. M. Lattimer, F. Mackie, D. G. Ravenhall \&
  D. N. Schramm, 
ApJ {\bf 213}, 225 (1977)

\bibitem{MEYER89} B. S. Meyer, ApJ {\bf 343}, 254 (1989)

\bibitem{COLPI93} M. Colpi, S. L. Shapiro \& S. A. Teukolsky, 
 ApJ {\bf 414}, 717 (1993)

\bibitem{RPL05}S. Ratkovic, M. Prakash \& J. M. Lattimer, 
submitted to ApJ, astro-ph/0512133 (2005)

\bibitem[\protect\citeauthoryear{ (Peters 1964)}{}]{Peters64}
P. C. Peters, Phys. Rev. {\bf B136}, 1224 (1964)

\bibitem[\protect\citeauthoryear{ (Kopal 1959)}{}]{Kopal59}
Z. Kopal, {\it Close Binary Systems}, Wiley, New York (1959)

\bibitem[\protect\citeauthoryear{ (Eggleton 1983)}{}]{Eggleton83}
P. P. Eggleton, ApJ {\bf268}, 368 (1983)

\bibitem{Alcock}C. Alcock \& A. Olinto, 
Ann. Rev. Nucl. Part. Sci., {\bf 38}, 161 (1988)

\bibitem{PBP90} M. Prakash, E. Baron \& Manju Prakash, 
Phys. Lett. B. {\bf 243}, 175 (1990)

\bibitem{JMad} J. Madsen, 
J. Phys. G. {\bf 28}, 1737 (2002)

\bibitem{CUTLER94}C.Cutler \& E. E. Flanagen, 
Phys. Rev. {\bf D49}, 2658 (1994)

\bibitem{FABER02}J. A. Faber, P. Grandclement, F. A. Rasio \& K. Taniguchi,
Phys. Rev. Lett. {\bf 89}, 231102 (2002)



\bibitem[\protect\citeauthoryear{ (Bethe \& Bacher)}{}]{bvw}
H. A. Bethe \& R. F. Bacher, Rev. Mod. Phys. {\bf8}, 82 (1936);
C. F. von Weiz\" acker, Z. Physik {\bf96}, 431 (1935)

\bibitem{Myers69} W. D. Myers \& W. J. Swiatecki,
Ann. Phys. {\bf 55}, 395 (1969) 

\bibitem{Lattimer85} J. M. Lattimer, C. J. Pethick, D. G. Ravenhall 
\& D. Q. Lamb, 
Nucl. Phys. {\bf A 432}, 646 (1985). 

\bibitem{Steiner05} A. W. Steiner, M. Prakash, J. M. Lattimer \& 
P. J. Ellis,
Phys. Rep. {\bf 411}, 325 (2005)

\bibitem{Danielewicz03} P. Danielewicz,
Nucl. Phys. {\bf A 727}, 233 (2003)

\bibitem{nmasses}M. Hausmann et al., 
Hyperfine Interactions, {\bf 132}, 289 (2001); 
D. Lunney, J. M. Pearson \& C. Thibault, 
Rev. Mod. Phys. {\bf 75}, 1021 (2003); 
U Hager et al.,  
Phys. Rev. Lett. {\bf 96}, 042304 (2006)  


\bibitem{shlomo06} S. Shlomo, V.M. Kolomietz \& G. Colo, 
Eur. Phys. J. {\bf A30}, 23 (2006); 
T. Sil, S. Shlomo, B. K. Agarwal \& P.-G. Reinhard,
Phys. Rev. C {\bf 73}, 034316 (2006); 
B. K. Agarwal, S. Shlomo \& V. Kim Au, 
Phys. Rev. C {\bf 72}, 014310 (2005);
B. K. Agarwal, S. Shlomo \& V. Kim Au, 
Phys. Rev. C {\bf 68}, 031304 (2003)
\bibitem{Lipparini89}E. Lipparani \& S. Stringari,
Phys. Rep. {\bf 103}, L175 (1989)

\bibitem{Bergere77} R. Bergere, in {\it Lecture Notes in Physics} {\bf
61}, eds. S. Costa and C. Scharf, Springer, Berlin, 1 (1977)

\bibitem{Laszewski79} R. M. Laszewski \& P. Axel, 
Phys. C {\bf 19}, 342 (1979)

\bibitem{Bohigas81} O. Bohigas, N. Van Giai \& D. Vautherin,
Phys. Lett. {\bf B 102}, 105 (1981)

\bibitem{Lipparini82} E. Lipparini \& S. Stringari, 
Phys. Lett. {\bf B 112}, 421 (1982)

\bibitem{Krivine82} H. Krivine, C. Schmit \& J. Treiner, 
Phys. Lett {\bf B 112}, 281 (1982)

\bibitem{Lipparini88} E. Lipparini \& S. Stringari, 
Nucl. Phys. A {\bf 482}, 205c (1988)


\bibitem{Khamerdzhiev97} S. Khamerdzhiev, J. Speth \& G. Tertychny, 
Nucl. Phys. {\bf A 624}, 328 (1997)



\bibitem{Clark03} B. C. Clark, L. J. Kerr \& S. Hama,
Phys. Rev. {\bf C 67}, 054605 (2003)

\bibitem{Trzcinska01} A. Trzcinska et al., 
Phys. Rev. Lett. {\bf 87}, 082501 (2001)

\bibitem{Karataglidis02} S. Karataglidis, K. Amos \& B. A. Brown, 
Phys. Rev. {\bf C 65}, 044306 (2002)

\bibitem[\protect\citeauthoryear{ (Horowitz et al. 2001)}{}]{Horowitz01}
C. J. Horowitz, S. J. Pollock, P. A. Souder \& R. Michaels, R., 
Phys. Rev. {\bf C63}, 025501 (2001)

\bibitem{Michaels00} R. Michaels, P. A. Souder \& G. M. Urciuoli, 
Jefferson Laboratory Proposal {\bf PR-00-003} (2000).

\bibitem{Brown00} B. A. Brown, 
Phys. Rev. Lett. {\bf 85}, 5296 (2000)

\bibitem{Typel01} S. Typel \& B. A. Brown, 
Phys. Rev. {\bf C 64}, 027302 (2001)

\bibitem{Horowitz01b} C. J. Horowitz \& J. Piekarewicz, 
Phys. Rev. Lett. {\bf 86}, 5647 (2001)


\bibitem{Gutbrod89} H. H. Gutbrod, A. M. Poskanzer \& H. G. Ritter, 
Rep. Prog. Phys. {\bf 52}, 267 (1989)

\bibitem{Danielewicz85} P. Danielewicz \& G. Odyniec, 
Phys. Lett. {\bf 157}, 146 (1985)

\bibitem{Gustafsson84}H. A. Gustafsson et al., 
Phys. Rev. Lett. {\bf 52}, 1590 (1984)

\bibitem{Welke88} G. M. Welke, M. Prakash, T. T. S. Kuo \& S. Das Gupta, 
Phys. Rev. {\bf C 38}, 2101 (1988)

\bibitem{Siemens79} P. Siemens \& J. O. Rasmussen, 
Phys. Rev. Lett. {\bf 42}, 880 (1979)

\bibitem{Bertsch88} G. F. Bertsch \& S. Das Gupta, 
Phy. Rep. {\bf 160}, 189 (1988)

\bibitem{Gale90} C. Gale et al.,
Phys. Rev. {\bf C 41}, 1545 (1990)

\bibitem{Youngblood99} D. H. Youngblood, H. L. Clark \& Y.-W. Lui, 
Phys. Rev. Lett. {\bf 82}, 691 (1999)

\bibitem{Danielewicz02} P. Danielewicz, R. Lacey \& W. G. Lynch,
Science {\bf 298}, 1592 (2002)

\bibitem{Das03} C. B. Das, S. Das Gupta \& C. Gale, 
Phys. Rev. {\bf C 67}, 034611 (2003)

\bibitem{Li04} B.-A. Li, C. B. Das Gupta \& C. Gale, 
Phys. Rev. {\bf C 88}, 192701 (2004)

\bibitem{Li04b} B.-A. Li, C. B. Das \& S. Das Gupta,
Nucl. Phys. {\bf A 735}, 563 (2004) 

\bibitem{Chomaz04} P. Chomaz, M. Colonna \& J. Randrup, 
Phys. Rep. {\bf 389}, 263 (2004)

\bibitem{Li97} B. A. Li \& C. M. Ko,
Nucl. Phys. {\bf A 618}, 498 (1997)

\bibitem{Xu00} H. S. Xu et al., 
Phys. Rev. Lett. {\bf 85}, 716. (2000)

\bibitem{Tsang01} M. B. Tsang et al.,
Phys. Rev. Lett. {\bf 86}, 5023 (2001) 

\bibitem{Ono03} A. Ono et al.,
Phys. Rev. {\bf C 68}, 051601 (2003)

\bibitem{Das04}C. B. Das et al., Phys. Rep. {\bf 406}, 1 (2005)

\bibitem{Tsang04} M. B. Tsang et al.,
Phys. Rev. Lett. {\bf 92}, 062701 (2004)

\bibitem{Shetty04} D. V. Shetty et. al.,
Phys. Rev. {\bf C70}, 011601 (2004) 

\bibitem{Chen05} L. W. Chen, C. M. Ko \& B.-A. Li, 
Phys. Rev. Lett. {\bf 94}, 032701 (2005)

\bibitem{akmal98} A. Akmal, V. R. Pandharipane \& D. G. Ravenhall,
Phys. Rev. {\bf C58}, 1804. (1998)

\bibitem{Rami00} F. Rami et al., 
Phys. Rev. Lett. {\bf 84}, 1120 (2000)

\bibitem{Steiner05a} A. W. Steiner \& B.-A. Li,
Phys. Rev. {\bf C 72}, 041601 (2005)

\bibitem{Li05} B.-A. Li \& A. W. Steiner,
Phys. Lett. {\bf B642}, 436 (2006) 

\end{thebibliography}
\end{document}